\documentstyle[aps,prd,epsfig,euler,12pt]{revtex}

\newcommand{\DD}{{\bf D}}
\newcommand{\mbold}[1]{\mbox{\boldmath $#1$}}

\newcommand{\half}{{\textstyle{\frac{1}{2}}}}
\newcommand{\third}{{\textstyle{\frac{1}{3}}}}
\newcommand{\forth}{{\textstyle{\frac{1}{4}}}}

\newcommand{\imag}{{\rm i\hspace{0.13ex}}}
\newcommand{\trc}{{\rm Tr\hspace{0.2ex}}}
\renewcommand{\Re}{{\rm Re \hspace{0.2ex}}}

\newcommand{\ket}[1]{| #1 \rangle}
\newcommand{\bgeq}{\begin{equation}}
\newcommand{\bgeqa}{\begin{eqnarray}}
\newcommand{\edeq}{\end{equation}}
\newcommand{\edeqa}{\end{eqnarray}}
\newcommand{\void}[1]{\multicolumn{#1}{c}{---}}

\newcommand{\ainv}{a^{-1}}
\newcommand{\tmin}{t_{\rm min}}

\newcommand{\lqcd}{\Lambda_{\rm QCD}}
\newcommand{\gsmear}[1]{\phi_{{\rm G},#1}}
\newcommand{\hsmear}[2]{\phi_{{\rm H{}#1},#2}}
\newcommand{\nrqcdord}[1]{{\cal O}(({\lqcd}/{m_Q})^{#1})}

\newcommand{\PRD}[1]{Phys.\ Rev.\ \textbf{D{}#1}}

\begin{document}
\title{\rule{0in}{1in}%
Scaling of the $B$ and $D$ meson spectrum in lattice QCD}
\author{Joachim Hein\thanks{Member of the UKQCD
collaboration}}
\address{Newman Laboratory of Nuclear Studies, Cornell University,
Ithaca, NY 14853, USA}
\author{Sara Collins$^*$, Christine T.H.~Davies$^*$}
\address{Dept.\ of Physics \& Astronomy, University of Glasgow, Glasgow G12 8QQ, Scotland, UK} %
\author{Arifa Ali Khan}
\address{Center for Computational Physics, University of Tsukuba, Ibaraki 305-8577, Japan}
\author{Harry Newton$^*$}
\address{Dept.\ of Physics \& Astronomy, University of Edinburgh, Edinburgh EH9 3JZ, Scotland, UK}
\author{Colin Morningstar}
\address{Physics Department, Florida International University, Miami, FL 33199, USA}
\author{Junko Shigemitsu}
\address{Physics Department, The Ohio State University, Columbus, OH 43210, USA}
\author{John Sloan\thanks{Present address: Spatial Technologies, Boulder, CO, USA} }
\address{Dept.\ of Physics \& Astronomy, University of Kentucky, Lexington, KY 40506, USA}

\maketitle 

\centerline{\today}

\makebox[0in][l]{
 \raisebox{5.9in}[0in][0in]{
  \parbox[t]{3in}{\textbf{\textsf{%
\noindent%
CLNS 00-1665\\[-1ex]
GUTPA/98-12-1\\[-1ex]
OHSTPY-HEP-T-99-021\\[-1ex]
UTCCP-P-78\\[-1ex]
hep-ph/0003130
}}}}}

\begin{abstract}
We give results for the $B$ and the $D$ meson spectrum using NRQCD on
the lattice in the quenched approximation. The masses of radially and
orbitally excited states are calculated as well as $S$-wave hyperfine
and $P$-wave fine structure. Radially excited $P$-states are observed
for the first time. Radial and orbital excitation energies match well
to experiment, as does the strange-non-strange $S$-wave splitting. We
compare the light and heavy quark mass dependence of various
splittings to experiment. Our $B$-results cover a range in lattice
spacings of more than a factor of two.  Our $D$-results are from a
single lattice spacing and we compare them to numbers in the
literature from finer lattices using other methods. We see no
significant dependence of physical results on the lattice spacing.

\end{abstract}

\newpage
\section{Introduction} 

Mesonic bound states consisting of a single heavy quark, $b$ or $c$,
and a light quark, $u$, $d$ or $s$, as well as gluons, provide an
interesting laboratory to study strong interactions. The typical
momentum within such states is much lower than the mass of the heavy
quark. This leads to a situation where the heavy quark becomes
non-relativistic and the properties of the bound state are essentially
determined by the light quark and the glue. At leading order the
splittings within the spectrum become independent of the properties of
the heavy quark, such as its mass $m_Q$ and spin $s_Q$, so that
orbital and radial excitation energies are expected to match between
the $B$-system and the $D$-system.  The resulting approximate
SU($2N_h$) symmetry, with $N_h$ denoting the number of heavy flavours,
is usually referred to as {\em heavy quark symmetry},
see~\cite{neubert} and the references therein.  At the next order,
$1/m_Q$ effects give rise to fine structure in the spectrum, several
times larger in the $D$-system than for the $B$, see e.g.~\cite{cschlad}
for a review.

The spectrum of $B$ and $D$ states is not yet well established experimentally
\cite{pdg} although several new results have been reported recently
\cite{bradref1,bradref2,Dradial,D1wide,D1wide2}. 
Here we study the spectrum theoretically and from first principles
using lattice QCD. This will aid the experimental search for new
states.  In the case of well-established states it will provide a test
for the theory and/or the systematic errors in our calculation.  Of
key interest are decay matrix elements for $B$-factory experiments.
Knowing how well the spectrum has been obtained gives confidence that
we understand how to simulate $B$ and $D$ mesons reliably. This is
important for the analysis of systematic errors in matrix element
determinations.

To formulate heavy $b$ and $c$ quarks on the lattice, a na\"{\i}ve
discretisation is inappropriate since the lattice spacings currently
available are not small compared to the Compton wave length of those
quarks ($m_Qa > 1$). Presently there are two different formulations
available to simulate heavy quarks, non-relativistic QCD (NRQCD)
\cite{thalep,LepD46} and the heavy Wilson approach
\cite{fermilab}. For the $b$-quark on present lattices both approaches
become essentially the same. However, in this regime, NRQCD is to be
preferred since the inclusion of higher order correction terms is
easily implemented.

In this publication we report on our calculations of the $B$-meson
spectrum for two different values of the lattice spacing $a$. Together
with the results of \cite{arifalat98}, which were obtained with the
same methods at another value of the lattice spacing, we can
investigate the dependence on $a$ of our results.  Physical
results must be independent of $a$ and hence we can perform a test of
systematic errors inherent in our calculation.
We find no such errors at a
significant level.  In addition, on our coarsest lattice, we were able
to simulate the $D$-meson spectrum and compare to results using heavy
Wilson methods on finer lattices (where NRQCD does not work well since
$am_c < 1$).  Early results on our coarse lattice have already been
published in \cite{jhlat97}.

Section~\ref{simdetsec} gives details of the simulations we performed
and section~\ref{simmasssec} gives details of our fit
procedure. Section~\ref{qmasssection} gives our determination of the
bare $b$ and $c$ quark masses. Section~\ref{splitsec} discusses the
behaviour of the splittings in the spectrum that we obtain. This
includes fits to the dependence of the splittings on the mass of the
heavy quark.  Section~\ref{spectrum_section} compares the results in
physical units at different values of the lattice spacing and with
previous results as well as with experiment. Readers interested in our
results for the physical meson spectrum could jump directly to this
section.  Section~\ref{conclusec} contains our conclusions and our
best estimate for the $B$-spectrum, based on the combined input from
three different values of the lattice spacing.

\section{Simulation details}\label{simdetsec}

\subsection{Gauge field action}
Our calculation was performed on two sets of gauge field
configurations, which were generated using the Wilson gauge action
\bgeq
S_G = \beta \sum_{x,\mu<\nu} [1-\third \Re \trc 
(U_{x,\nu}U_{x+\hat\nu,\mu}U^+_{x+\hat\mu,\nu}U^+_{x,\mu})] \,.
\edeq
This action has lattice artifacts of ${\cal O}(a^2)$.
For the bare gauge coupling $\beta$, we used $5.7$ and $6.2$. The
lattice volumes and the number of configurations are given in
table~\ref{gpartab}. We will refer to these configurations by their
respective $\beta$-values. 

\subsection{Light quark propagators}
The light quark propagators have been generated with the use of the
Sheikholeslami-Wohlert action, also known as the clover action
\cite{swaction},
\bgeqa
S_L &=& a^4 \sum_x \bigg[\bar \psi_x \psi_x 
      + \kappa \sum_\mu \left[\bar\psi_{x-\hat\mu}(\gamma_\mu-1)
             U_{x-\hat\mu,\mu}\psi_{x}
        - \bar\psi_{x+\hat\mu}(\gamma_\mu+1)
             U^+_{x,\mu}\psi_{x}\right]\nonumber\\
&& \hspace{10ex}-a\half\imag c_{\rm sw}\kappa\sum_{\nu,\rho}\bar\psi_x 
       F_{\nu\rho,x} \sigma_{\nu\rho}\psi_x\bigg]\,.
\edeqa
On the configuration set with $\beta = 5.7$ the clover coefficient
$c_{\rm sw}$ is set to its tadpole-improved tree level value $c_{\rm
sw}= 1.5667$, as determined from the 4th root of the plaquette
\cite{tadpole}.  This reduces the lattice spacing artifacts in the
light quark propagators to ${\cal O}(\alpha_s a,a^2)$.  At
$\beta = 6.2$ we used the non-perturbative determined value of $c_{sw}
= 1.6138$, which  removes the ${\cal O}(\alpha_s a)$ artifacts
from the light quark propagator as well\cite{nonpert}. 

In reference~\cite{rowland} the light hadron spectrum at $\beta = 6.2$
has been calculated using the non-perturbative as well as the
tadpole-improved tree level value for $c_{\rm sw}$. 
No significant differences in the meson
and baryon spectrum could be resolved between the two values of
$c_{\rm sw}$. From this we expect the difference between tadpole and
non-perturbatively improved light quarks at $\beta=6.2$ to be well
covered by the size of the statistical errors in our case as well.
This allows us to compare our $\beta = 6.2$ results to the
tadpole-improved results at $\beta =5.7$ and in
reference~\cite{arifalat98}.

For each value of $\beta$ we used 3 different values for the hopping
parameter $\kappa$. The actual values are detailed in
table~\ref{kappa_tab}. The table also contains the values of
$\kappa_c$ and $\kappa_s$ from the UKQCD collaboration
\cite{rowland,hpsprd55} used in our
calculation. The use of these values is appropriate for the analysis
in terms of chiral extrapolations and scale setting that we have
done. We also carefully include systematic errors from different
chiral extrapolations and associated uncertainties in setting the
scale. A recent re-analysis by UKQCD of their light hadron spectrum
\cite{Ukqcdlhw} gives somewhat different values for $\kappa_c$
and $\kappa_s$. Our errors encompass any changes this would produce in
our physical results.

\subsection{Heavy quark propagators}
The typical momentum scale inside a heavy light meson such as a $B$ or
$D$ meson is of the ${\cal O}(\lqcd)$, which is small compared to the
mass of the heavy quark. Therefore the mass of the heavy quark $m_Q$
represents an irrelevant scale for the dynamics of the mesonic bound state
and it is possible to simulate these states on lattices with a
lattice spacing larger than the Compton wavelength of the heavy quark.

In our simulation we use a non-relativistic expansion of the 
heavy quark Hamiltonian, which is known as NRQCD
\cite{thalep,LepD46}. 
\begin{mathletters}
\label{hamilton}
\bgeqa 
H &=& H_0 + \delta H\,,\\[1ex] \label{kinhamilton}
H_0 &:=& - \frac{\DD^2}{2m_Q}\,,\\[1ex]
\delta H &:=& -c_4 \frac{g}{2m_Q}{\mbold{\sigma}\cdot \mbold{B}} +
c_2 \frac{\imag g}{8m_Q^2}(\DD\cdot\mbold{E} - \mbold{E}\cdot \DD) 
- c_3 \frac{g}{8m_Q^2}\mbold{\sigma}\cdot(\DD \times \mbold E - \mbold E
\times \DD)  \nonumber \\
&&-c_1 \frac{(\DD^2)^2}{8m_Q^3}+c_5a^2 \frac{\DD^{(4)}}{24m_Q}
-c_6a\frac{(\DD^2)^2}{16nm_Q^2}\,.\label{deltah}
\edeqa
\end{mathletters}
Please note that the rest mass term of $H$ has been omitted, resulting
in a shift of the Hamiltonian, which is discussed in
section~\ref{qmasssection}.  In the case of a heavy-light meson the NRQCD
expansion has to be organised in powers of $\lqcd/m_Q$
\cite{nrqcdpower}. Here this expansion is used up to 
${\cal O}((\lqcd/m_Q)^2)$. We also include the $\mbold{p}^4$ term,
which is believed to be the leading term in ${\cal O}((\lqcd/m_Q)^3)$. 
The last two terms correct for discretisation
errors from finite lattice spacing in respectively the spatial and
temporal derivatives. $n$ is a stability parameter used in the
evolution equation~(\ref{evolution}).  The matching coefficients
$c_1$,
\dots, $c_6$ are set to their tadpole-improved tree level values 
\cite{tadpole}.

With the Hamiltonian $H$ and $\delta H$ the propagator of the heavy
quark can be obtained from a Schr\"odinger-type evolution equation
\begin{mathletters}
\label{evolution}
\bgeqa
G_{t+1}&=& \left(1-a \half \delta H\right)
\left(1-a {\textstyle\frac{1}{2n}}H_0\right)^n U^+_4
  \left(1-a {\textstyle\frac{1}{2n}}H_0\right)^n \left(1-a \half
  \delta H\right)G_{t} \mbox{\hspace*{3ex}for\ }t>1\,,\\
G_1 &=& \left(1-a \half \delta H\right)
\left(1-a {\textstyle\frac{1}{2n}}H_0\right)^n U^+_4
  \left(1-a {\textstyle\frac{1}{2n}}H_0\right)^n \left(1-a \half
  \delta H\right)\phi_x \,.
\edeqa
\end{mathletters}
With $\phi_x$ we denote the source smearing function used on the
initial time slice. At $\beta=5.7$ we use 20 different values for
$m_Q$ in the range $0.6 \le am_Q \le 20.0$ and at $\beta=6.2$ we use
10 values in the range $1.1 \le am_Q \le 6.0$. Details, including the
$n$ values, are given in the tables~\ref{hmasstab57} and
\ref{hmasstab62}.  For each value of $\beta$ we performed 3 different
runs.  At $\beta = 5.7$ we label them A, C and S; for $\beta = 6.2$
they are labeled H, N and P.

For the $S$-wave mesons at $\beta=5.7$ we used up to three different
smearing functions, $\gsmear{0}$, $\gsmear{1}$ and $\gsmear{2}$, in the
different runs. These are convolutions of Gaussian functions for the
light and the heavy quark with radii as detailed in
table~\ref{smeartab}. The configurations were fixed to Coulomb gauge. A
local sink will be denoted with $\phi_{\rm L}$. In most cases our
final $\beta=5.7$ results were obtained with both sink and source smearing.

For $\beta=6.2$ we use smearing for the heavy quark propagators only.
In run H at $\beta=6.2$ we applied a hybrid procedure of Jacobi
smearing \cite{baxter} and fuzzing \cite{lacock}.  For runs N and P we
fixed the configurations to Coulomb gauge. We used hydrogenic wave
functions $\hsmear{g}{1}$, $\hsmear{g}{2}$ and $\hsmear{e}{1}$
for run N.  The indices `g' and `e' denote wave functions of
the ground and first excited state. The details are given in
table~\ref{smeartab62}. In the P run we used Gaussian smearing with
two different radii, $ar_Q = 2.5$ and $5.0$.

The spin operators applied to construct mesonic states with the
correct quantum numbers are detailed in table~1 of reference \cite{Upsilon}.

\subsection{Lattice spacing}\label{ainvsubsec}
In the quenched approximation one obtains different values for the
lattice spacing, depending on the quantity it is determined
from. This is expected to be caused by the 
strong coupling $\alpha_s$ running differently in the \textit{real
world} and the quenched theory.

We use the physical mass of the $\rho$-meson \cite{pdg} to fix the
lattice spacing. This procedure is justified from the typical gluon
momentum in a $B$ or $D$ meson being of similar size to the momentum in
a light meson such as the $\pi$ and $\rho$. Since heavyonium states probe
a higher physical scale these are not appropriate to fix the scale for
a heavy-light system in the quenched approximation. Using the $\rho$-scale
should take care of most of the quenching effects.

The determination of $m_\rho$ is complicated by the chiral
extrapolation required, see reference \cite{burkhalter} for a review. At
$\beta=5.7$ we use the result of \cite{hpsprd55}.  The result of the
linear extrapolation in the light quark mass $m_q$ is quoted as the
central value and the deviation of the quadratic fit is treated as a
systematic uncertainty. At $\beta = 6.2$ a linear extrapolation is
reported in reference \cite{rowland}. We treat the difference to the
3rd order extrapolation from \cite{gockspec} as a systematic
uncertainty. The numbers are compiled in
table~\ref{ainvtab}. We use:
\begin{mathletters}\label{ainveq}
\bgeqa 
\label{ainveq57}
&\beta=5.7:\hspace*{3ex}
\ainv =  1.116(12)(^{+56}_{-0})\mbox{\ GeV}\,,\hspace*{3ex}
    a = 0.1768(19)(^{+0}_{-88})\mbox{\ fm}\,,&\\
&\beta=6.2:\hspace*{3ex}
\ainv =  2.59(^{+6}_{-10})(^{+9}_{-0})\mbox{\ GeV}\,,\hspace*{3ex}
    a = 0.0762(_{-18}^{+29})(^{26}_{-0})\mbox{\ fm}\,.&
\edeqa
\end{mathletters}
For comparison, table~\ref{ainvtab} also shows the lattice spacing as
obtained from the string tension $\sigma$ and the bottomonium
splitting $\overline{\chi}_b-\Upsilon$. As a physical value for
$\sigma$ we choose a result obtained from a potential model fit to the
charmonium spectrum \cite{eichten80}. The lattice numbers originate
from \cite{stscri,stteper}. These results are in agreement with the
outcome of the $m_\rho$ analysis. As explained above, the bottomonium
system probes a different scale and the values obtained using it do
not agree with the result from light spectroscopy \cite{Upscale}.

\section{Fitting techniques} \label{simmasssec}
\subsection{Parametrisations}
At $\beta = 5.7$ we used several different smearings at source and sink. 
For hadron correlators with a local sink, we applied simultaneous vector fits,
requiring the fitted mass(es) $m_k$ to agree for all propagators:
\bgeqa\label{vecfiteq}
\langle \phi_{\rm L}(t)\ket{\phi_i(0)}
&=& \sum_{k=1}^n A_{i,k}\exp(-m_k t)\,,\hspace{3ex} 1\le i \le m\,,\\ 
A_{i,k}&=&\langle \phi_{\rm L}\ket{\psi_k} \langle \psi_k \ket{\phi_i} \,.
\edeqa
In the case of sink and source smearing, we used simultaneous matrix
fits. In matrix fits, the fitted amplitudes are constrained in their relationship
 with each
other as well:
\bgeqa\label{matfiteq}
\langle \phi_j(t)\ket{\phi_i(0)}
&=& \sum_{k=1}^n B^*_{j,k}B_{i,k}\exp(-m_k t)\, ,
                              \hspace{3ex} 1\le j,i \le m\,,\\ 
B_{i,k}&=& \langle \psi_k \ket{\phi_i}\, .
\edeqa
The fitting techniques are described in more detail in reference
\cite{Upsilon}. We found matrix fits to be more precise with respect
to statistical errors. Due to the omission of the rest mass in
eq.~(\ref{hamilton}) the fitted mass is shifted with respect to the
bound state mass. We denote the result of the fit as the simulation
mass, $m_{\rm sim}$. The determination of the shift will be discussed
in section \ref{qmasssection}.

To extract mass splittings we applied two different procedures. One is
to fit the masses as above, take their difference, and then calculate
the error from the bootstrap or jackknife samples of the difference.
With this procedure one can easily take advantage of using different
smearings. In the case of a single smearing function, a ratio-fit
provides an alternative \cite{nrqcdpower}. For this one divides the
bootstrap or jackknife samples of the two propagators and fits the
outcome with an exponential ansatz. The mass shift cancels out of the
difference in both procedures.

\subsection{Pseudo-scalar and vector meson} \label{psvecmesonsubsec}
On the $\beta = 5.7$ configurations the simulation masses for
pseudo-scalar and vector mesons have been determined most accurately in
run A. In this run we only used the smearing functions $\gsmear{1}$
and $\gsmear{2}$.  We found the double exponential matrix fit with
sink and source smearing to deliver the most precise result. For the
fit range we choose the initial time slice $t_{\rm min}$ two time
slices larger than the first time slice delivering a reasonable
$\chi^2$. In general we choose the number of dropped time slices
multiplied by the number of propagators used for the fit to be larger
than or equal to the number of fit parameters. The reason for this
procedure is as follows. The first reasonable value of $\chi^2$ is
observed once the residual excitations are just masked by the
statistical uncertainties, which allows for them to be still of
similar size. Each excited data point can eat up one fit parameter.
Dropping as many data points as fit parameters delivers a fit which is
entirely dominated by statistical fluctuations.  The residual fit
range dependence of those fits becomes negligible against the
statistical uncertainties.  We judge $\chi^2$ values resulting in a $Q
\ge 0.1$ as reasonable, where $Q$ denotes the probability of a fit
having an even higher value of $\chi^2$.  The final result is given in
table~\ref{swavetab57}.

In run H at $\beta=6.2$ we only had one smearing function available.
We extracted the final results from single exponential fits to the
propagators with source and sink smearing. Their fit results turned
out to be more precise than the ones from using a local sink. The
final fit range was determined such that we observed a reasonable
$\chi^2$ and achieved stability of the fitted result against variation
of the fit range. The results are displayed in table~\ref{swavetab62}.

In the run N we used hydrogenic wave functions of
different radii. We generated smeared local and smeared smeared meson
propagators. However no cross correlators, e.g.\ $\hsmear{g}{1}$ at
sink and $\hsmear{e}{1}$ at source, were calculated. Hence
eq.~(\ref{matfiteq}) was inapplicable and we had to use vector fits in
the case of smearing at sink and source as well. 

\begin{figure}
\centerline{\epsfig{file=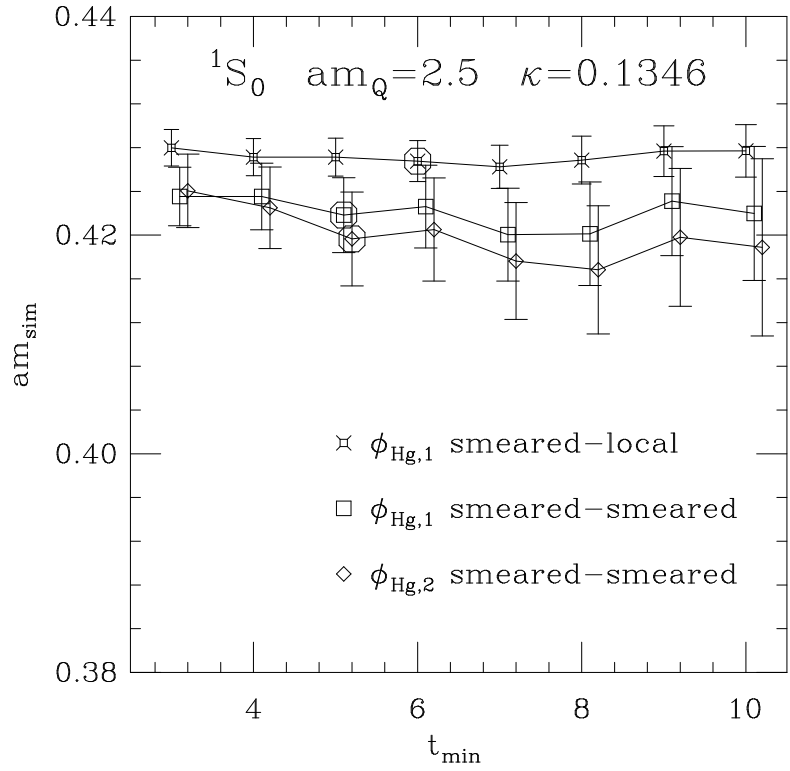,width=11cm}}
\caption{\label{b62smearfig} Dependence of the fitted pseudo-scalar
simulation mass on the starting point $\tmin$ of the fit range at
$\beta=6.2$. The results are shown for three different propagators
with smearing at source and sink or at the source only.  The octagons
give those values of 
$\tmin$ which, we determined in the $Q$-value analysis, to
give the final result. The connecting lines are for guidance only.}
\end{figure}

In double exponential vector fits to two smearing functions, we
observed extremely low values of $Q$. We observed that this is
connected to unfortunate statistical fluctuations on certain time
slices. However the fit parameters turned out to be stable with
respect to variations of the fit range.  These fits will be discussed
in subsection~\ref{radialsubsect} in more detail. To obtain a more
precise result for the $S$-wave ground states, we resorted to single
exponential fits to single propagators and compared the outcome for
the different smearing functions. This is shown in
figure~\ref{b62smearfig} for the pseudo-scalar propagator at $am_Q =
2.5$. The octagons indicate the final result for each propagator, as
determined from the $Q$-value after dropping two time slices.  Within
statistical errors all results are in reasonable agreement with each
other. For the final result, which is also included in
table~\ref{swavetab62}, we choose the smeared-smeared $\hsmear{g}{1}$
propagator. In the end these deliver the more accurate hyperfine
splitting, due to superior noise cancellation between the
pseudo-scalar and the vector meson state.

In this context it is interesting to note that the propagators with
local sink and $\hsmear{g}{2}$ source smearing plateau much later than
the others, but the results are in agreement with those from other
propagators.

For the $S$ wave states in the run P we only had the Gaussian smearing
at the source with radius $ar_0=2.5$ and local sink. Since these
propagators plateau quite late, we used $\tmin = 16$ for the final
result, the error bars are not competitive with those above. Since they
are needed for the later analysis of the $P$ states we include them as
well in table~\ref{swavetab62}.

To describe physical bound states involving light $u$ and $d$ quarks,
the results of tables~\ref{swavetab57} and \ref{swavetab62} have to be
extrapolated in the light quark hopping parameter.  On both sets of
configurations, the difference between the critical and normal hopping
parameter is smaller than the uncertainty we assigned to $\kappa_c$ in
table~\ref{kappa_tab} \cite{rowland} and we use $\kappa_c$ in our
extrapolations. The normal hopping parameter is the one for which the
extrapolations deliver the physical $m_\pi/m_\rho$ ratio.

Due to the high statistical accuracy we achieved at $\beta =5.7$ in the
pseudo-scalar case, a linear ansatz in $am_q :=
\half(\frac{1}{\kappa}-\frac{1}{\kappa_c})$ in a full covariant fit to
all three data points, results in a fit with $\chi^2/{\rm d.o.f.} >
8/1$ for $am_Q < 10$. This corresponds to $Q < 0.004$.  The resulting
curves do not describe the data.  We carefully checked whether this is
caused by a residual fit range dependence and found all the fit
parameters including the \textit{would-be} strange to non-strange
meson splitting to be stable against variation of the fit range. This
was done for an initial time slice $t_{\rm min}$ in a range from $3$
to $6$.

\begin{figure}
\centerline{\epsfig{file=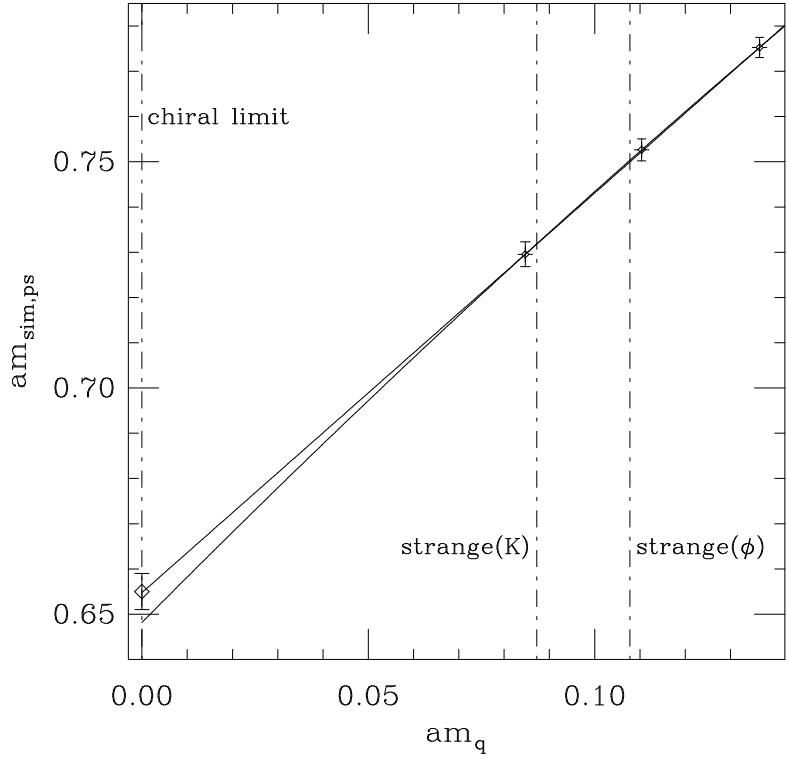,width=11cm}}
\caption{\label{chiralexpfig} Chiral extrapolation of the pseudo-scalar
simulation mass for $\beta=5.7$, $am_Q = 4.0$ in $am_q =
\frac{1}{2}(\frac{1}{\kappa}-\frac{1}{\kappa_c})$. Small symbols give
the simulation result. The curves give a linear and quadratic
extrapolation as described in the text. The diamond to the left gives
the outcome from the linear extrapolation.  }
\end{figure}

We therefore extracted our final result from a linear spline to the
points with highest and lowest $m_q$ and use the deviation of a
quadratic spline as a systematic uncertainty of the chiral
extrapolation. An example for the extrapolation is given in
figure~\ref{chiralexpfig}. From the figure it is obvious that
interpolations to extract the heavy-strange meson mass are insensitive
to the different ans\"atze and we do not assign an uncertainty due to
the different interpolations. However, in the case of the heavy
strange meson, we are faced with the problem that $\kappa_s$ is highly
sensitive to the quantity it is determined from.  Our central value is
interpolated to the $\kappa$ as determined from $m_K/m_\rho$, and the
difference to the outcome for $\kappa$ corresponding to
$m_\phi/m_\rho$ is treated as an uncertainty of the quenched
approximation.  The results are presented in table~\ref{chextrtab57}.

For $\beta = 6.2$ the statistical accuracy is not as high and our data
are well described by linear extrapolations. The results are presented
in table~\ref{chextrtab62}.

\section{Heavy quark masses}\label{qmasssection}
\subsection{Mass shift from dispersion relation}
The omission of the rest mass term $m_Q$ in the Hamiltonian
eq.~(\ref{hamilton}) causes most of the shift of the simulation mass
$m_{\rm sim}$ with respect to the physical meson mass.
The mass, $m_{\rm rel}$, of the meson can be determined from the
relativistic dispersion relation of the meson $E({\vec p}) =
\sqrt{m^2_{\rm rel} + {\vec p}^2}$, which gives
\bgeq
m_{\rm rel} = 
\frac{{\vec p}^2-[E({\vec p})-E(0)]^2}{2 [E({\vec p})-E(0)]}\,.
\edeq
Here $E({\vec p})$ denotes the total energy of the meson.
The mass shift $\Delta_{\rm rel}$ is defined as the difference
\bgeq
\Delta_{\rm rel} := m_{\rm rel} - m_{\rm sim}\,.
\edeq
This shift per heavy quark
should be universal for all hadronic states simulated at the bare heavy
quark mass $m_Q$.

\begin{figure}
\centerline{\epsfig{file=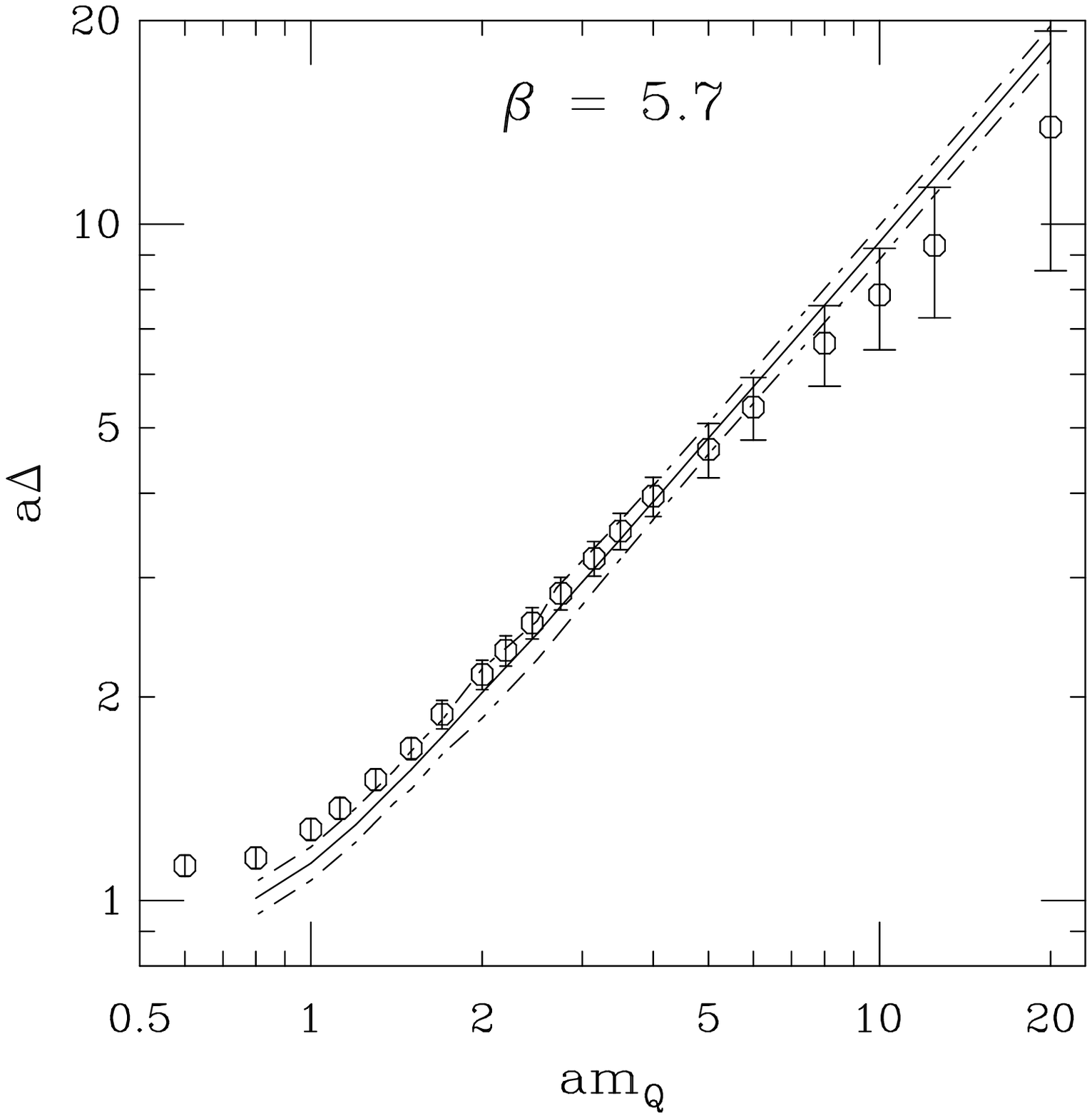,width=8cm}
\epsfig{file=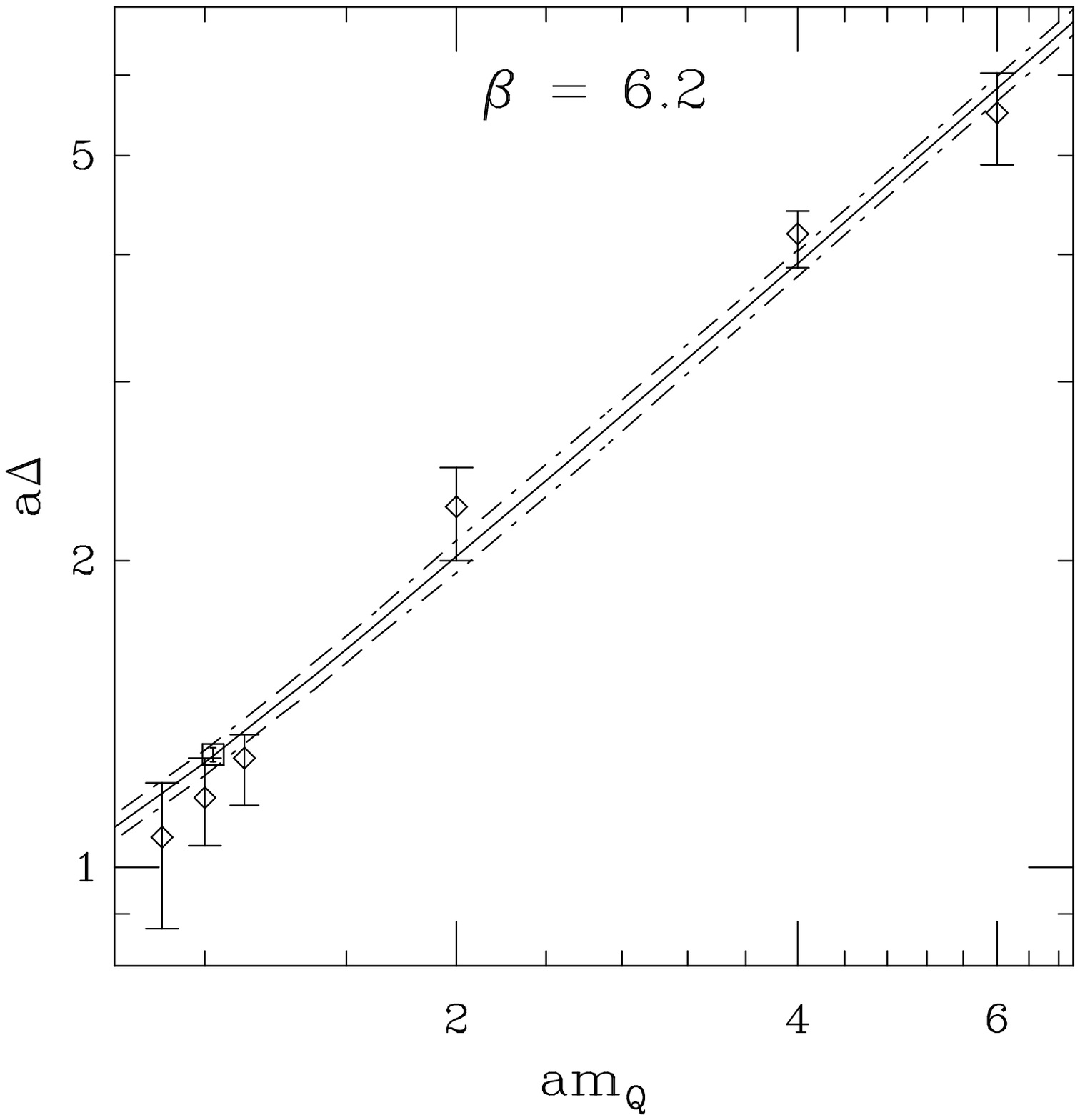,width=8cm}}
\caption{\label{shift57fig} Comparison of the mass shift $a\Delta$ from
the simulation and lattice perturbation theory as a function of the 
bare heavy quark mass $am_Q$.
The octagons give the outcome of the dispersion relation for heavy
light mesons, diamonds from heavyonium. The lines 
represents the perturbative outcome. The square gives the 
heavyonium shift from reference~\protect\cite{Upscale} for $\beta = 6.2$
for comparison.
}
\end{figure}

In our calculation at $\beta=5.7$ we determined the mass shift from
the difference in energy of the pseudo-scalar meson propagators with
$a|\vec p| = 0$ and $2\pi/12$. This was done in run C at
$\kappa=0.1400$ with source smearing $\gsmear{1}$ and a local sink. At
large values of $m_Q$, we found a single exponential fit to the ratio
of the correlators to plateau much later than the fits to the
individual propagators. This is reflected in a large fit range
dependence of the jackknife difference of the masses of the individual
fits, for time slices in which no plateau was observed in the
ratio-fit. For our final result we choose a minimal $t$-value two
time slices larger than the first $t$-value for which we obtained a
decent $\chi^2$ in a fit to the ratio of propagators. The final result
is presented in table~\ref{shifttab57} and figure~\ref{shift57fig}.

We also tried simultaneous vector fits according to eq.~(\ref{vecfiteq})
with two exponents. We used propagators with source smearing
$\gsmear{1}$ and $\gsmear{2}$. The jackknifed difference of the fitted
ground state mass is in agreement with the above procedure; however,
the statistical uncertainties, especially for large values of $m_Q$,
are larger.

For $\beta = 6.2$ we calculated the mass shift in heavy quarkonia,
since the statistical precision for heavy-light correlators at finite
momentum was not sufficient. In the following, mass shifts
from heavy quarkonia will be denoted by $\Delta_H$.  We simulated the
vector-meson for $a|\vec p| \le 2\frac{2\pi}{24}$.  
The kinetic mass $m_1$ was obtained from fits to the dispersion relations:
\begin{mathletters}
\bgeqa
\label{hdisp1}
E_{\rm sim}(\vec p) &=& m_0 + \frac{\vec p ^2}{2m_1} - \frac{\vec p ^4}{8m_2^3} \,,\\ 
\label{hdisp2}
E_{\rm sim}(\vec p) &=& m_0 + \frac{\vec p ^2}{2m_1} - \frac{\vec p ^4}{8m_1^3} \,,\\ 
\label{hdisp3}
E_{\rm sim}(\vec p) &=& m_0 + \frac{\vec p ^2}{2m_1} \,, 
\edeqa
\end{mathletters}
with parameters $m_0$, $m_1$ and $m_2$. $E_{\rm sim}(\vec p)$ denotes the 
simulation energy as determined from the propagator falloff.
In the case of
$am_Q \le 1.3$ we used
the ans\"atze~(\ref{hdisp1}) and (\ref{hdisp2}); for the three heavier
$m_Q$-values, (\ref{hdisp2}) and (\ref{hdisp3}).  All fits gave fit
parameters which were consistent within half of the statistical
error. To obtain the shifts required for heavy-light spectroscopy
we subtracted the simulation mass of the quarkonium vector-meson and
divided by two. The final results are displayed in
table~\ref{shifttab62} and figure~\ref{shift57fig}.  It is interesting
to compare to the result from reference~\cite{Upscale} --- $a\Delta_H =
1.29(2)$ obtained at $am_Q = 1.22$. Due to  higher statistics, this
result is much more precise. This value is included as a square into
figure~\ref{shift57fig} and agrees well with the newer results.

\subsection{Mass shift in perturbation theory}
The mass shift $\Delta$ can also be calculated in lattice perturbation
theory \cite{perturb}:
\bgeq
\Delta_{\rm pert} = Z_mm_Q - E_0\,.
\edeq
Here $Z_m$ denotes the renormalisation constant connecting the bare
lattice mass $m_Q$ with the pole mass and $E_0$ denotes the heavy
quark self energy constant. In the perturbative expansion the 1-loop
contributions from $Z_m$ and $E_0$ cancel each other to a large extent
and the direct perturbative expansion of $\Delta_{\rm pert}$ is much better
behaved than either perturbative series on its own. The Lepage
Mackenzie scale $aq^*$ \cite{tadpole} 
has been determined separately for $\Delta$ and
it is larger than for $Z_m$ or for $E_0$.  The coefficients for
\bgeq
\Delta_{\rm pert} = m_Q [1 + \alpha_s(aq^*)\cdot \Delta^{(1)}]
\edeq
can be found in table~\ref{deltaperttab}. We use the
$\alpha_P(aq=3.4)$ values as determined from the $1\times 1$ Wilson
loop \cite{alphas} with 2-loop running in order to evolve to the
respective $aq^*$. For the final mass shift we assign a relative
uncertainty of $\alpha_s^2(aq^*)$. Since $\Delta^{(1)}$ is small, this
is more conservative than the squared 1-loop contribution. The final
results are displayed in table~\ref{shifttab57} for $\beta = 5.7$ and
table~\ref{shifttab62} for $\beta = 6.2$. For values of $m_Q$ not
included in table~\ref{deltaperttab} we interpolated linearly between
the nearby values, which is completely sufficient within the claimed
accuracy. The results for $a\Delta$ from perturbation theory and the
lattice simulation are compared in figure~\ref{shift57fig}.
Apart from possibly the low $m_Q$ region at $\beta = 5.7$, the figure shows
excellent agreement between the two ways of calculating the mass
shift.

For $\beta=5.7$ the stability parameter $n$ differs in some cases
between the perturbative results and the simulation.  However for
$am_Q = 4$, where perturbative results exist for $n=1$ and $2$, the
effect of $n$ is completely negligible: we obtain $\Delta_{\rm pert} =
3.88(22)$ vs $3.89(24)$.  From a comparison of the simulation result of
the runs A and C at $\beta=5.7$ we can also obtain evidence of the
effect of the different $n$ on the simulation mass $m_{\rm sim}$. For
$am_Q=1.0$ and $\kappa = 0.1400$ we measure $am_{\rm
sim,ps}=0.6265(21)$ and $0.6248(21)$ for $n=5$ and $6$
respectively. This difference is again completely negligible against
the uncertainty we assign to $a\Delta$, even if we enlarge it by a
factor of $3$ to allow for a larger effect between $n=4$ and $5$. The
former $n$ was used in the perturbation theory. Note also that this
difference tends to be in the opposite direction to that in $a\Delta$
implying that the effect of $n$ on the physical mass is reduced
when compared to the shift.

Here it is interesting to note that physical mass differences like the
hyperfine splitting $m_{\rm hpf} = m_{\rm sim, v}-m_{\rm sim, ps}$ are
even less sensitive to $n$. At the above mass parameter of $am_Q =
1.0$ we measure $am_{\rm hpf} = 0.0833(20)$ for $n=4$ and $0.0835(20)$
for $n=5$.

In summary the differences in $n$ between the different runs as well
as the perturbative shifts can be neglected safely even at the high
level of accuracy we achieved here. This leaves us with a discrepancy
between $\Delta_{\rm pert}$ and $\Delta_{\rm rel}$ for our lowest
$m_Q$-values, which is roughly twice as large as the uncertainty we
assign to the perturbative result.

On the other hand for $\beta = 6.2$ we observe excellent agreement
between the precise result of $\cite{Upscale}$ with the
perturbative calculation at the relatively low value $am_Q = 1.22$.

Given a value for the shift $\Delta$ and the simulation mass $m_{\rm
sim}$ from tables \ref{swavetab57}, \ref{swavetab62},
\ref{shifttab57} and \ref{shifttab62}, we can now calculate absolute
masses for all the states. We do this for the ground state vector and
pseudo-scalar mesons, both to fix the quark mass, as described in the
next subsection. Moreover, we use the meson mass rather than the quark
mass to discuss the $m_Q$ dependence, since it is more directly
comparable to experiment. 
We frequently plot results against $1/m_{\rm
sav}$, where $m_{\rm sav}$ is the spin-average of the ground state
vector and pseudo-scalar mesons
\bgeq
m_{\rm sav} = \forth ( 3 m_{\rm v} + m_{\rm ps} )\,.
\edeq
This is preferable to $m_{\rm ps}$ alone since
the spin-averaging reduces the dependence on sub-leading
spin-dependent terms.

\subsection{Bare heavy quark mass}\label{bareqsubsect}
To determine the bare quark mass $m_Q$ corresponding to the $b$ and
$c$-quark, we compared the mass of the spin-averaged $S$-wave meson
denoted with an overbar, with the experimental result.  We
used 5313~MeV for the $\bar B$, 5405~MeV for the $\bar B_s$, 1973~MeV
for the $\bar D$ and 2076~MeV for the $\bar D_s$ \cite{pdg}. For the
interpolations we used spline-fits to three neighbouring points. The
fits were done quadratically in $m_Q$ and $1/m_Q$ and no significant
difference was observed between the two.  From the strange and
non-strange meson we obtained identical results for the quark masses:
\begin{mathletters}
\bgeqa
\label{mbval62}
am_b &=& 1.64(5)(^{+8}_{-5})(^{+0}_{-7})\,,\hspace{3ex}\beta=6.2\\
\label{mbval57 }
am_b &=& 4.20(25)(5)(^{+0}_{-24})\,,\hspace{3ex}\beta=5.7\\
\label{mcval}
am_c &=& 0.87(6)(3)(^{+0}_{-13})\,,\hspace{3ex}\beta=5.7
\edeqa
\end{mathletters}
The errors as indicated in the parentheses give the uncertainty
arising from the mass shift and the statistical and systematic
uncertainty of the $a$~determination. The uncertainties associated
with the simulation mass $m_{\rm sim}$ are completely negligible
here. For $m_b$ we used the perturbative shifts $\Delta_{\rm pert}$.
Using $\Delta_{H}$ at $\beta=6.2$ delivers
$am_b=1.59(^{+14}_{-5})(^{+6}_{-3})(^{+0}_{-5})$ and using
$\Delta_{\rm rel}$ at $\beta = 5.7$ gives
$am_b=4.16(^{+53}_{-31})(7)(^{+0}_{-30})$, which is agreement with
$\Delta_{\rm pert}$ but with larger error. For $m_c$ we used the
simulation result $\Delta_{\rm rel}$. Here $\Delta_{\rm pert}$ would
give $am_c = 1.02(8)(2)(^{+0}_{-10})$.  The deviation from the result
eq.~(\ref{mcval}) reflects the difference between the $\Delta$-values
at low $am_Q$ discussed above.

This careful analysis to fix the bare heavy quark mass is particularly
necessary for fine structure splittings in the spectrum to be
discussed in the next section. These are very sensitive to the heavy
quark mass, generally as $1/m_Q$. In addition, any errors in the heavy
quark mass must be fed into errors in the fine structure splittings in
order to avoid underestimating those errors.

The bare masses do not scale with the lattice spacing as expected, because they
are unphysical. A better quantity to consider would be the mass
in the $\overline{\rm MS}$-scheme. This will be discussed in a future
publication \cite{qmasspap}.

\section{Mass dependence of level splittings} \label{splitsec} 

In this section we describe how the results for the level splittings
are extracted from our data.  The dependence of the different
splittings on the light and heavy quark mass is also discussed.

\subsection{Flavour dependent splittings}
The mass difference between heavy-light states distinguished only by
their strangeness survives into the static limit. Based on the ideas
of heavy quark symmetry such splittings are expected to depend weakly
on the mass of the heavy quark.  If a spin-averaged combination is
taken, the leading heavy quark mass dependence arises purely from the
kinetic term in eq.~(\ref{kinhamilton}). The size of the slope in
$1/m_{\rm sav}$ then gives information on the difference in $\langle
p_b^2\rangle$ for the $b$-quark in the strange and non-strange states.

We calculated this splitting from the ground state S-wave results for
$\kappa_s$ and $\kappa_c$. The result is highly sensitive to the
reported uncertainties in the chiral extrapolation and the
determination of the strange hopping parameter $\kappa_s$.  At
$\beta=5.7$ we determine the statistical uncertainties in a jackknife
procedure applied to the difference of the individual results and at
$\beta=6.2$ we use the bootstrap.

\begin{figure}
\centerline{\epsfig{file=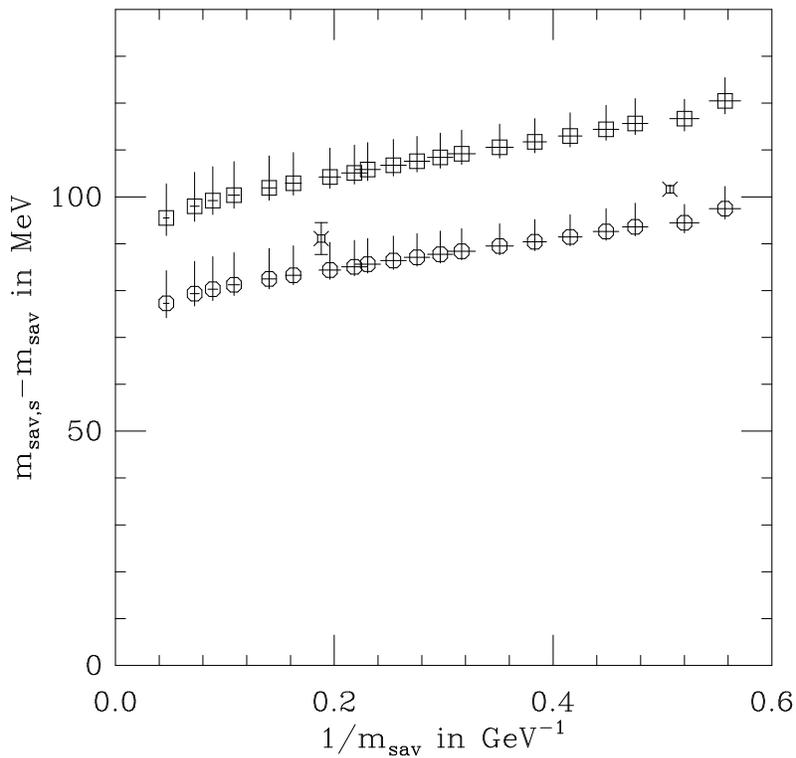,width=11cm}}
\caption{\label{strangesplitfig}Mass splitting between the spin-averaged
ground state $S$-wave heavy-light
strange and non-strange meson. The octagons give the result for
$\kappa_s$ determined from the $K$-meson, squares from the
$\phi$-meson. For both data sets the upper errors  give the uncertainty from the
different chiral extrapolations, the lower ones the statistical
uncertainty. For simplicity this figures does not consider
uncertainties arising from the value of $a$, see eq.~(\protect\ref{ainveq57}).
Experimental results are displayed by the fancy squares.
}
\end{figure}

For $\beta = 5.7$ our statistical errors are very small and we
consider additional systematic uncertainties for the chiral
extrapolation and $\kappa_s$. For $\beta=6.2$ the quality of our data
is not as good and we give statistical errors only. Using the
$\kappa_s$-value determined from the $\phi$ would lead to a 9\%
increase of the result, which is small compared to our statistical
errors.  The results are displayed in tables~\ref{strangesplit} and
\ref{strangesplit6.2}.  In figure~\ref{strangesplitfig} we plot the
result for the spin-averaged splitting at $\beta = 5.7$ versus the
inverse of $m_{\rm sav}$, in order to display its heavy quark
dependence.

The figure displays a clear increase of the splitting with decreasing
heavy quark mass. To quantify the slope of this dependence we perform
a linear fit of the splitting result versus $1/m_{\rm sav}$. The
result, converted into physical units, is detailed in
table~\ref{hqettab}. The slope corresponds to a $\langle p^2_b
\rangle$ difference 
of $\approx [0.25(3)\mbox{ GeV}]^2$, which is of the size of
$\lqcd^2$, as expected.

Because of the larger statistical uncertainties, we do not observe a
significant slope at $\beta = 6.2$. The data can be fitted nicely to a
constant.

\begin{figure}
\centerline{\epsfig{file=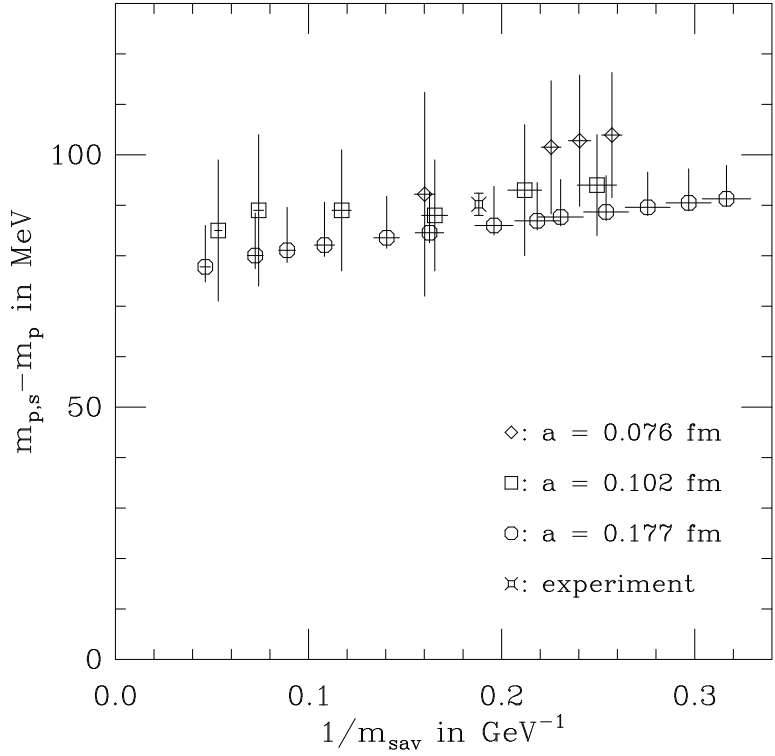,height=11cm}}
\caption{\label{strangevgl} Comparison of the strange
non-strange splitting for the pseudo-scalar state at different values
of the lattice spacing $a$. Octagons give our
result for $\beta=5.7$ and diamonds for $\beta=6.2$.
The squares give the result from \protect\cite{arifalat98} at $\beta=6.0$
and the fancy square the experimental outcome for the $B$-meson 
\protect\cite{pdg}.
For $a=0.102$~fm and $0.076$~fm 
we give the statistical errors only, the 
errors for $a=0.177$~fm have been described in 
figure~\protect\ref{strangesplitfig}. }
\end{figure}

A comparison of our results with the ones of \cite{arifalat98} for the
pseudo-scalar case is plotted in figure~\ref{strangevgl}. In this plot
we show the result for the strange quark as determined from the
$K/\rho$ mass ratio only. Due to the large error bars we do not
include the results obtained at the larger values of $m_Q$ for $\beta
= 6.2$.  Within the accuracy of around 12\% in the case of
$\beta=6.2$ or better, no sign of scaling violations shows up
in the plot.  We also observe excellent agreement with the experimental
result.

\subsection{Radial excitations} \label{radialsubsect}
In order to obtain a reasonably stable and long plateau for the
radially excited $S$-states on the coarse lattice at $\beta=5.7$ 
we applied triple
exponential matrix fits to the three smearing functions $\gsmear{0}$,
$\gsmear{1}$ and $\gsmear{2}$. This was done for the
run S for a single $\kappa$ of $0.1400$ only, which is approximately
equal to the strange as determined from the $K$-meson. Since in
reference \cite{arifalat98} the dependence of the $2^1S_0-1^1S_0$
splitting on the light quark mass was found to be very small, a
variation of less than 2\% when fixing $\kappa_s$ from the $K$ or
$K^*$-meson, we can ignore any mismatch in our $\kappa$ vs $\kappa_s$
compared to the statistical uncertainties.  We therefore treat our
result as the answer for this splitting with $\kappa_s$ as determined
from the $K$.

\begin{figure}
\centerline{\epsfig{file=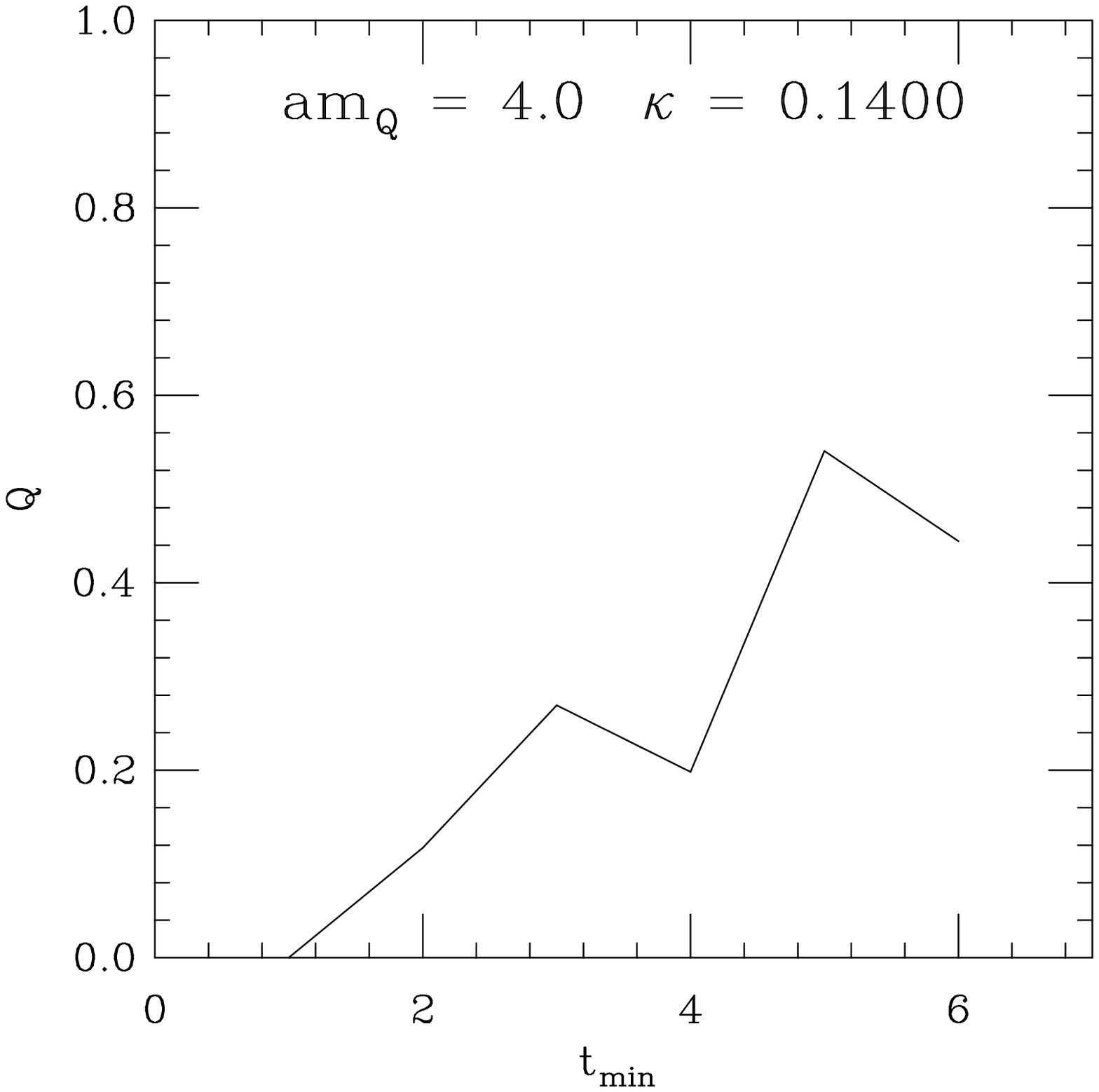,width=8cm}
\epsfig{file=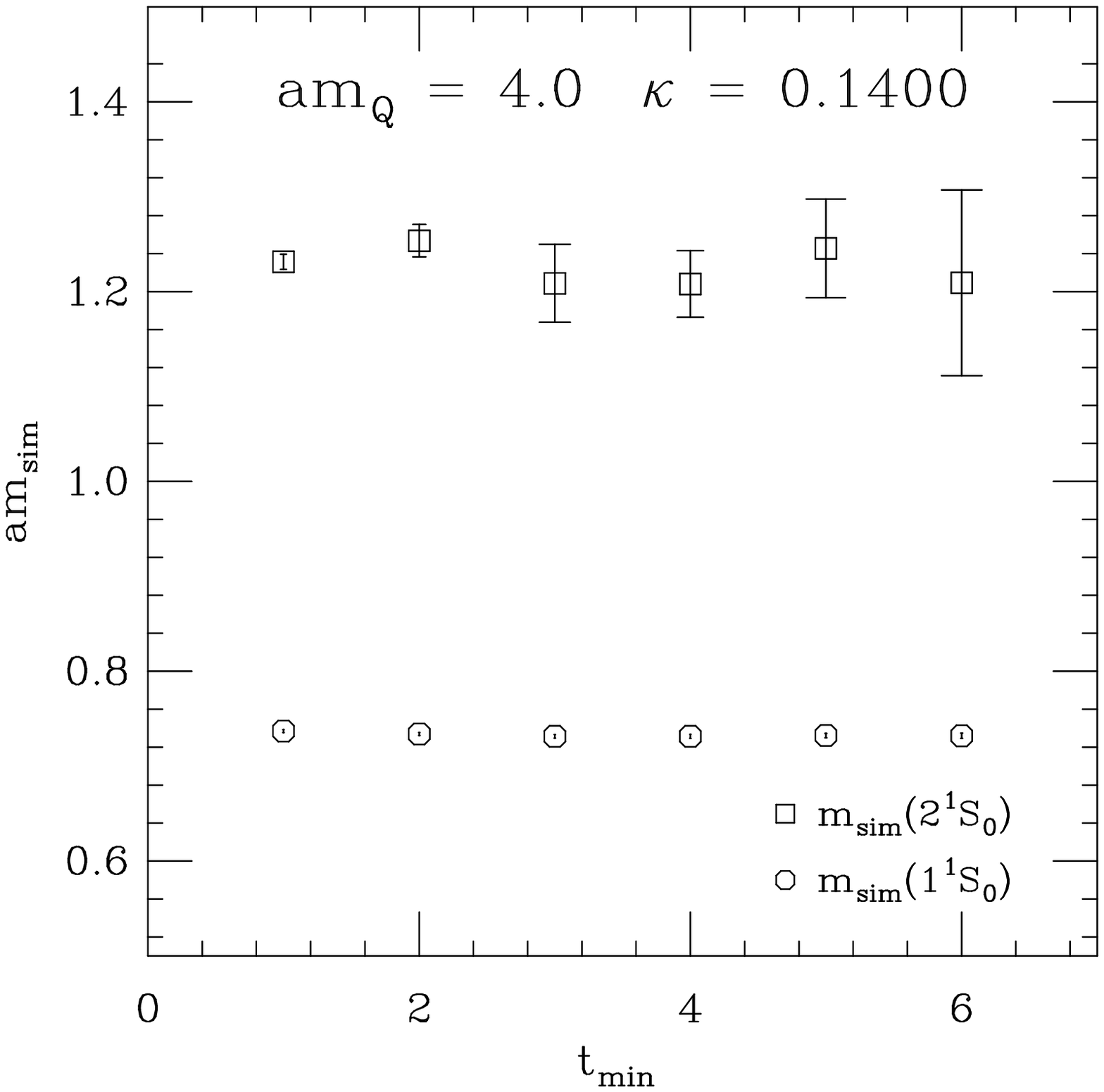,width=8cm}}
\caption{\label{radsplitfit} Fitting the radially excited $S$-state at
$\beta=5.7$. The left hand side gives the $Q$-values and right hand
side the fitted simulation masses of the pseudo-scalar ground state and the
first radially excited state.}
\end{figure}

In figure~\ref{radsplitfit} we show a typical example for the
excellent stability of the simulation masses $am_{\rm sim}$ against
variation of the starting point $t_{\rm min}$ of the fit range. The
extent in $t_{\rm min}$ for which we can resolve the excited state is
5 time slices or $0.28$~GeV$^{-1}$. The rate of its disappearance is
set by the $2S - 1S$ splitting of $600$~MeV.

In the figure the first good value of $Q$ is observed for $t_{\rm
min}=2$. To be safe with respect to residual excitations we quote
the final result for a fit range starting at time slice $4$, which is
the procedure described in subsection~\ref{psvecmesonsubsec}. The peak
in $Q$ at $t_{\rm min}=5$ results from the fit becoming insensitive to
the third exponential at this point.

\begin{figure}
\centerline{\epsfig{file=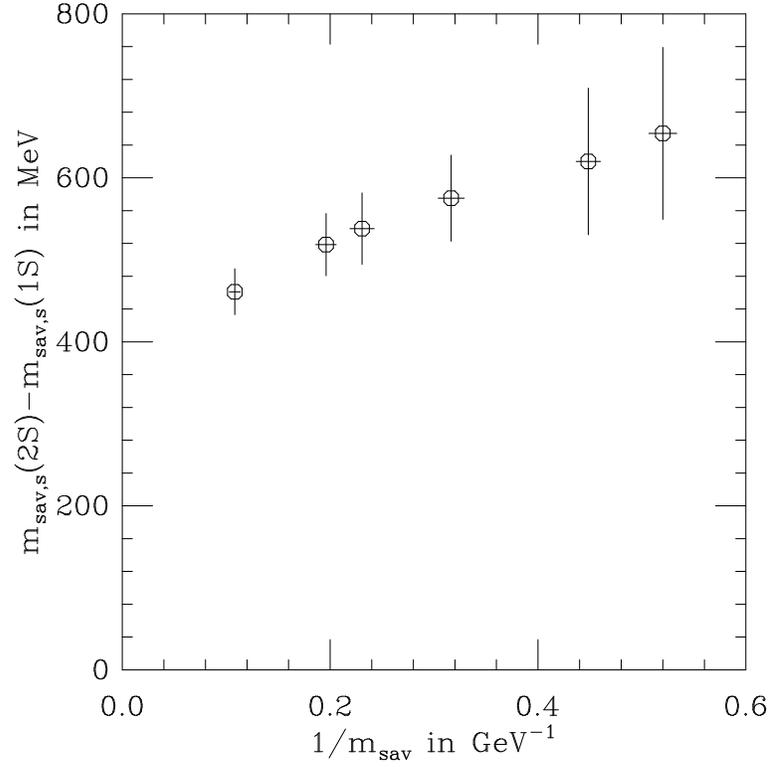,width=11cm}}
\caption{\label{radsplitfig} Splitting between radially excited and
ground state $S$-wave splitting at $\beta=5.7$. The results are for
$\kappa = 0.1400 \approx \kappa_s(K)$ and the spin-average.}
\end{figure}

The results for all 6 heavy quark masses are given in
table~\ref{radial}. In figure~\ref{radsplitfig} we plot the heavy
quark mass dependence of the spin-averaged splitting. The result shows
a clear increase as the heavy quark mass is reduced. In
table~\ref{hqettab} we detail fit results for the offset and slope of
this splitting with respect to $1/m_{\rm sav}$.  From the assumption
that the increase of the splitting with $1/m_{\rm sav}$ is caused by
the difference in the kinetic energy $p^2/2m_Q$ between ground and
radially excited states, the fitted slope gives a $\langle
p_b^2\rangle$ difference of $\approx [0.95(15) \mbox{ GeV}]^2$. This
is of the size of a few times $\lqcd^2$ as would be expected.

On our fine lattice, since no cross-correlators between the different
smearings had been calculated, we used simultaneous vector fits
in all cases.  The differences between the smearings $\hsmear{g}{1}$,
$\hsmear{e}{1}$ and $\hsmear{g}{1}$ turned out to be too small for
simultaneous fits with three exponentials and we had to resort to fits
using 2 exponentials.  Again we choose the starting point $\tmin$ of
the fit range as described in subsection~\ref{psvecmesonsubsec}. With
this procedure it is possible to extract reliable information on the
excited state, as can be verified from the tables of reference
\cite{Upsilon} for the case of $\Upsilon$-spectroscopy. Using
propagators with sink and source smearing in vector fits, leads to a
suppression of the excited state contamination, which made it
impossible to extract a signal for the excited state. Therefore we
used propagators with smearing at the source and local sinks to
extract the radially excited states.

\begin{figure}
\centerline{\epsfig{file=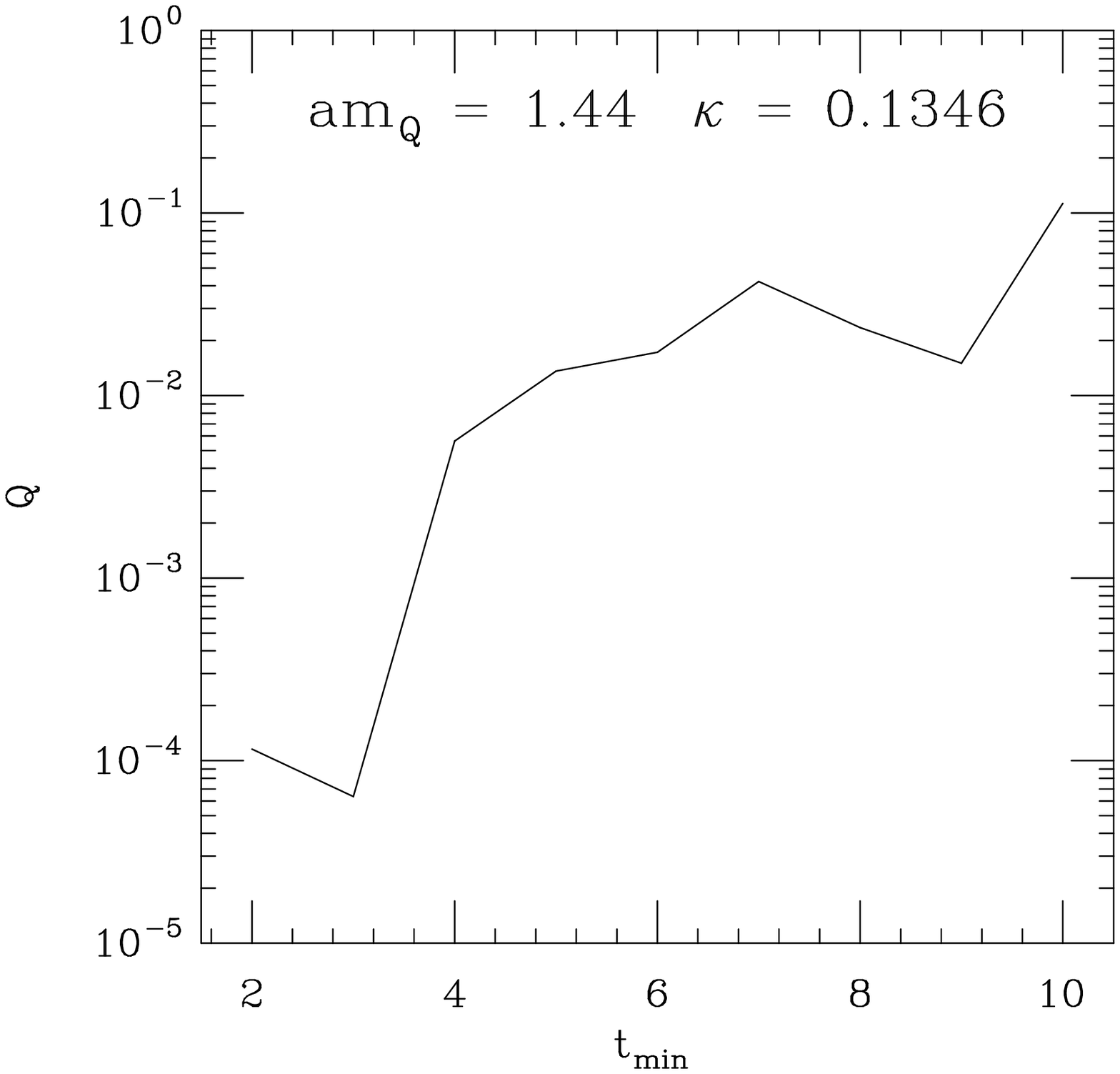,width=8cm}
\epsfig{file=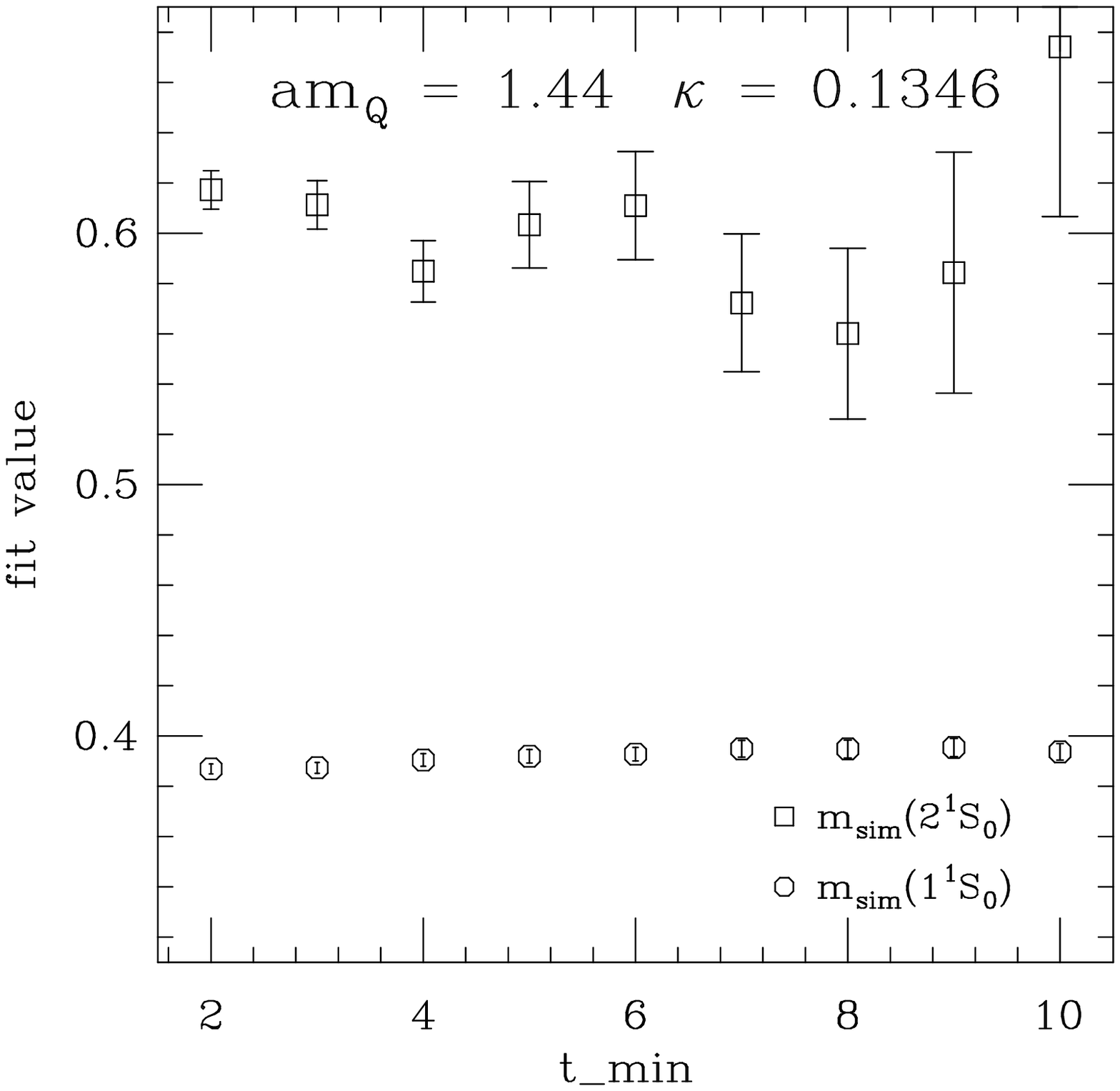,width=8cm}}
\caption{\label{radsplitfit62} Fitting the $2S$ radial excitation at
$\beta=6.2$. Again we give the $Q$-values and on the right the 
simulation masses of the pseudo-scalar ground state and the
first radially excited state.  This example uses
$\hsmear{g}{1}$ and $\hsmear{e}{1}$ at the sources.}
\end{figure}

As denoted earlier, the fits to these propagators are plagued
by statistical fluctuations, which lead to quite large $\chi^2$ and
low $Q$ values. However, the fits describe the data reasonably and the
fitted parameters are stable against variation of $\tmin$. This is
shown in figure~\ref{radsplitfit62}. For a fit range starting point
$\tmin \ge 5$ we obtain $Q > 1\%$, which is low compared to what we
obtained in the other fits. However it is not that low, that the fit
could be ruled out on statistical grounds. Together with the good
stability of the fitted masses against variations of $\tmin$ as
displayed on the right hand side of the figure, we believe that
our signal is genuine. In this example we extract the final result
from $\tmin=8$. The results for the radially excited $S$-wave for this
lattice spacing are compiled in table~\ref{radial6.2}.

\subsection{Orbital excitations}
Orbitally excited $P$-state mesons have been investigated at both of
our lattice spacings. The possible states consist of four
non-degenerate energy levels; total angular momentum  $J=0$ and $2$ 
as well as two $J=1$
states. In the heavy quark symmetry picture a $jj$ coupled basis is
appropriate. In the vicinity of the static limit, the $J=2$ and the
higher of the $J=1$ states are close and separated from the 
$J=0$ and lower $J=1$. The former correspond to a light quark total
angular momentum of $j_l = \frac{3}{2}$, the latter of $j_l =
\frac{1}{2}$. We use an $LS$ coupled basis to study the states, but
expect our $^1P_1$ and $^3P_1$ channels to mix, leading to the
observation of the lower $J^P = 1^+$ state with both operators.  We
will denote the state corresponding to $j_l=\frac{3}{2}$ with a prime.

At $\beta=5.7$ again we use one light hopping parameter,
$\kappa=0.1400$. As in the case of the radial excitations, we treat this
as the value corresponding to the strange quark as determined from the
$K$-meson and the simulations have been performed in the run S.  We
used the derivatives of the smearings $\gsmear{0}$ and $\gsmear{1}$ at
source and sink. The final results were obtained from double
exponential simultaneous matrix fits to both smearings and are listed
in table~\ref{ptab57}. 

The selection of the fit range proved to be very delicate for this
$a$-value. The statistical error grows rapidly when increasing $t_{\rm
min}$, since the signal to noise is exponentially related to the $P-S$
splitting \cite{thalep,daviesthacker}.  We give an example in
section~\ref{pfinesubsec}, where the fine structure is discussed.  For
the lightest values of $am_Q$ and correlators with $^3P_2$ wave
operators in the $T$-representation we obtained small values of $Q$ of
a few permille.  We include the corresponding mass values into the
table for the sake of completeness and mark them in
\textit{italics}. However we disregard them in the further
evaluation. The results are always in agreement with the ones obtained
in the $E$-representation and we do not believe there is a serious
problem with this, simply statistical fluctuations.

\begin{figure}
\centerline{\epsfig{file=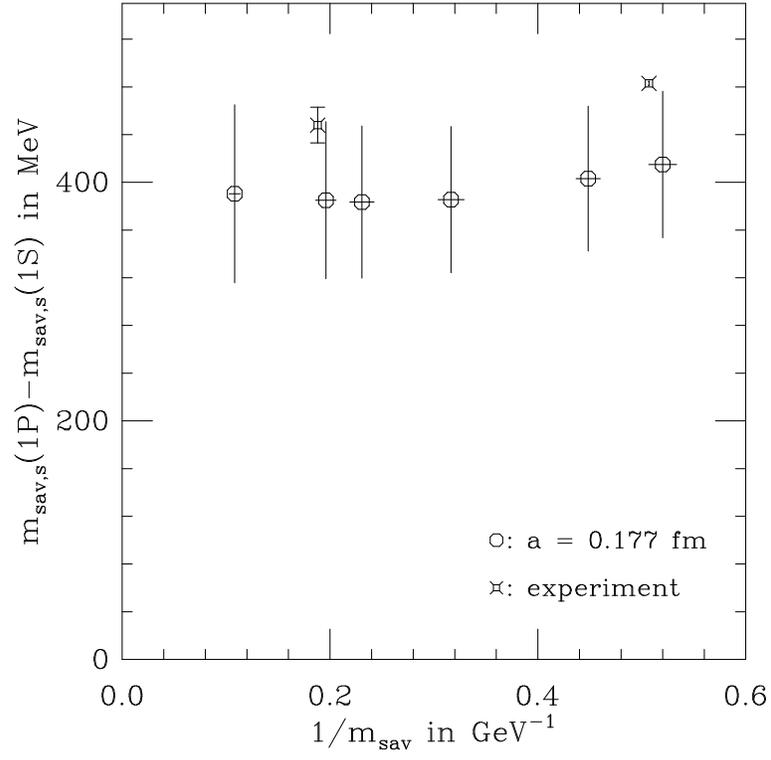,width=11cm}}
\caption{\label{psavsplitfig} Splitting of spin-averaged $P$ to
$S$-wave. The results are 
for strange light quarks and the error bars give the
statistical uncertainties only. The experimental result gives the
$B^*_{sJ}(5850)$ resonance and the spin-average of the $D_{s1}$ and $D^*_{s2}$.}
\end{figure}

In the second to last column of table~\ref{ptab57} 
we give the spin-averaged $P$-state
result which we calculate as
\bgeq \label{psaveq57}
m(P_{\rm sav}) = \frac{1}{12} \left[ 
1\cdot m(^3P_0) + 3\cdot m(^1P_1) + 3\cdot m(^3P_1) + 5\cdot m(^3P_2E) 
\right] \,.
\edeq
The result is also shown in figure~\ref{psavsplitfig}.  For comparison
we include the experimental result for the $B^*_{sJ}(5850)$ resonance
and the spin-average of the $D_{s1}$ and $D^*_{s2}$ \cite{pdg}. The
figure displays at most a mild heavy quark mass dependence. To
quantify this, we report in table~\ref{hqettab} on the offset and
slope of this splitting in physical units.  The slope is consistent
with a $\langle p_b^2\rangle$ difference of ${\cal O}(\lqcd^2)$, but
is also consistent with zero.

$P$-states were also investigated in the run P for $\beta = 6.2$. We
choose $am_Q = 1.6$, directly corresponding to the $b$-quark in
eq.~(\ref{mbval62}).  For the light quarks we use the strange hopping
parameter $\kappa = 0.1346$.  Since the results on our coarse lattice
depend only weakly on $m_Q$ we take the outcome as the final answer
for $B_s$. 

In the simulations we used two different smeared sources together with
local sinks. Again we used derivatives of Gaussian smearing functions.
The masses were extracted from double exponential
vector-fits to both propagators. We observe reasonable $Q$-values for
all applied operators and include all channels into the spin-average
\bgeq \label{psaveq62}
m(P_{\rm sav}) = \frac{1}{12} \left[ 
1\cdot m(^3P_0) + 3\cdot m(^1P_1) + 3\cdot m(^3P_1) + 2\cdot m(^3P_2E)
+ 3\cdot m(^3P_2T) 
\right] \,.
\edeq
The results for the fitted masses are displayed
in table~\ref{ptab62}. The splitting between the spin-averaged $P$ and
$S$ waves is given in tables~\ref{psplittab57} and \ref{psplittab62}.

\subsection{Radially excited $P$-states}
\begin{figure}
\centerline{\epsfig{file=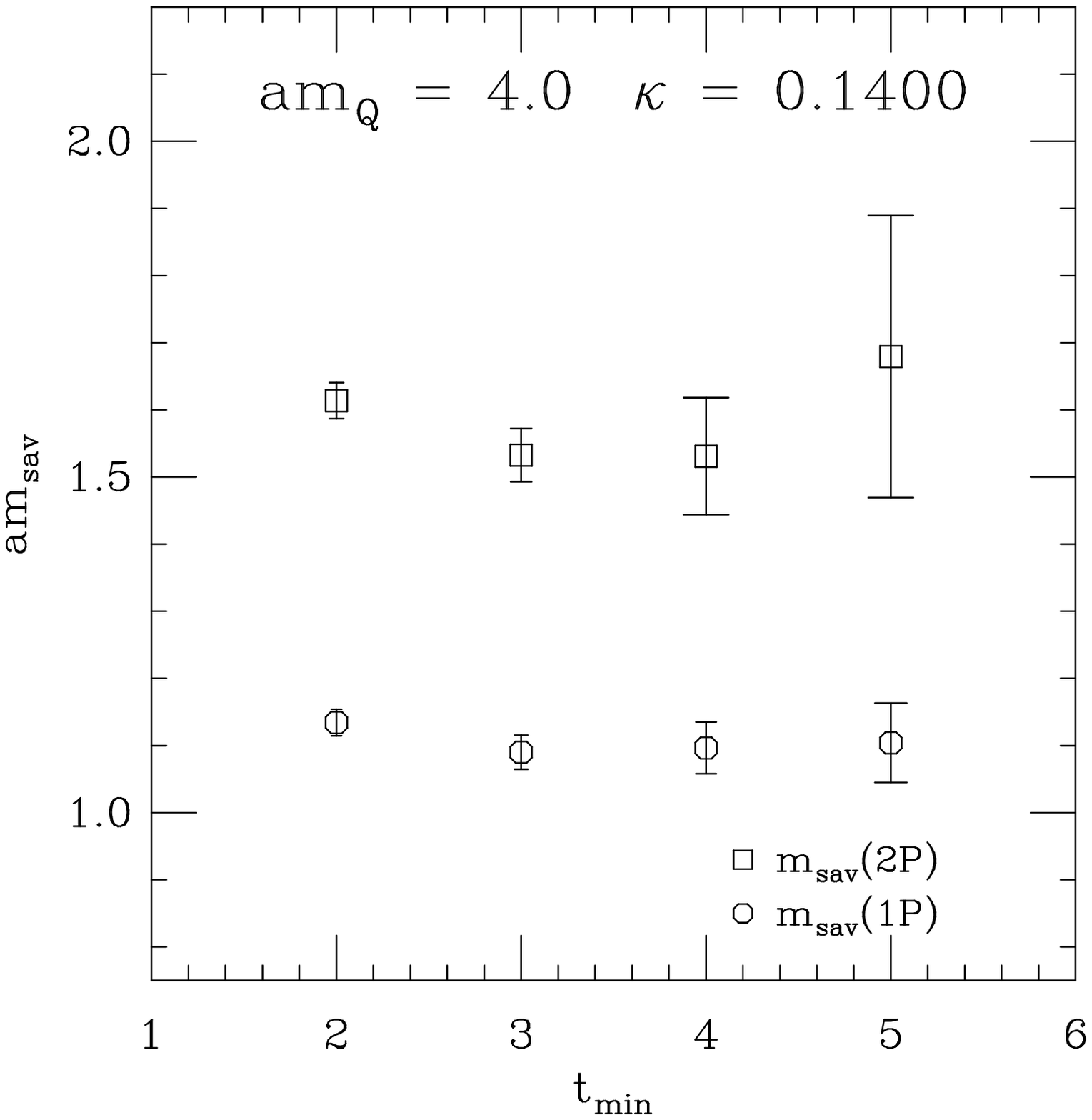,width=8cm}
\epsfig{file=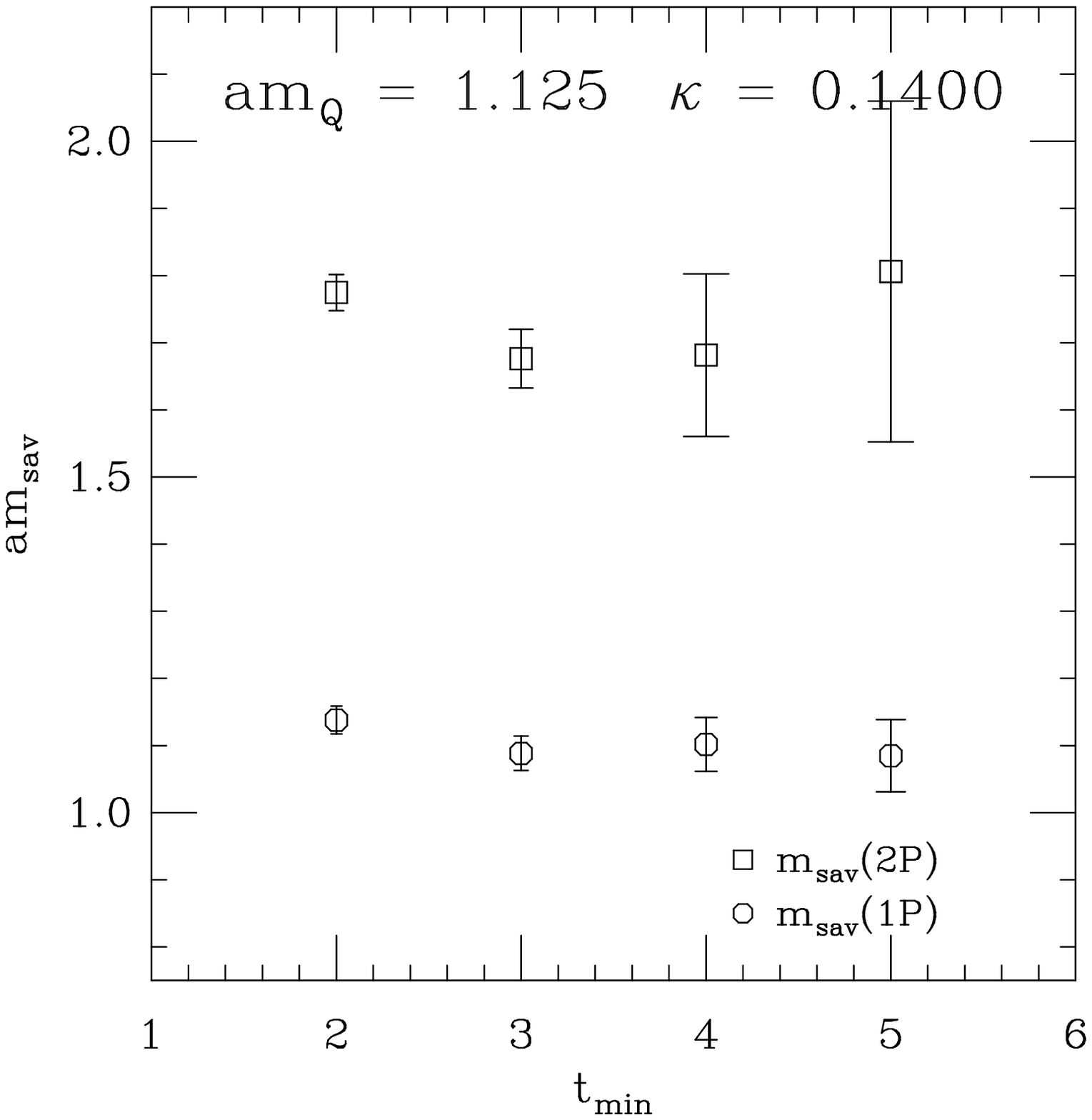,width=8cm}}
\caption{\label{psavradtminfig57} Fitting radially excited $P$-states at
$\beta=5.7$. The plots give the spin-averaged $1P$ and $2P$ states. 
The final answer is extracted from $t_{\rm min}=5$ in both cases.}
\end{figure}

Having available 2 different smearing functions at both of our lattice
spacings, it is possible to obtain information on the radially excited
$P$-states as well. In figures~\ref{psavradtminfig57} we show the
dependence of the fitted masses of the spin-averages of the $1P$ and
the $2P$ on the starting point $t_{\rm min}$ of the fit range for two
different values of $m_Q$. The figure displays a clear signal for an
excited state and reasonable stability with respect to variations of
$\tmin$. However the error grows rapidly with increasing $t_{\rm
min}$.  The $Q$-values of the $^3P_2E$-fit, which is the last of the
individual states included in the spin-average to reach a 
plateau, are included in
figure~\ref{pfinetminplot}. Decent $Q$-values are observed for $t_{\rm
min}=3$.  Since this is a 6 parameter matrix fit to four propagators,
we take our final result from $t_{\rm min}=5$.  The results for
the spin-averaged $2P$-state are summarised in table~\ref{ptab57}. 
We give the splitting to the spin-averaged
$1S$ and $1P$-states in
table~\ref{psplittab57}. We do not observe a significant slope for the
splitting with respect to $1/m_{\rm sav}$.

\begin{figure}
\centerline{\epsfig{file=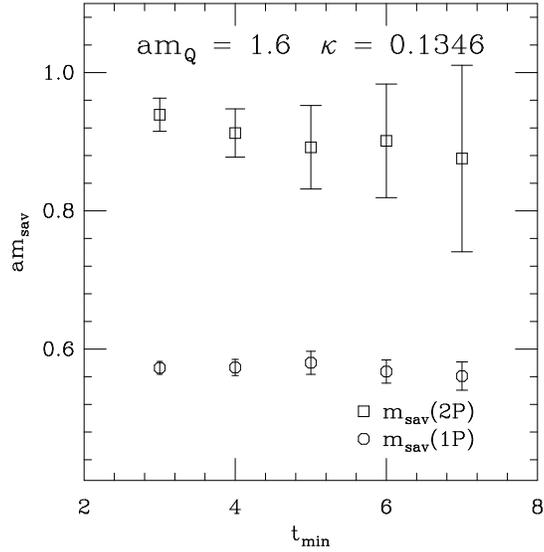,width=8cm}}
\caption{\label{psavradtminfig62} Fitting radially excited $P$-states at
$\beta=6.2$. The plots give the spin-averaged $1P$ and $2P$ state.
We take our final result from $t_{\rm min} = 6$.
}
\end{figure}

For $\beta=6.2$ we show the plateau in
figure~\ref{psavradtminfig62}. Due to the finer lattice, the growth in
the error with increasing $\tmin$ is smaller than before. For the mass
parameters used here, an example of a $Q$-plot will be given in
figure~\ref{pfinetminplot62} below. Here the first decent $Q$ is
observed for $\tmin =2$. Since this is a vector fit to 2 propagators,
we take our final results from $\tmin = 6$. The result and splittings are
included in tables~\ref{ptab62} and \ref{psplittab62}.

As noted when discussing radially excited $S$-states, due to our
conservative selection of the fit range, we expect residual excitations
to be negligible within the quoted statistical errors.

\subsection{Hyperfine splittings} \label{subsecthpf}
The mass difference between a pseudo-scalar and a vector $S$-wave
meson is caused by the spin of the heavy quark. This hyperfine
splitting is expected to vanish in the limit of infinitely heavy quark
mass.

On our $\beta=5.7$ lattice we determined the hyperfine splitting
$m_{\rm hpf}$ from the difference of the results in
table~\ref{swavetab57}. A crucial ingredient in obtaining a small
statistical error is to choose identical fitting ranges for both
correlators. If one of them has a plateau at a larger value of $t_{\rm
min}$ than the other, we took this larger value to obtain $m_{\rm
hpf}$ from the difference of the fitted masses. The error in this
procedure is estimated with a jackknife.  The results are displayed
in table~\ref{hpftab57}.

The chiral extrapolation of the hyperfine splitting turned out to be
less difficult than that for the simulation mass of the
pseudo-scalar and vector mesons. The curvature seems to cancel out
between them and we have been able to perform linear fits to
extrapolate to the chiral limit. However, in order to be
consistent, we assign a systematic uncertainty to the result.  This
uncertainty is obtained from the difference to the outcome of first
extrapolating the individual mesons to the chiral limit and then
determining the hyperfine splitting. In this case we use the quadratic
extrapolations to the chiral limit to take a possible curvature into
account.

\begin{figure}
\centerline{\epsfig{file=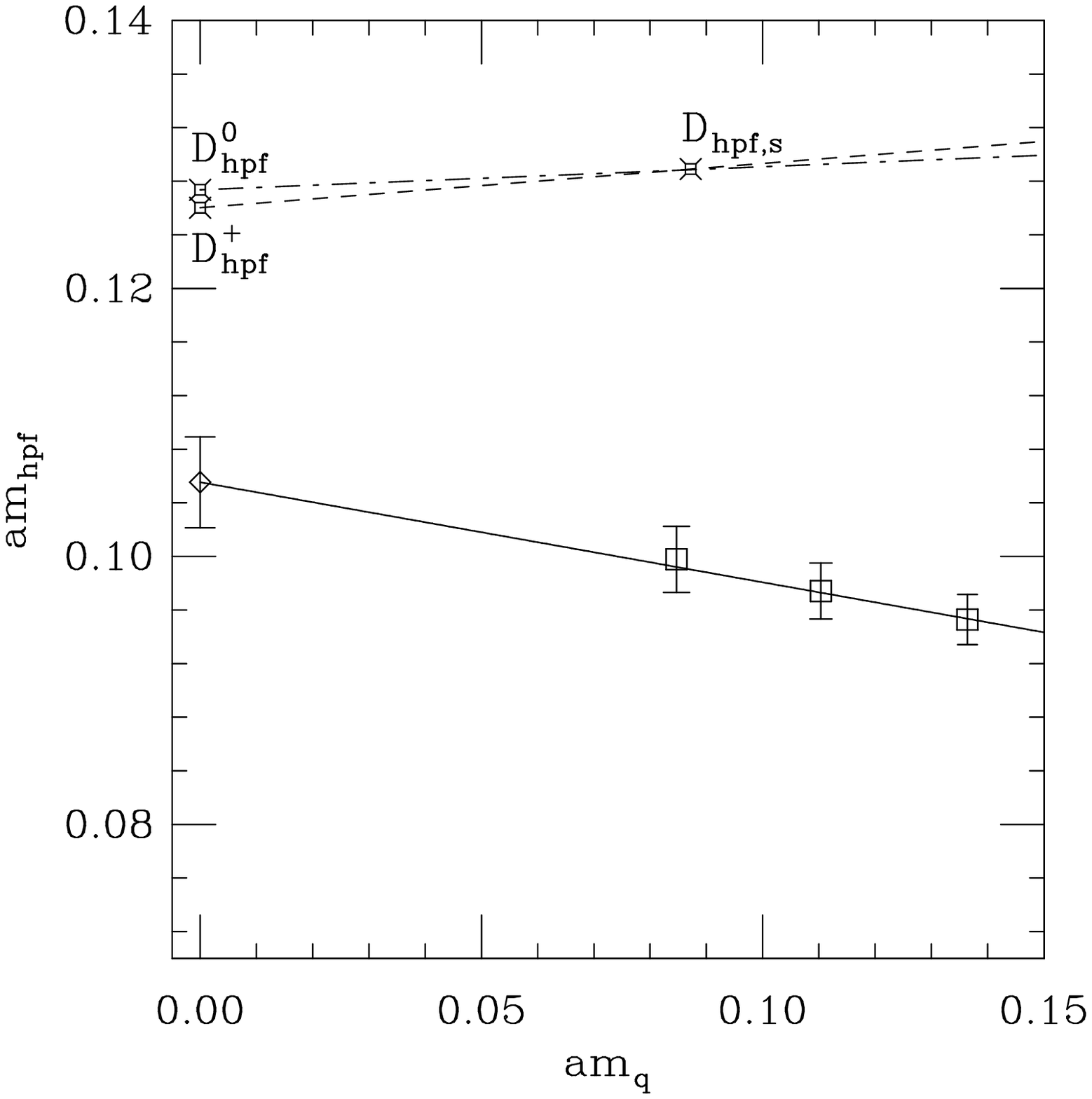,height=8cm}
            \epsfig{file=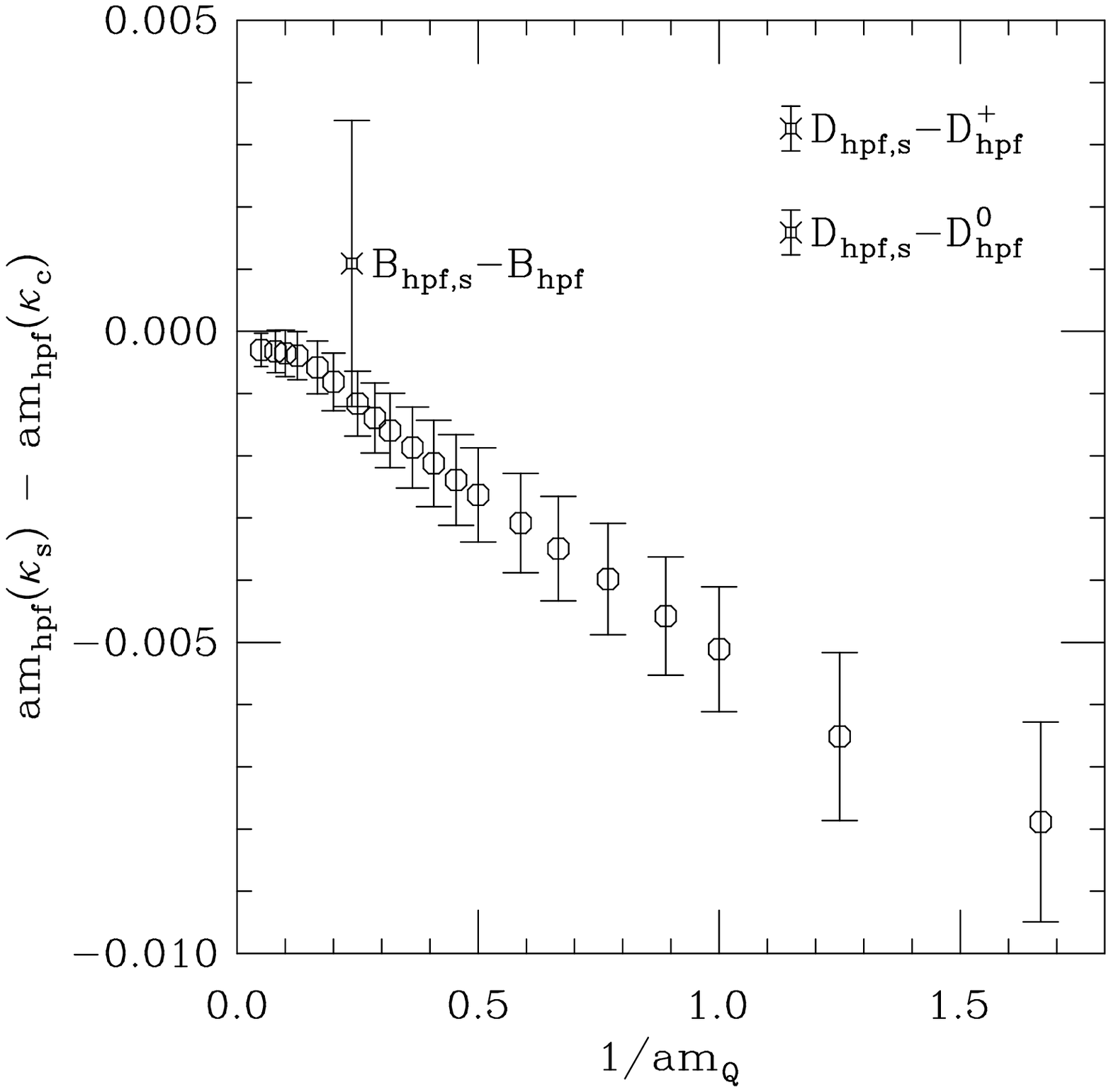,height=8cm}}
\caption{\label{hpfslope} Light quark mass dependence of the hyperfine
splitting at $\beta = 5.7$. On the left hand side we show a linear fit
to all 3 data points for $am_Q=0.8$. This corresponds approximately to
the $D$ meson. The fancy squares give the experimental result. The
right hand side gives the fitted slope multiplied by $am_{s}$,
as determined from the $K$, for all $m_Q$.  In order not to disguise
the significance of our findings, the error bar gives the statistical
errors of the fit parameter only. Experimental results are given by the
fancy squares.  }
\end{figure}

In our chiral extrapolations of the hyperfine splitting we observe a
negative slope with respect to the mass $am_q$ of the light quark, which
is illustrated in figure~\ref{hpfslope}. The left hand side gives an
example of our chiral extrapolations and the right hand side shows
the slopes measured at each value of $am_Q$. In order to construct a
physically meaningful quantity, the latter has been multiplied by the
strange quark mass, such that we can compare with experimental results
for the difference between the strange and non-strange hyperfine
splittings. 

For $B$-mesons the light quark dependence of the hyperfine splitting
is not well resolved experimentally, because of large uncertainties in
the $B_s$ hyperfine splitting. For $D$-mesons the situation is much
clearer and one observes an increase with the light quark
mass. However the magnitude of the slope is largely dependent on
whether you compare the $D_s$ hyperfine splitting with the hyperfine
splitting of the $D^+$ or the $D^0$. We expect this difference in the
experimental results to be mainly due to QED-effects, since these come
in with opposite signs in the $D^+$ and the $D^0$. Since the $D_s$ is
positively charged as well, QED effects should largely cancel when
comparing the hyperfine splittings of the $D_s$ and the $D^+$ and one
obtains a positive slope for the $D$-meson from the experiment.

Comparing our data to the experimental results, one observes our
hyperfine splittings to be too small. This will be discussed in more
detail in section~\ref{spectrum_section}.  With respect to the slope,
the result at the $D$ has clearly the wrong sign and its magnitude is
approximately twice as large as that from the experiment. We did not
observe this effect in our $\beta=6.2$ results, neither was it
observed in \cite{arifalat98}. Both of these results did not achieve
the high statistical accuracy we have at $\beta = 5.7$ and also use
values of the heavy quark mass at around the $B$ or heavier. For those
values of $m_Q$ the light quark mass dependence at $\beta=5.7$ is also
not that significant.

A slope of similar sign and size has been observed in the calculations
of \cite{lewiswolo,peterhl}, although the authors did not comment on
this. Reference~\cite{lewiswolo} used a highly improved gluonic action
with NRQCD heavy quarks on even coarser lattices and \cite{peterhl} a
heavy clover action for the heavy quarks on a finer lattice
$\beta=6.0$. A detailed comparison with these results will be given
below in section~\ref{spectrum_section}.  In this context it is
interesting to note, the slope of the hyperfine splitting as a function of the
quark mass turns out to be to small in light hadron spectroscopy as
well \cite{yoshieed97}.

The calculations listed above are performed in the quenched
approximation, which could be a factor contributing to 
the wrong slope. In
potential model language, which is not necessarily appropriate here,
the hyperfine splitting is related to the square of the wave-function
at the origin. This in turn depends on the light quark mass and is
independent of $m_Q$ as $m_Q \to \infty$.  The wrong slope could then
reflect the fact that the wave-function at the origin is not
increasing rapidly enough as the light quark mass increases. This is
natural in the quenched approximation as the potential at the origin
is weakened by the coupling constant running too quickly to zero at
short distance.

An alternative scenario is one in which the coefficients of the
relevant terms in the action, here $c_{\rm sw}$ in the light quark
action, effectively carry some quark mass dependence that has not been
included, leading to an underestimate of the hyperfine splitting at
large $am_q$. In this case the effect would disappear as $a$ is
reduced and this seems to be contradicted by results on finer lattices
\cite{peterhl}. 

Another cause could be a problem in the chiral extrapolation
itself. The experimental result for hyperfine splitting $J/\psi -
\eta_c$ in the charmonium system is smaller than the hyperfine
splitting for the $D$ and $D_s$-mesons. If one considers charmonium as
a $D_c$-meson, one has to conclude that there is a maximum of the
hyperfine splitting as a function of the light quark mass for $m_q <
m_c$. If this maximum is attained for $m_q < m_s$, our observation of
a negative slope of the hyperfine for $m_q \approx m_s$ would be in
agreement with nature. In this case extrapolations from the strange
region to lighter $u$ and $d$-quarks as well as the chiral limit would
be impossible.

\begin{figure}
\centerline{\epsfig{file=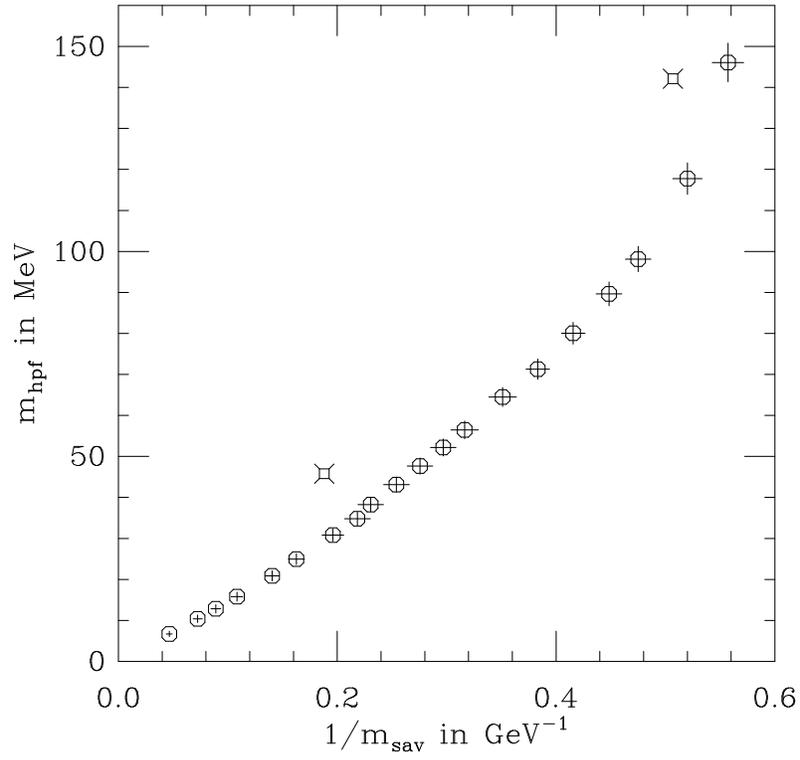,height=11cm}}
\caption{\label{hpfmsavfig} The hyperfine splitting as a function of
the spin-averaged heavy-light meson 
mass. Error bars give statistical uncertainties only. The fancy squares
give the experimental result \protect\cite{pdg}.}
\end{figure}

Our final results for the hyperfine splitting at $\kappa_c$ and
$\kappa_s$ are given in table~\ref{hpftab57}. In
figure~\ref{hpfmsavfig} we display the dependence of the hyperfine
splitting on the spin-averaged heavy-light meson mass. A linear fit in
$m^{-1}_{\rm sav}$ for the five heaviest $m_{\rm sav}$ values
gives reasonable values of Q. In table~\ref{hqettab} we give the
numerical outcome of this fit for the strange and non-strange
hyperfine splittings. As expected from HQET the intercept always turns
out to be zero within statistical errors.

In the H run for $\beta=6.2$ we determine the hyperfine splitting from
ratio-fits. In order to determine this without excited state
contamination we use a fit interval for which both of the individual
correlators have reached a plateau. As noted above, no significant dependence
on the light quark mass was observed and we were able to fit the
results to a constant with reasonable values of $Q$. The result is
given in table~\ref{hpftab62}.

In case of the N and P run we determined the hyperfine splitting from
the jackknife difference of masses obtained from the pseudo-scalar and
vector meson propagator. In the N run we compared the outcome for the
different smearings available, for different values of the starting
point $\tmin$ of the fit range. An example is shown in
figure~\ref{b62hpfsmearfig}. The different smearing functions lead to
compatible answers for the hyperfine splitting. We use the outcome
from the propagators with sink and source smearing $\hsmear{g}{1}$ for
our final result. In figure~\ref{b62hpfrunfig} we compare the outcome
of the different runs at $\beta=6.2$. Clearly the outcome from the run
N is the most precise. The result for the physical $B_s$ hyperfine
splitting will be extracted from this results.

\begin{figure}
\centerline{\epsfig{file=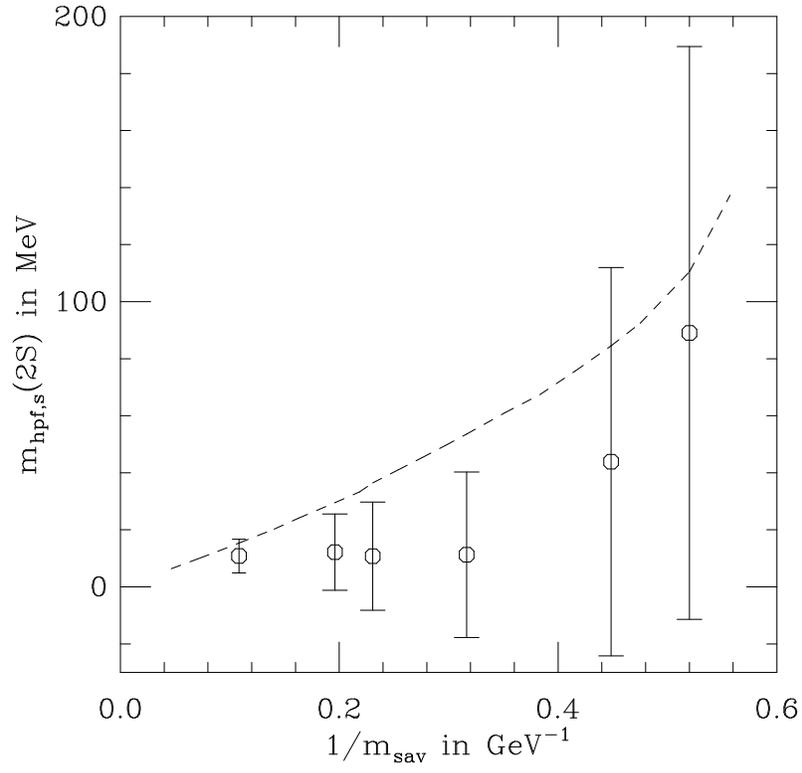,height=11cm}}
\caption{\label{hpf2Smsavfig} The hyperfine splitting of the radially
excited $S$-wave state is given by the octagons. The error bars give
statistical uncertainties only. The dashed line gives the ground state
hyperfine splitting of the strange meson for comparison. This line is 
not a fitted curve.}
\end{figure}

\begin{figure}
\centerline{\epsfig{file=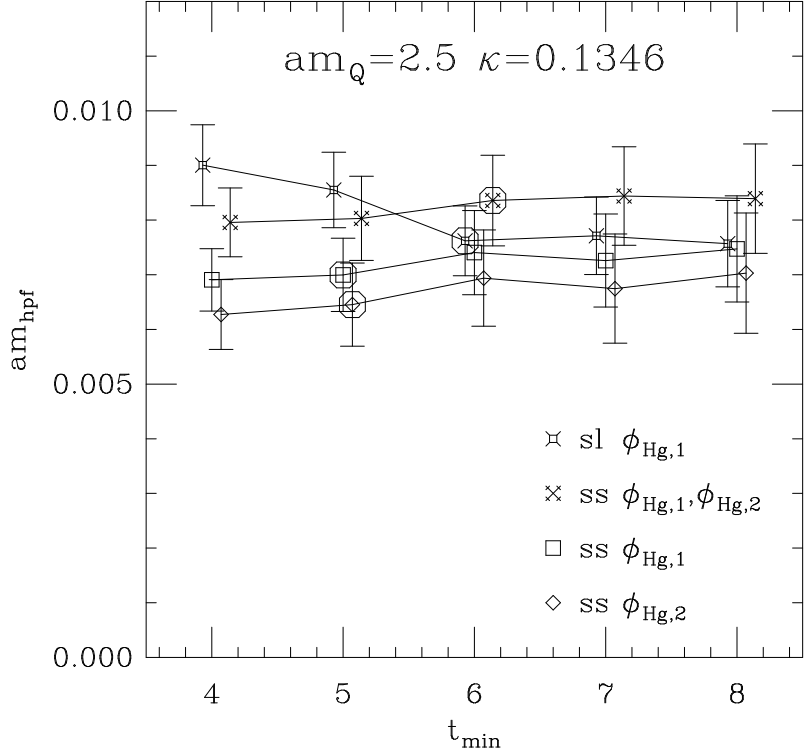,width=11cm}}
\caption{\label{b62hpfsmearfig} Hyperfine splitting in the N run at
$\beta=6.2$. The results are extracted from the difference of fitted
masses for the $^3S_1$ and $^1S_0$ propagators. We display the
dependence on the starting point $\tmin$ of the fit range of the
propagators. In all cases we used single exponential fits.
With `sl' we denote results obtained from smeared-local
propagators, `ss' refers to smearing at source and sink. The octagons
give those $\tmin$, which we determined in the $Q$-value analysis to
give the final result. The connecting lines are for guidance only.}
\end{figure}

\begin{figure}
\centerline{\epsfig{file=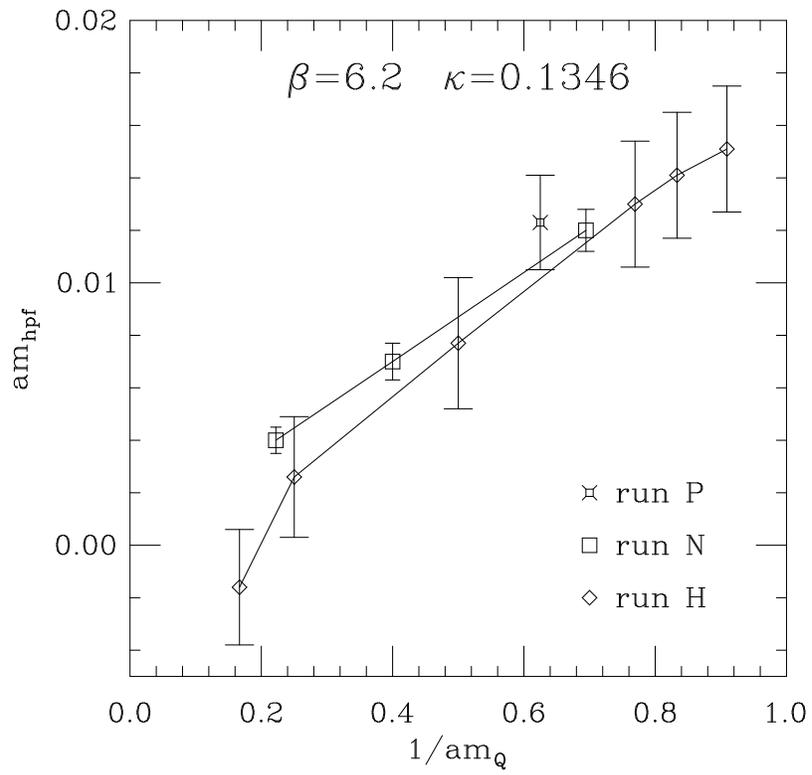,width=11cm}}
\caption{\label{b62hpfrunfig} Comparison of the outcome for the
hyperfine splitting from the different runs at $\beta=6.2$. The lines
are for guidance only and connect the points of the different runs.  }
\end{figure}

\begin{figure}
\centerline{\epsfig{file=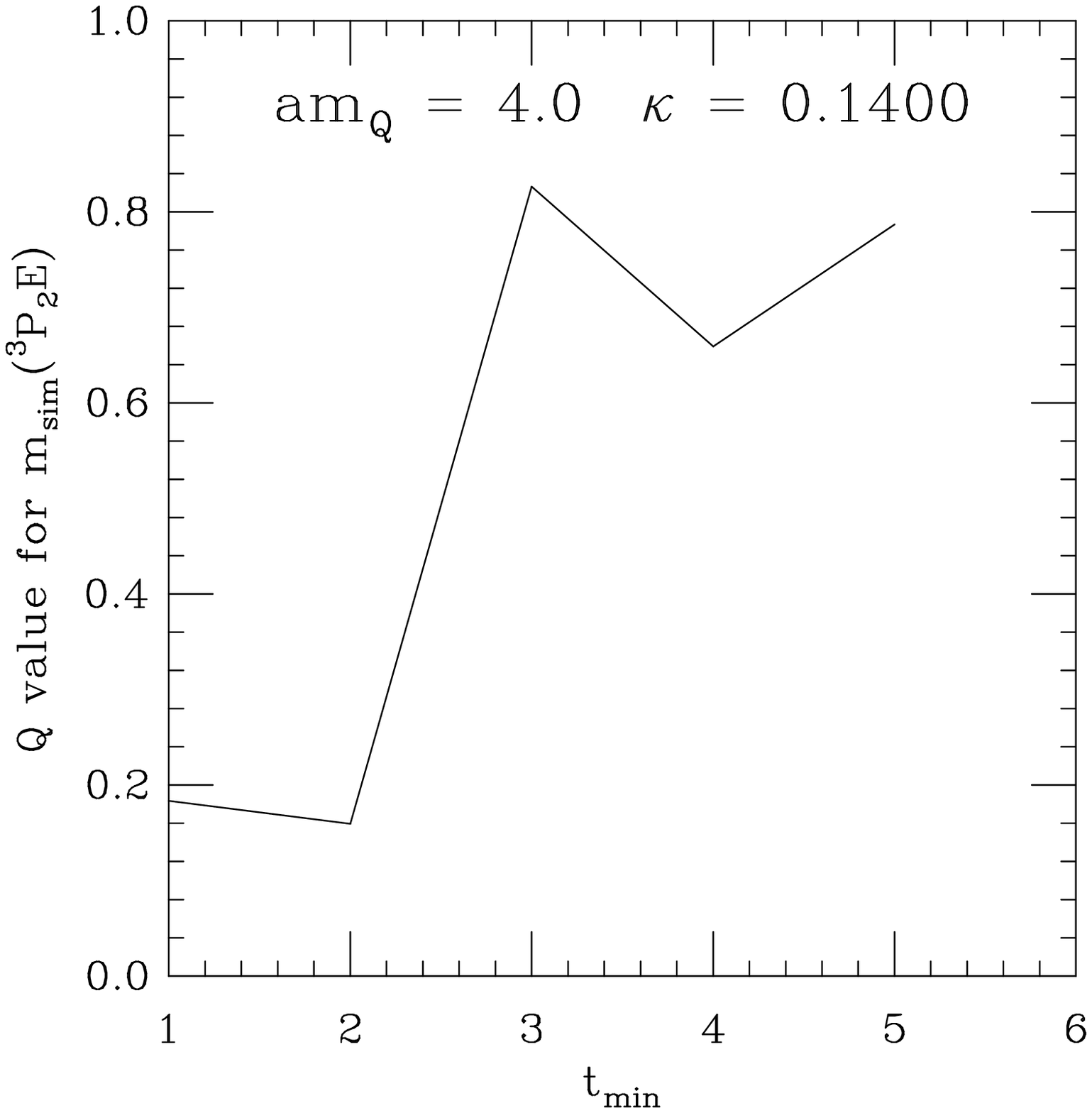,width=8cm}
\epsfig{file=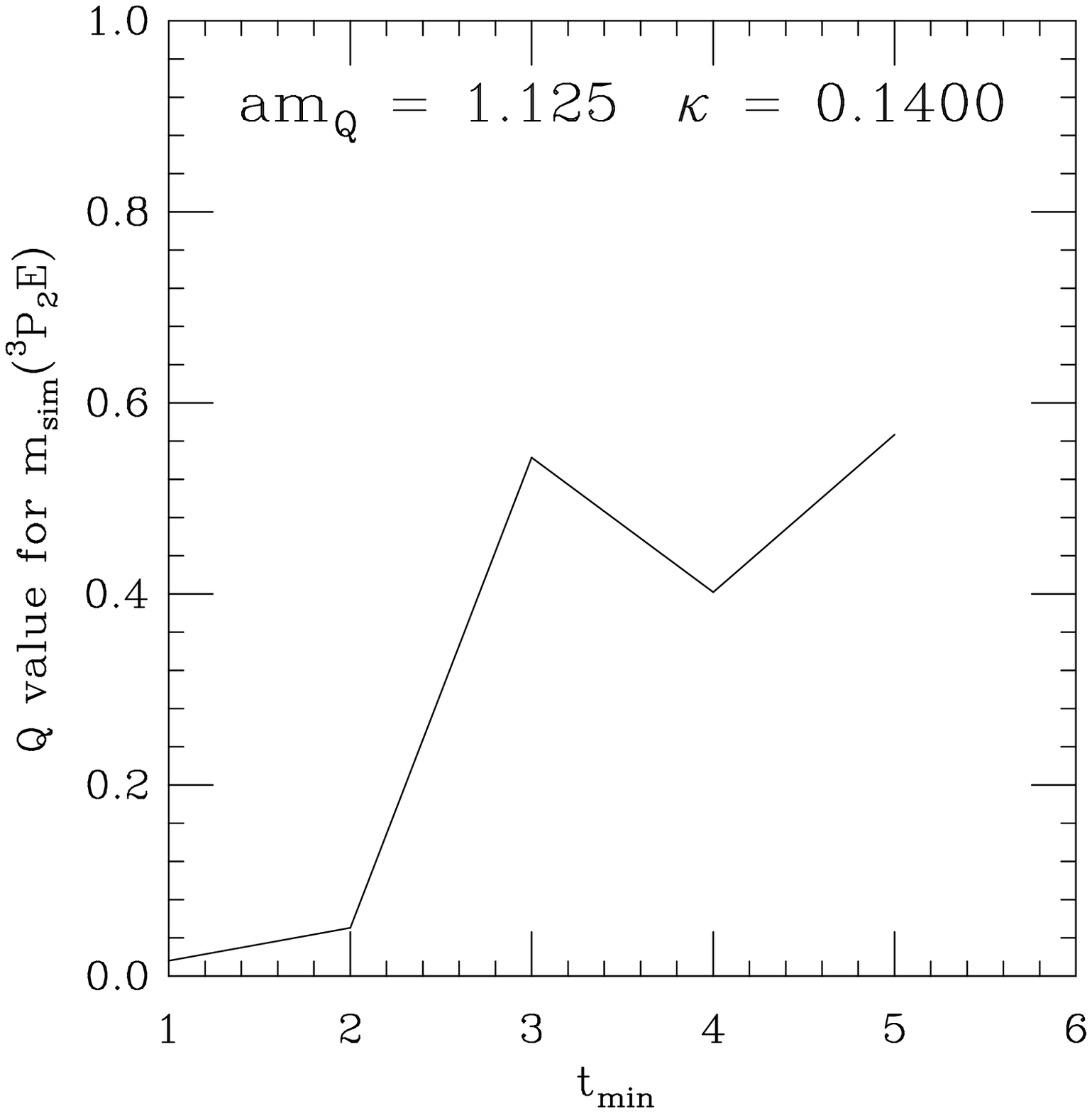,width=8cm}}
\centerline{\epsfig{file=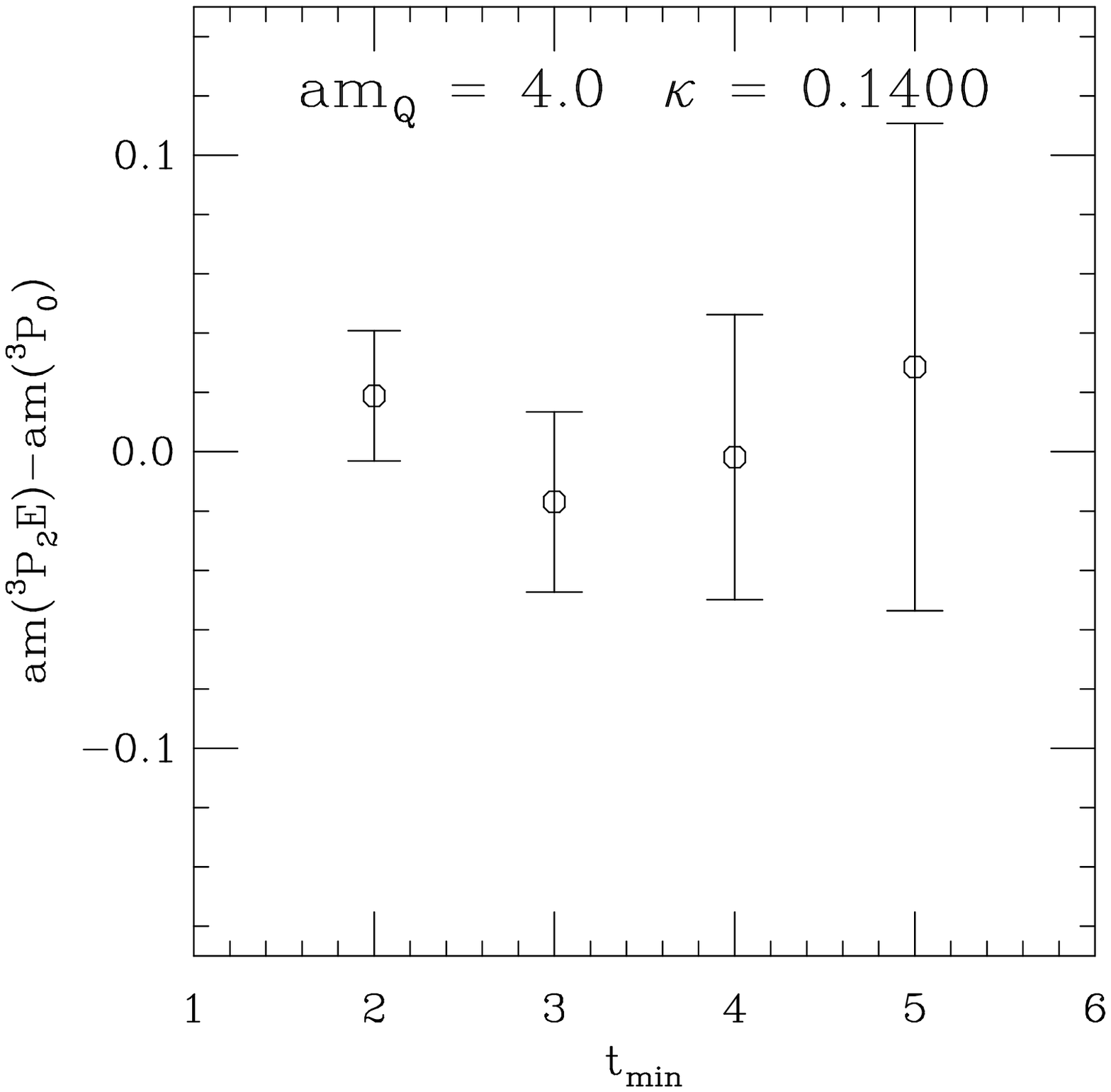,width=8cm}
\epsfig{file=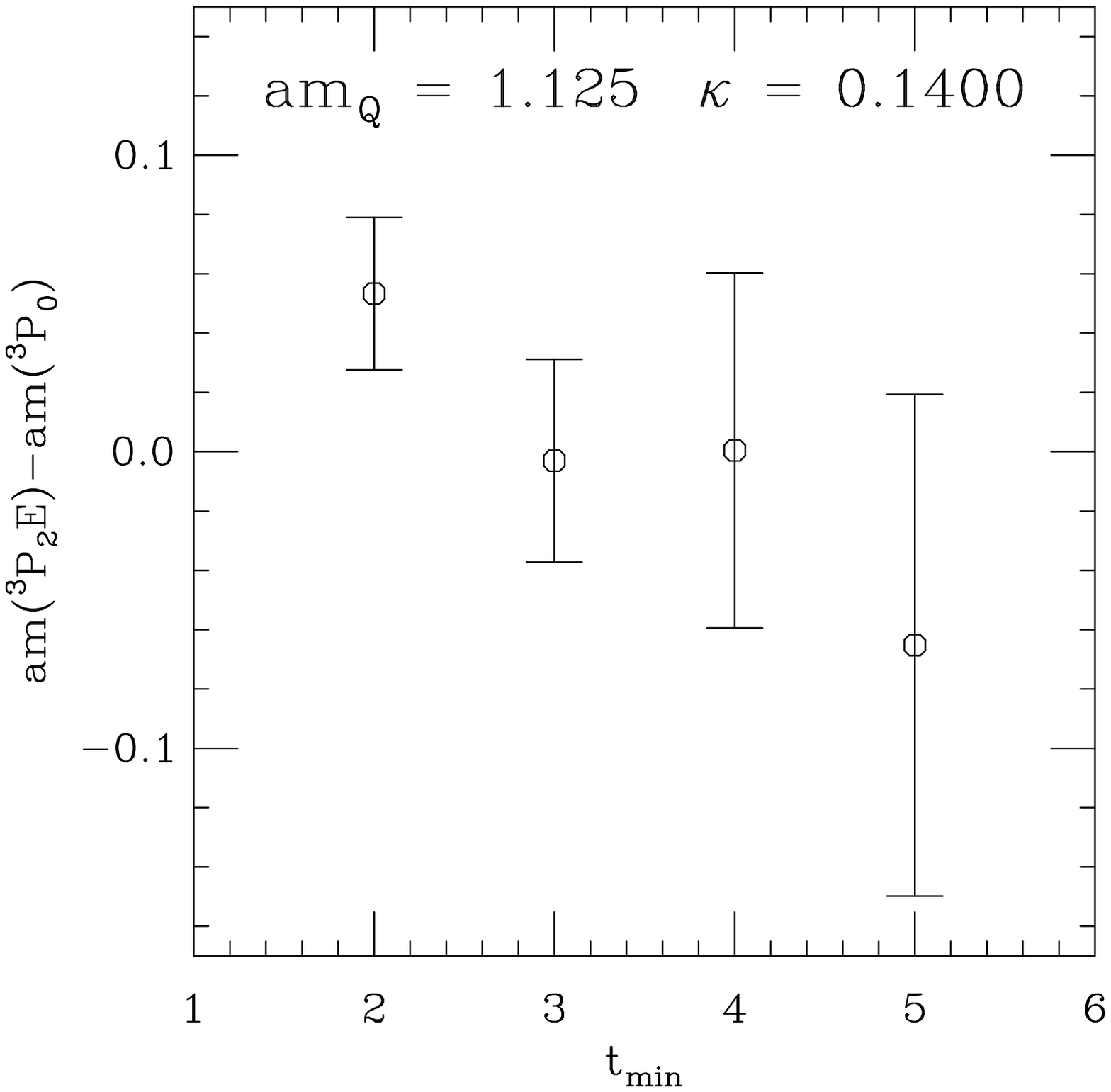,width=8cm}}
\caption{\label{pfinetminplot} Fit range dependence of the
$m(^3P_2E)-m(^3P_0)$ $P$-state fine structure from a double exponential
matrix fit at $\beta=5.7$. The upper line displays the $Q$-value of
the fit to the $^3P_2E$-state. The fit to the $^3P_0$-state gives
$Q > 0.6$ even for $t_{min}=1$. The bottom line gives the splitting
as determined from the difference of the individual fit results.}
\end{figure}

Having observed clear signals for the radially excited $S$-wave states
on our coarse lattice, we also studied their hyperfine
splittings. Unfortunately the statistical noise grows rapidly and we
observe no clear signal for a non-zero splitting.  Our results are
given in figure~\ref{hpf2Smsavfig}, comparing the radially excited
state hyperfine splitting to that of the ground state. Although we
cannot give a value for the hyperfine splitting of the radially excited
$S$-state, our results support the expectation that it should be equal to
or smaller than the ground state splitting.

\subsection{$P$-state fine structure}\label{pfinesubsec}
To extract the $P$-state fine structure we investigate the jackknife
difference of the masses of the individual channels reported in
tables~\ref{ptab57} and \ref{ptab62}. Because the statistical noise
grows rapidly as $\tmin$ increased, this proved to be delicate. For
$\beta = 5.7$ this is illustrated in figure~\ref{pfinetminplot}. For
the matrix-fit to the $^3P_2E$ propagators we observe a jump in $Q$
for $t_{\rm min} = 3$. However in the plot of the fit range dependence
of the $^3P_2E - {^3P_0}$ splitting the statistical uncertainty
doubles between $t_{\rm min} = 3$ and $5$. Therefore it is hard to
tell whether there is a plateau or not.

We quote final results for $t_{\rm min} = 5$, which corresponds to
dropping 2 timeslices from the first reasonable $Q$-value. With this
procedure we obtain a large statistical error and no significant
splitting can be resolved. More aggressive fitting would have led to
a result compatible with zero but with a statistical error of $\approx
30$~MeV. We give our final numbers in table~\ref{psplittab57}.  For
the splitting we used the same fit range for both channels, which
leads to slight deviations from the direct difference of the results
in table~\ref{ptab57}.

For $\beta = 6.2$ the situation is easier, as shown
in figure~\ref{pfinetminplot62}. The noise on the splitting does not
grow as fast as on the coarse lattice, because the $P-S$ splitting is
smaller in lattice units. We observe the first
reasonable $Q$ values for $t_{\rm min}=3$. Since this is a 6
parameter fit to two propagators, we drop 3 time slices and quote the
final result for $t_{\rm min} = 6$. The results are given in
table~\ref{psplittab62}. Here we also quote results for the splitting
of the $J=1$ channels to the $J=0$ state.
We reiterate that no significance should be attached to any difference
in the results between the $^3P_1$ and the $^1P_1$ operators.

\section{The physical meson spectrum} \label{spectrum_section}
In this section we determine the physical $B$ and $D$-meson spectrum
and investigate scaling by comparing results at different values of
the lattice spacing. We also compare with experimental results and other
lattice calculations. 

\subsection{$B$-meson spectrum}\label{bmesonsubsection}
At both of our lattice spacings we can simulate the $b$-quark
directly. Here we discuss our results for the physical
$B$-spectrum. Together with the findings of \cite{arifalat98} we want
to investigate the dependence of the individual splittings on the
lattice spacing. The findings are compared both to the existing
experimental results and lattice investigations performed by other
groups within a similar framework using NRQCD
\cite{lewiswolo,ishikawa}.

The results of reference~\cite{lewiswolo} are useful in that they work
at a larger lattice spacing than we do here. There are a number of problems,
however that make their results not directly comparable. For example,
they do not use either smeared correlators or standard fitting
techniques, and this will give rise to an unknown systematic error. In
addition they do not see a difference between fixing the lattice spacing from
$m_\rho$ using the clover action and from the charmonium $1P-1S$
splitting. This is clearly seen on finer lattices
\cite{rowland,hpsprd55,jpsinrqcd}. If this arises from overestimating
$\ainv$ from $m_\rho$ because of discretisation errors, then this is
another source of systematic error. In particular, this feeds
into the fixing of the bare $b$ or $c$ quark mass and
into hyperfine splittings.  Their final result for the splitting does
not take into account the effect of any of the uncertainties in the
bare quark mass determination. This is particularly important for the
hyperfine splitting and causes their errors to be heavily
underestimated.

The results of \cite{ishikawa} overlap with, but are not as complete
as ours.

Unfortunately there are no results for heavy clover fermions available
that we can use. References \cite{peterhl,fermispec} quote numbers for
the $B$ spectrum. The first reference still uses extrapolations from
the lighter quark masses into the $b$-region.  Reference
\cite{fermispec} determines the bare $b$-quark mass from heavyonium,
which is not suitable for the heavy clover approach at the lattice
spacings used \cite{sarashift,kronfeld}. However we will later compare
to their findings for the $D$-spectrum, since this problem is not so
severe for charmonium at the lattice spacings used.

Results from taking the $b$-quark as a static source also exist for
spin-independent and flavour
 splittings, which survive in the infinite mass limit, see for example
\cite{michaelpeisa}. However we restrict the discussion here to a
comparison with results simulated at the physical $b$-quark mass directly.

In the following we denote spin-averaged states by an overline.

We summarise our results for the $B$-spectrum in the
tables~\ref{Bsplittab} and \ref{Bsplittab6.2}.  As an example of the
splitting between a strange and a non-strange meson we discuss the
difference of the pseudo-scalar $B_s - B$ splitting in
figure~\ref{bscalefig}. We observe no scaling violations between our
results and the agreement with the experimental value is excellent.

The results from \cite{ishikawa} given in figure~\ref{bscalefig} include
the statistical errors only and are taken from their figure~15. In
\cite{ishikawa} additional uncertainties for the average of the
results at the two finest lattices are mentioned in the text. The
overall agreement with our results is good.  They notice an upward
jump, however, for their result on their finest lattice. Unfortunately
our result from our finest lattice comes with large uncertainties so
that we are unable to clarify whether there is any real effect
here. Given the lack of scaling violations in the rest of the results,
it seems unlikely to us.  Table~\ref{ainvtab} confirms that the scale
from $m_\rho$ used by us and the scale from $\sigma$ used in
\cite{ishikawa} are very close. The mismatch of scales of
$\approx 3\%$ can be neglected safely. 

The results from \cite{lewiswolo} are also in agreement with
ours. They use the $K^*/K$ ratio to fix the strange quark mass. This
reduces their results compared to that using the $K/\rho$
ratio. Assuming a shift of $10$ to $20$~MeV from this would increase
the agreement. This is the size of the effect we observe on our coarse
lattice from fixing the strange from $\phi/\rho$ instead of the $K$.

Figure~\ref{bradscalefig} shows the scaling of the radially excited
$B_s$-meson.  As discussed in subsection~\ref{radialsubsect} already,
the extraction of a
result for our finest value of $a$ turned out to be more problematic
than anticipated, and we are left with quite large statistical
uncertainties.  However the final result is in good agreement with the
result from our coarser lattice as well as the result of
\cite{arifalat98}.

We also included a preliminary experimental result from the DELPHI
collaboration for an admixture of the non-strange $B'-B$ and $B^*{}'-
B^*$ splitting \cite{bradref1,bradref2}. Assuming the hyperfine
splitting of the two states to be of similar size, which our findings
support, we observe reasonable agreement here. Table~\ref{Bsplittab}
also contains results for the radial excitation energy of the vector
state and the spin-averaged $S$-wave.

The orbital excitations are compared in figure~\ref{borbscalefig}. The
lattice results for the spin-averaged strange $P$ state scale very
well. The magnitude agrees nicely with the $B_{sJ}^*(5850)$ resonance,
which is expected to be an admixture of the two $j_l=\frac{3}{2}$
states.

Our results for the radially excited $P$-states are compared in
figure~\ref{borbradscalefig}. This is the first observation of a
signal for these states in a lattice calculation.  As the figure shows
we get consistent results from the two different lattice spacings
investigated.  To the best of our knowledge radially excited
$P$-states have not been observed yet experimentally.

The splittings discussed above are all essentially light quark
quantities, which survive into the static limit. Their scaling or
non-scaling says more about the light quark action than the heavy
quark sector. The hyperfine splitting is of a different nature and
from its scaling behaviour one can learn about how well the heavy
quarks are being described on the lattice. We display this in
figure~\ref{bhypscalefig}. Our result for the strange hyperfine
splitting together with the findings from \cite{arifalat98} shows good
scaling.

However the result is much smaller than the experimental value. Since
the leading term in the hyperfine splitting arises from the
${\mbold{\sigma}\cdot \mbold{B}}$ term in the action,
eq.~(\ref{deltah}), the result for the splitting is sensitive to the
coefficient $c_4$ and the inclusion of radiative corrections beyond
tadpole improvement is required.  Preliminary
calculations \cite{c4pap} indicate that the inclusion of the 1-loop
corrections would increase the hyperfine splittings on the order of
$10\%$ for the lattice spacings used. The quenched approximation
might also play a r\^ole here, since in light spectroscopy the
hyperfine splittings turn out to be too small as well, see
\cite{yoshieed97} for a review. This effect increases with increasing
quark mass.  Unfortunately reference
\cite{saradyna}, which investigates the effect of the inclusion of two
flavours of dynamical quarks on the $B$-meson, does not give any
evidence for an increase of the $B^*-B$ splitting due to sea-quark
vacuum polarisation effects.

From the experience \cite{Upscale} in $\Upsilon$-spectroscopy using NRQCD, 
one could have expected to observe scaling violations in the hyperfine 
splitting. Using the same heavy quark Hamiltonian as we do, \cite{Upscale}
reports an increase of $50\%$ for the $\Upsilon-\eta_b$ splitting, within 
the range from 
$\beta=5.7$ up to $\beta=6.2$. The leading discretisation correction 
for the hyperfine splitting is ${\cal O}((a p_{\rm gluon})^2)$ \cite{LepD46}.
Typical gluon momenta for the $\Upsilon$-system are $\approx 1$~GeV, while
for the $B$-system they are ${\cal O}(\lqcd)$. From this one expects reduced
scaling violations of $\approx 10\%$ in the $B$-system for our range of
lattice spacings. This is the same size as our uncertainties on the
hyperfine splitting and therefore consistent with the fact that no
scaling violations show up in figure~\ref{bhypscalefig}.

Results from \cite{ishikawa} are for the chirally extrapolated
splitting $B^*-B$ with statistical errors only, taken from their
figure~17.  In the text, the authors quote a result for the strange
hyperfine splitting from the average of the two finest lattices, which
is $3$~MeV higher than the same average for the non-strange hyperfine
splitting. An upwards shift of $3$~MeV increases the already excellent
agreement even further.

The results of \cite{lewiswolo}, on the other hand, exhibit a clear
disagreement to our findings as well as the findings of \cite{ishikawa}. We
believe that this is because they have determined the
bare $b$-quark mass $am_b$ using heavyonium. 

The fine structure of the $P$-states is the last topic of this
section. Unfortunately we have not been able to resolve this clearly
on our coarsest lattice. The situation is displayed in
figure~\ref{pfinescale}, for the three sublevels which were resolved
at $\beta=6.0$ and $6.2$. To investigate whether there
is evidence for scaling violations in the fine structure we calculate
the jackknifed difference of the highest and lowest state. This is
shown in figure~\ref{bpfinescalefig}. The error bars turn out to
be large and the figure is inconclusive. A more aggressive fit on the 
coarser lattice, as discussed in subsection~\ref{subsecthpf} would 
lead to the conclusion that scaling violations were seen, but we believe 
that further work is needed to resolve this question.

Our results for the $B$-meson spectrum, together with those of
\cite{arifalat98} do not show signs of residual lattice spacing
dependence within the achieved accuracy. Therefore we can average the
results for the different values of the lattice spacing $a$ to obtain
our final results on the quenched $B$-meson excitation spectrum. The
averages were obtained in the following way, for each value of $a$ we
add the different uncertainties in quadrature to obtain a single
value. Here we omitted those sources of uncertainty which are
associated with the quenched approximation. These are the uncertainty
arising from fixing the strange quark mass from different physical
quantities and in the case of the results of \cite{arifalat98} the
additional uncertainty of the lattice spacing $a$ associated with the
physical quantity used to fix $a$.  At this step we also symmetrised
with respect to unsymmetric uncertainties. The central values have
been obtained from fitting the results to a constant in an
uncorrelated fit with the above described uncertainties. This puts
more weight on the more precise results than a simple average.

Our analysis at the individual values of $a$ does not include an
uncertainty for the residual effect of the lattice spacing.  Our
actions are improved to ${\cal O}(\alpha_sa,a^2)$. Therefore for
each value of $a$ we add the maximum of $\alpha_s a \lqcd$ and $a^2
\lqcd^2$ in quadrature to the uncertainty used in the fit. We quote
the smallest of these three so obtained 
uncertainties as our final uncertainty for
the quenched $B$-meson spectrum. This way we quote an accuracy which
is of the same size as the one we checked for scaling
violations. Determining the final uncertainty from the $\chi^2$ of the
fit would reduce the uncertainty beyond this level. This procedure
also ensures that residual lattice spacing artifacts are properly
included if the achieved accuracy differs over the three individual
results and the average is largely determined by the coarser lattices.

Our final result on the $B$-meson splitting spectrum in the quenched
approximation is given in table~\ref{Bsplittabsum} and
figure~\ref{bspectrumfig}.

The question of the effect of quenching on the spectrum goes beyond
the scope of this paper. We refer the reader to \cite{saradyna}. There
the effects of the inclusion of 2 flavours of dynamical quarks on the
spectrum in NRQCD have been investigated and compared to the findings
of \cite{arifalat98}.  No significant difference between the quenched
and $n_f=2$ results for the $1S$, $2S$ and $1P$-states was found. In
particular, as mentioned earlier, no significant sea quark effects
were seen in the $1S$-hyperfine splitting. Since our investigations
confirm scaling in the quenched heavy-light spectrum the conclusions
of \cite{saradyna} are unchanged.

\subsection{$D$-meson spectrum}
In this section we discuss the $D$-meson spectrum and compare our
result to existing lattice results as well as to experiment.

The convergence of the NRQCD expansion is particularly important in the
$D$-range, where the expansion parameter $\lqcd/m_Q \approx
\frac{1}{4}$.  Useful results on the 
question of the convergence are contained in
\cite{lewiswolo}. There the authors study the complete NRQCD action to 
$\nrqcdord{3}$. Here we include all terms up to $\nrqcdord{2}$ and the
relativistic correction to the kinetic energy in $\nrqcdord{3}$.  The
authors of \cite{lewiswolo} calculate the heavy light kinetic masses
using eq.~(\ref{hdisp3}) and show that the difference arising from
$\nrqcdord{3}$-terms is consistent with the expectation that they are
sub-sub-leading in a $\lqcd/m_Q$ expansion. The changes to the
spin-averaged meson mass that we use to fix the quark mass
are dominated by the $\frac{p^4}{m_Q^3}$
relativistic correction that we include. From this we conclude that
remaining $\nrqcdord{3}$ and higher order terms in the NRQCD expansion
would only change the physical masses by at most a few percent. This
allows us to use the results of \cite{lewiswolo} to estimate the
changes in the hyperfine splitting which would be produced by these
additional terms at fixed bare quark mass.

The authors of \cite{lewiswolo} find that the $\nrqcdord{2}$ terms
produce an effect somewhat smaller than a $\lqcd/m_Q$ expansion might
suggest, since they affect the hyperfine splitting indirectly. The
only spin-dependent term at $\nrqcdord{2}$ is a spin-orbit type
interaction.
At $\nrqcdord{3}$ most terms produce a change of a few percent, but
the one which is directly related to the
$\mbold{\sigma}\cdot\mbold{B}$-term: $\{\DD^2,\mbold{\sigma}\cdot
\mbold{B}\}$ reduces the hyperfine splitting at the charm by
$20\%$. However when including the other operators of $\nrqcdord{3}$,
the second largest effect comes from the $\mbold{\sigma}\cdot(\mbold{E}\times 
\mbold{E} + \mbold{B}\times \mbold{B})$ operator, which is also
spin-dependent
and works in the opposite direction to the other one.  The total
effect of $\nrqcdord{3}$ is below $10\%$. This is of the size of the
na\"\i{}ve expectation for the suppression with respect to the leading
term and not at all inconsistent with good convergence of the NRQCD
expansion. Since we do not include these terms, we conclude, that we
may be overestimating the quenched lattice hyperfine splitting of the
$D$-meson by $10\%$.

Our results on the $D$-meson spectrum are summarised in
table~\ref{Dsplittab} and in figure~\ref{dspectrumfig}. The overall
agreement to the experimentally observed spectrum is good. We will now
discuss the individual splittings in more detail. We also compare our
results to the lattice studies of
\cite{lewiswolo,peterhl,fermispec}. All of these results use the
quenched approximation as well. The publications
\cite{peterhl,fermispec} apply the heavy clover
approach~\cite{fermilab}, which has quite different systematic
uncertainties from NRQCD for charm quarks.

The flavour dependent $D^{(*)}_s - D^{(*)}$ splittings
are in good agreement with the experimental results. Here it is
interesting to note that our results reflect the increase
of $\approx 10$~MeV from the $B$ to the $D$-meson system, which can
already be expected from the good agreement of the slope with the 
experimental outcome in table~\ref{hqettab}.  

In figure~\ref{dscalefig} we compare our result for the strange
to non-strange 
spin-averaged splitting
to other lattice calculations. In order not to disguise 
other possible
effects, we excluded the uncertainty of the strange quark mass from
the plot. The results from \cite{lewiswolo} use the $K^*/K$ ratio to
define the strange quark. This should shift the results downwards, compared
to fixing $\kappa_s$ from the $K/\rho$ ratio as used for the other
results. The implications have already been discussed in the previous
sub-section~\ref{bmesonsubsection}. 
The uncertainties again allow for an upwards
shift of these results by $10$ to $20$~MeV. It should be noted that
we combined the results for the hyperfine splittings and the
pseudo-scalar $D_s-D$ splitting from \cite{lewiswolo} to obtain the
spin-averaged splitting.

Because of different systematic uncertainties, it is particularly
interesting to compare to the heavy clover results of
\cite{peterhl}. The results obtained with the use of the
$m_\rho$-scale, which is the same as what we use, agree very well with
ours. This agreement is expected since \cite{peterhl} uses the same
light quark and gauge field action and this quantity is essentially
determined by the light quarks and the gluon field.  We conclude that
the results for the flavour dependent $D_s-D$ splitting agree well
between the different approaches and, within the accuracy achieved,
agree well with the experimental result.

For 
radially excited $D_s^{(*)}{}'$-mesons no experimental results are
known to us. However the DELPHI collaboration reports on the
observation of the non-strange
$D^{*}{}'$ \cite{Dradial}. This result is still awaiting confirmation
by the OPAL and the CLEO collaboration and its interpretation is disputed
on the ground of its small experimentally 
observed width \cite{melikhov,rpage}.
The splitting between the DELPHI result and the
$D^*$ has a similar size to our $D_s^{*}{}' - D^*_s$ splitting.
Reference~\cite{fermispec} reports lattice results from the
heavy-clover approach. From their plot we read $D_s'-D_s \approx
840(160)$~MeV, which is in agreement with our findings. However this
includes what the authors call ``\emph{continuum}'' extrapolation out
of a regime where the expansion parameter $am_Q = {\cal O}(1)$ is not
small. We would prefer to compare to the unextrapolated results
at the individual values of $a$.

Experimentally the only well established charmed $P$-states in the 
particle data book~\cite{pdg} are those which are expected to correspond to
the states of total light angular momentum $j_l=\frac{3}{2}$.
Recently the CLEO collaboration \cite{D1wide,D1wide2} claimed the observation
of the $D_1$ state corresponding to $j_l = \frac{1}{2}$. 
CLEO gives a preliminary result of 
$D_1 = 2461(^{+41}_{-34})(10)(32)$~MeV, which
is slightly heavier but compatible in error bars to the  $D'_1 = 2425$~MeV
\cite{pdg}\footnote{We quote the charge-average.}. 
Our lattice calculation delivers the mass of the lighter of the two states.
We did not observe a signal for an excited state slightly heavier than this. 

In table~\ref{Dsplittab} our result for the $\bar D_s(1P) - \bar D_s$
splitting is compared to the spin-average of the $D'_{s1}$ and
$D^*_{s2}$, the $j_l= \frac{3}{2}$ states.  The agreement is
reasonable. Reference \cite{peterhl} reports on the $D_{s1} - \bar
D_s$ splitting from a lattice study with the heavy clover approach.  A
comparison to our result for this splitting is given in
figure~\ref{dpscalefig}. When using the same scale obtained from
$m_\rho$ both lattice results agree very well with each other. The
agreement with experiment is also good.

It should be noted however, that the experimental result included in
figure~\ref{dpscalefig} is not necessarily 
the same as ours. The experimental $P$-state corresponds to
$j_l=\frac{3}{2}$. If the CLEO trend is confirmed and the $D_1$ is indeed
heavier than the $D'_1$ and the same holds for the $D_{s1}$ states, then
the lattice result also corresponds to $j_l=\frac{3}{2}$. If not the lattice
result will correspond to $j_l=\half$, but the two states will be so close,
that any mismatch is well covered by the error bars.

Table~\ref{Dsplittab} contains our final result for the radially
excited $2P$-state. This is the first result for this state from a
lattice simulation.

As in the $B$-system the hyperfine splittings are too small when
compared to the experimental result. Whereas in the $B$-system they
were too small by $\approx 40 \%$, here they are low by $\approx 25
\%$.  This could reflect a more severe quenching error for $B$-mesons.
$B$-mesons are somewhat smaller states than $D$ mesons and probe
slightly different scales. This is a sub-leading effect in a heavy
quark symmetry picture, however. Alternatively, if the error comes from
radiative corrections to the $c_4$ coefficient, that would need to
increase with $m_Q$.  That is seen by the authors of \cite{c4pap}.

The large uncertainty of $\approx 20$~MeV on our result arises from
the chiral extrapolations used in the lattice spacing determination
and the way in which this feeds into the fixing of the bare quark
mass.  Naively we expect a doubling of the relative error, because a
larger value of $a$ requires a smaller value of $am_Q$ to deliver the
same physical $m_{\rm sav}$. This smaller value of $am_Q$ gives a
larger hyperfine splitting $am_{hpf}$. When converting to physical
units it picks up the uncertainty of $a$ for the second time.  In fact
a factor of four is seen because of the flattening of the relation
between the mass shift $\Delta$ and bare heavy quark mass $m_Q$, see
figure~\ref{shift57fig}, as well as the steepening up of the hyperfine
splitting curve for large values of $m_{\rm sav}$ in
figure~\ref{hpfmsavfig}.

In figure~\ref{dhypscalefig} we compare our results to the results
from~\cite{lewiswolo} obtained in NRQCD. We choose their result in
$\nrqcdord{2}$ as most relevant for this comparison. The good
agreement with our results is in fact misleading. They fix their
$c$-quark mass from charmonium instead of the $D$. Fixing from the $D$
would lead to a larger value of $am_c$ and lower hyperfine splitting,
see subsection~\ref{bareqsubsect} and reference~\cite{jpsinrqcd}.

It is interesting to compare the result for the hyperfine splitting
between NRQCD and heavy clover quarks. This is also done in
figure~\ref{dhypscalefig} for the $D_s$ hyperfine splitting.  The
heavy clover results of \cite{fermispec} appear to be higher than our
result. However this is a result of their higher choice of scale
coming from $J/\psi$ instead of from $m_\rho$. This is confirmed by
the findings of \cite{peterhl}. Using a scale from $m_\rho$ gives a
result which agrees with ours, using a scale from $J/\psi$ agrees with
reference \cite{fermispec}. It should be noted that in
\cite{peterhl} the bare quark mass is determined from the $D$, where as
in \cite{fermispec} it is determined from charmonium. For the heavy
clover approach at $\beta=6.0$ these differences are negligible within
statistical errors \cite{boylepriv}. 

As discussed at the beginning of this subsection, the inclusion of the
terms $\nrqcdord{3}$ contributing to the hyperfine splitting would
decrease our result by $\approx 10\%$. However the heavy
clover approach requires similar correction terms in the Hamiltonian
to achieve this level of accuracy.

The agreement of the NRQCD and the heavy clover result for the
spin-dependent hyperfine splitting is encouraging, since the
systematic uncertainties are quite different.  The light quark content
plays only a minor r\^ole for the hyperfine splitting, which depends
essentially only on the heavy quark Hamiltonian.  In NRQCD
the leading contribution to the hyperfine splitting comes from the
${\mbold{\sigma}\cdot \mbold{B}}$ term in the action, whereas for
heavy clover this is split between the kinetic hopping term and the
clover term $\sigma_{\nu\rho} F_{\nu\rho}$, with the latter becoming
more important as the lattice spacing becomes coarser. Both these
actions give rise to systematic errors in the hyperfine splitting from
mass-dependent radiative corrections to coefficients and neglected
higher order terms, each at the $10\%$~level, so the differences could
have been significantly larger than observed.

\section{Discussion} \label{conclusec}
We present an extensive study of the $B$ and $D$-meson spectrum using
NRQCD heavy quarks and clover light quarks in the quenched approximation.

Our results include the splitting between
the strange and the non-strange meson, hyperfine splittings, radially
and orbitally excited states. For the first time in a lattice
calculation we obtained a result on radially excited $P$-wave states.
For spin-independent splittings we observe good agreement with
experimental results. However, our result for the spin-dependent
hyperfine splitting turns out to be too low in comparison to
experiment. This is a well known effect in quenched hadron
spectroscopy. Furthermore, in the present calculation hyperfine
splittings are also affected by the neglect of radiative corrections
in the matching of lattice NRQCD to continuum QCD.

Using two different values of the lattice spacing in the $B$-spectrum
together with the results of \cite{arifalat98} allows for a detailed
investigation of the residual lattice spacing dependence of our final
results. No scaling violations are observed within the achieved
accuracy. Of particular interest is the scaling of the $B_s^*-B_s$
splitting, which depends heavily on the properties of the heavy quark
content of the theory. Here scaling violations could be ruled out with
an accuracy of $\approx 10\%$. The $P$-fine structure has not been
resolved for all values of the lattice spacings and further work is
needed for this quantity.

Our results on the $B$-meson spectrum are summarised in
table~\ref{Bsplittabsum} and figure~\ref{bspectrumfig} together with
the findings of \cite{arifalat98}.  In addition to the uncertainties
considered in the analysis at the individual values of $a$, the
quoted uncertainties also contain an estimate for the residual lattice
spacing artifacts of ${\cal O}(\alpha_sa,a^2)$. The table gives our final
results for the $B$-meson spectrum in the quenched approximation.

Our final results on the $D$-meson spectrum are shown in
table~\ref{Dsplittab} and figure~\ref{dspectrumfig} above. This is our
final result for a lattice spacing of $\ainv \approx 1.1$~GeV and does
not include an estimate of the residual lattice spacing artifacts of
${\cal O}(\alpha_s a,a^2)$. For the above value of $a$, this
corresponds to $13\%$, which is of similar size to or smaller than the
otherwise achieved accuracy.

We compared our results to lattice results of other collaborations
obtained with NRQCD or in the heavy clover framework. In general we
observe good agreement. Discrepancies which appear at first sight
could be traced to underestimated errors in these other results or the
use of different scales when converting the lattice results into
physical units. The excellent agreement of our results with the
results obtained in the heavy clover approach is noteworthy because of
the different systematics of these approaches.

These results are the most complete lattice results on the $B$ and
$D$-meson spectrum to date.

\subsection*{Acknowledgements}
We would like to acknowledge useful discussions with Peter Boyle and
Peter Lepage.

J.H.~was supported by a Marie Curie research fellowship by the
European commission under ERB FMB ICT 961729, by PPARC and 
the National Science Foundation. S.C.~acknowledges fellowships from the
Royal Society of Edinburgh and the Alexander von Humboldt
Stiftung. A.A.K.~was supported by the Research for the Future Program
of the Japanese Society for the Promotion of Science. J.Sh.~would like
to thank members of the theoretical physics group at the University of
Glasgow for their hospitality during an extended visit.  Support from
an UK PPARC Visiting Fellowship PPA/V/S/1997/00666, is gratefully
acknowledged.

This work was supported by the DOE under DE-FG02-91ER40690, 
PPARC under GR/L56343 and NATO under CRG/94259.

The gauge configurations at both values of $\beta$ and the light quark
propagators at $\beta=6.2$ have been generously provided by the
UKQCD-collaboration.  The simulations at $\beta = 5.7$ have been
performed at NERSC supported by DOE and the ones at $\beta=6.2$ at
EPCC in Edinburgh supported by PPARC.

\newpage
\begin{table}
\caption{\label{gpartab} Simulation parameters of the gauge field
configurations. For $\beta=6.2$ there have been three different runs
H, N and P with different numbers of configurations. All
configurations are generously provided by the UKQCD collaboration.}
\begin{tabular}{cccc}
$\beta$      &    volume           & box size & \# configurations\\ \hline
5.7          &   $12^3 \times 24$  &   2.1 fm &  278\\
6.2          &   $24^3 \times 48$  &   1.8 fm &  H: 68; N,P: 144\\
\end{tabular}
\end{table}

\begin{table}
\caption{\label{kappa_tab} The hopping parameters used in the simulation 
are denoted by $\kappa_1$ to $\kappa_3$. The values of $\kappa_c$ and
$\kappa_s$ are taken from \protect\cite{rowland,hpsprd55}. For $\kappa_s$ we give the results as determined from $K$, $K^*$ and $\phi$.}
\begin{tabular}{c r@{.}l r@{.}l r@{.}l r@{.}l r@{.}l r@{.}l r@{.}l}
$\beta$ & \multicolumn{2}{c}{$\kappa_1$} 
             & \multicolumn{2}{c}{$\kappa_2$} 
                      & \multicolumn{2}{c}{$\kappa_3$} 
                               & \multicolumn{2}{c}{$\kappa_c$} 
                                           & \multicolumn{2}{c}{$\kappa_s(K)$}
                                                       & \multicolumn{2}{c}{$\kappa_s(K^*)$}  
                                                                   & \multicolumn{2}{c}{$\kappa_s(\phi)$} \\ \hline
5.7 & 0&1380 & 0&1390 & 0&1400 & 0&1434(1) & 0&1399(1) & 0&1393(2) & 0&1391(2) \\
6.2 & 0&1346 & 0&1351 & 0&1353 & 0&13587($^{+2}_{-5}$) & 0&13466(7) & 0&13461($^{+9}_{-21}$) & 0&13455($^{+10}_{-21}$) \\
\end{tabular}
\end{table}

\begin{table}
\caption{\label{hmasstab57} Bare heavy quark masses used in the different runs
at $\beta=5.7$. In the second, third and fourth line we give the stability 
parameter~$n$ used in the evolution equation~(\protect\ref{evolution}) of the 
runs A, C and S.}
\begin{tabular}{lcccccccccccccccccccc}
$am_Q$ & 20.0 & 12.5 & 10.0 & 8.0 & 6.0 & 5.0 &  4.0  & 3.5 & 3.15 & 2.75 & 
         2.45 &  2.2 &  2.0 & 1.7 & 1.5 & 1.3 & 1.125 & 1.0 & 0.8  & 0.6 \\ \hline
A      & 1    &  1   &   1  &  1  &  1  &  2  &  2    & 2   &  2   &  3   &
         3    &  3   &   3  &  3  &  3  &  5  &  5    &  5  &  6   &  7 \\ 
C      & 1    &  1   &   1  &  1  &  1  &  2  &  2    & 2   &  2   &  3   &
         3    &  3   &   3  &  4  &  4  &  5  &  6    &  6  &  8   & 10 \\ 
S      &  --  &  --  &   -- &  1  &  -- &  -- &  2    &  -- &  2   &  --  &
         --   &  --  &   3  &  -- &  -- &  -- &  6    &  -- &  8   &  -- \ 
\end{tabular}
\end{table}

\begin{table}
\caption{\label{hmasstab62} Bare heavy quark masses and stability 
parameters $n$
used in the runs H, N and P at $\beta=6.2$.
}
\begin{tabular}{lcccccccccc}
$am_Q$  & 6.0 & 4.5 & 4.0 & 2.5 & 2.0 & 1.6 & 1.44& 1.3 & 1.2 & 1.1 \\ \hline
H       &  1  & --  &  1  & --  &  2  & --  & --  &  3  &  3  &  4  \\ 
N       & --  &  1  & --  &  3  & --  & --  &  3  & --  & --  & --  \\
P       & --  & --  & --  & --  & --  &  3  & --  & --  & --  & --  \\
\end{tabular}
\end{table}

\begin{table}
\caption{\label{smeartab} Smearing radii applied at $\beta=5.7$ to the
heavy quark $Q$ and the light quark $q$.}
\begin{tabular}{cccc}
        & $\gsmear{0}$ & $\gsmear{1}$ & $\gsmear{2}$  \\ \hline
$ar_Q$  &   $1.0$      &   $2.0$      &    $3.0$      \\
$ar_q$  &   local      &   local      &    $3.0$      \\
\end{tabular}
\end{table}

\begin{table}
\caption{\label{smeartab62} Smearing radii applied to the
heavy quarks in the run N at $\beta=6.2$. The subscript `g'
denotes a ground state hydrogenic wave function, the `e' an excited state. 
Throughout this run we used local
light quarks.}
\begin{tabular}{lccc}
$am_Q$ &  $\hsmear{g}{1}$ &  $\hsmear{e}{1}$ &  $\hsmear{g}{2}$ \\ \hline
4.5  & $ar_0=5.0$ & -- & --          \\
2.5  & $ar_0=5.0$ & -- & $ar_0=8.0$  \\
1.44 & $ar_0=4.0$ & $ar_0=4.0$ & $ar_0=8.0$ \\
\end{tabular}
\end{table}

\begin{table}
\caption{\label{ainvtab} Determination of the inverse lattice spacing
$\ainv$ from the $\rho$-meson mass
\protect\cite{pdg,rowland,hpsprd55,gockspec}.  The first parenthesis gives the
uncertainties arising from statistical fluctuations, the second the
uncertainty resulting out of the chiral extrapolation. For comparison
we also give the scales as obtained from the string tension $\sigma$
\protect\cite{eichten80,stscri,stteper} and the bottomonium
$\overline{\chi}_b-\Upsilon$ splitting \protect\cite{pdg,Upscale}.
}

\begin{tabular}{ccr@{.}lr@{.}lr@{.}lr@{.}l}
 & &  \multicolumn{4}{c}{$\beta=5.7$} &\multicolumn{4}{c}{$\beta=6.2$} \\
\cline{3-6}\cline{7-10}
Quant. & {phys. [MeV]} &
\multicolumn{2}{c}{lattice}&\multicolumn{2}{c}{$\ainv$ [GeV]} &
\multicolumn{2}{c}{lattice}&\multicolumn{2}{c}{$\ainv$ [GeV]}\\ \hline
$m_\rho$ & 770.0(8)
      &0&690(8)($^{+0}_{-35}$) & 1&116(12)($^{+56}_{-0}$) 
      &0&297($^{+12}_{-7}$)($^{+0}_{-10}$)& 2&59($^{+6}_{-10}$)($^{+9}_{-0}$)\\
$\sqrt{\sigma}$ & $\approx$430 
      &0&3879(39) &  \multicolumn{2}{c}{$\approx$1.10 }
      &0&1608(10) &  \multicolumn{2}{c}{$\approx$2.67 }\\
$\overline{\chi}_b-\Upsilon$ & 440 & 0&311(6) & 1&41(4)(2)(5)&
          0&125(5) & 3&52(14)(4)(0)\\
\end{tabular}
\end{table}

\begin{table}
\caption{\label{swavetab57} Fitted simulation masses for the ground state
pseudo-scalar and vector mesons at $\beta=5.7$ from run A. 
This table has been obtained from double exponential matrix 
fits to correlators with
the smearing functions $\phi_{{\rm G},1}$ and $\phi_{{\rm G},2}$ at
source and sink.}
\begin{tabular}{*{7}{r@{.}l}}
\multicolumn{2}{c}{$am_Q$} &
  \multicolumn{6}{c}{$am_{\rm sim,ps}$} &\multicolumn{6}{c}{$am_{\rm sim,v}$}\\
\cline{3-8}\cline{9-14}
\multicolumn{2}{c}{} &
\multicolumn{2}{c}{$\kappa=0.1380$} &\multicolumn{2}{c}{$\kappa=0.1390$} &\multicolumn{2}{c}{$\kappa=0.1400$} &
\multicolumn{2}{c}{$\kappa=0.1380$} &\multicolumn{2}{c}{$\kappa=0.1390$} &\multicolumn{2}{c}{$\kappa=0.1400$} \\ \hline
 20&0&   0&765(4)&    0&745(4)&   0&724(4)&     0&771(4)&   0&750(4)&   0&730(4)\\    
 12&5&   0&769(3)&    0&748(3)&   0&727(4)&  	0&778(3)&   0&757(3)&  	0&736(4)\\   
 10&0&   0&7715(28)&  0&750(3)&   0&728(4)&  	0&7823(28)& 0&761(3)&  	0&740(4)\\   
  8&0&   0&7733(26)&  0&7518(28)& 0&730(3)&  	0&7868(28)& 0&765(3)&  	0&744(4)\\   
  6&0&   0&7752(23)&  0&7532(27)& 0&731(3)&  	0&7929(26)& 0&7712(28)&	0&749(3)\\   
  5&0&   0&7756(23)&  0&7534(26)& 0&7307(28)&	0&7966(25)& 0&7747(27)&	0&753(3)\\   
  4&0&   0&7753(22)&  0&7526(23)& 0&7296(27)&	0&8009(23)& 0&7787(26)&	0&756(3)\\   
  3&5&   0&7743(22)&  0&7514(23)& 0&7281(27)&	0&8031(23)& 0&7808(26)&	0&7582(28)\\ 
  3&15&  0&7730(21)&  0&7499(23)& 0&7264(26)&	0&8046(23)& 0&7821(26)&	0&7594(28)\\ 
  2&75&  0&7702(21)&  0&7468(23)& 0&7230(26)&	0&8057(23)& 0&7830(25)&	0&7602(28)\\ 
  2&45&  0&7670(20)&  0&7433(22)& 0&7194(25)&	0&8062(23)& 0&7833(25)&	0&7603(28)\\ 
  2&2&   0&7631(20)&  0&7392(22)& 0&7149(25)&	0&8059(23)& 0&7829(25)&	0&7598(28)\\ 
  2&0&   0&7586(20)&  0&7345(22)& 0&7101(23)&	0&8050(23)& 0&7818(25)&	0&7585(28)\\ 
  1&7&   0&7487(19)&  0&7242(21)& 0&6994(23)&	0&8016(23)& 0&7781(25)&	0&7546(28)\\ 
  1&5&   0&7386(19)&  0&7137(21)& 0&6886(23)&	0&7969(23)& 0&7732(26)&	0&749(3)\\   
  1&3&   0&7210(19)&  0&6957(20)& 0&6702(22)&	0&7864(23)& 0&7625(26)&	0&739(3)\\   
  1&125& 0&7000(18)&  0&6743(20)& 0&6484(22)&	0&7731(23)& 0&7490(26)&	0&725(3)\\   
  1&0&   0&6788(18)&  0&6528(19)& 0&6265(21)&	0&7587(23)& 0&7343(27)&	0&710(3)\\   
  0&8&   0&6238(18)&  0&5970(19)& 0&5706(24)&	0&7183(25)& 0&6934(28)&	0&668(3)\\   
  0&6&   0&5267(20)&  0&4989(21)& 0&4709(23)&	0&6445(28)& 0&619(3)&  	0&593(4)\\   
\end{tabular}
\end{table}

\begin{table}
\caption{\label{swavetab62} Fitted simulation masses for the 
pseudo-scalar and vector meson at $\beta=6.2$. The top section gives the 
results for run H, the middle one for N and in the bottom 
we give the result for run P. 
The results for the runs H and N have been obtained from propagators with 
source and sink smearing. In P we used source smearing with local sinks.
}
\begin{tabular}{*{7}{r@{.}l}}
\multicolumn{2}{c}{$am_Q$} &
  \multicolumn{6}{c}{$am_{\rm sim,ps}$} &\multicolumn{6}{c}{$am_{\rm sim,v}$}\\
\cline{3-8}\cline{9-14}
\multicolumn{2}{c}{} &
\multicolumn{2}{c}{$\kappa=0.1346$} &\multicolumn{2}{c}{$\kappa=0.1351$} &\multicolumn{2}{c}{$\kappa=0.1353$} &
\multicolumn{2}{c}{$\kappa=0.1346$} &\multicolumn{2}{c}{$\kappa=0.1351$} &\multicolumn{2}{c}{$\kappa=0.1353$} \\ \hline
6&0 & 0&449(10) & 0&438(12) & 0&419(21) & 0&460(13) & 0&431(20) & 0&415(24)\\
4&0 & 0&443(8)  & 0&425(10) & 0&417(12) & 0&447(11) & 0&426(11) & 0&417(14)\\
2&0 & 0&420(6)  & 0&405(8)  & 0&398(10) & 0&430(7)  & 0&412(9)  & 0&404(12)\\
1&3 & 0&383(5)  & 0&365(6)  & 0&357(7)  & 0&397(6)  & 0&379(8)  & 0&371(9) \\
1&2 & 0&373(5)  & 0&355(5)  & 0&347(7)  & 0&388(5)  & 0&370(8)  & 0&361(9) \\
1&1 & 0&358(4)  & 0&341(5)  & 0&333(7)  & 0&374(5)  & 0&353(8)  & 0&342(10)\\ \hline
4&5 &  0&435(4) & \void{2}  & \void{2}  & 0&440(4) & \void{2}  & \void{2} \\
2&5 &  0&421(4) & \void{2}  & \void{2}  & 0&429(4) & \void{2}  & \void{2} \\
1&44&  0&388(3) & \void{2}  & \void{2}  & 0&400(3) & \void{2}  & \void{2} \\
\hline
1&6 & 0&406(4)  & \void{2}  & \void{2}  & 0&419(4)  & \void{2}  & \void{2} \\
\end{tabular}
\end{table}

\begin{table}
\caption{\label{chextrtab57}Chiral extrapolation at $\beta = 5.7$. The 
first parenthesis gives the statistical uncertainty, the second one the
uncertainty in the respective hopping parameter and, in the case 
of $\kappa_c$, 
the third parenthesis gives the uncertainty arising from the 
chiral extrapolation.}
\begin{tabular}{*{5}{r@{.}l}}
\multicolumn{2}{c}{$am_Q$} &
  \multicolumn{4}{c}{$am_{\rm sim,ps}$} &\multicolumn{4}{c}{$am_{\rm sim,v}$}\\
\cline{3-6}\cline{7-10}
\multicolumn{2}{c}{} &
\multicolumn{2}{c}{$\kappa_c$} &\multicolumn{2}{c}{$\kappa_s$} &
\multicolumn{2}{c}{$\kappa_c$} &\multicolumn{2}{c}{$\kappa_s$} 
\\ \hline
 20&00&  0&656(6)(2)($^{+0}_{-7}$)& 0&726(4)($^{+17}_{-0}$)&       0&663(6)(2)($^{+0}_{-6}$)&  0&732(4)($^{+16}_{-0}$)\\  
 12&50&  0&657(6)(2)($^{+0}_{-7}$)& 0&729(4)($^{+17}_{-0}$)&       0&667(6)(2)($^{+0}_{-6}$)&	0&738(4)($^{+17}_{-0}$)\\  
 10&00&  0&658(5)(2)($^{+0}_{-7}$)& 0&731(4)($^{+17}_{-0}$)&       0&670(5)(2)($^{+0}_{-6}$)&	0&742(4)($^{+17}_{-0}$)\\  
  8&00&  0&658(5)(2)($^{+0}_{-7}$)& 0&732(3)($^{+18}_{-0}$)&       0&673(5)(2)($^{+0}_{-6}$)&	0&746(3)($^{+17}_{-0}$)\\  
  6&00&  0&658(5)(2)($^{+0}_{-7}$)& 0&733(3)($^{+18}_{-0}$)&       0&678(5)(2)($^{+0}_{-5}$)&	0&751(3)($^{+18}_{-0}$)\\  
  5&00&  0&657(4)(2)($^{+0}_{-7}$)& 0&7330(28)($^{+181}_{-0}$)&    0&680(5)(2)($^{+0}_{-5}$)&	0&755(3)($^{+18}_{-0}$)\\  
  4&00&  0&655(4)(2)($^{+0}_{-7}$)& 0&7319(27)($^{+184}_{-0}$)&    0&683(4)(2)($^{+0}_{-5}$)&	0&7586(29)($^{+179}_{-0}$)\\
  3&50&  0&652(4)(2)($^{+0}_{-6}$)& 0&7304(26)($^{+187}_{-0}$)&    0&685(4)(2)($^{+0}_{-5}$)&	0&7605(28)($^{+180}_{-0}$)\\
  3&15&  0&650(4)(2)($^{+0}_{-6}$)& 0&7287(25)($^{+188}_{-0}$)&    0&685(4)(2)($^{+0}_{-4}$)&	0&7617(28)($^{+181}_{-0}$)\\
  2&75&  0&646(4)(2)($^{+0}_{-6}$)& 0&7254(25)($^{+190}_{-0}$)&    0&686(4)(2)($^{+0}_{-4}$)&	0&7625(28)($^{+183}_{-0}$)\\
  2&45&  0&641(4)(2)($^{+0}_{-6}$)& 0&7218(24)($^{+192}_{-0}$)&    0&685(4)(2)($^{+0}_{-4}$)&	0&7626(28)($^{+184}_{-0}$)\\
  2&20&  0&636(4)(2)($^{+0}_{-6}$)& 0&7174(24)($^{+194}_{-0}$)&    0&684(4)(2)($^{+0}_{-4}$)&	0&7621(28)($^{+185}_{-0}$)\\
  2&00&  0&631(3)(2)($^{+0}_{-6}$)& 0&7126(23)($^{+195}_{-0}$)&    0&683(4)(2)($^{+0}_{-4}$)&	0&7609(28)($^{+186}_{-0}$)\\
  1&70&  0&619(3)(2)($^{+0}_{-5}$)& 0&7019(23)($^{+198}_{-0}$)&    0&678(4)(2)($^{+0}_{-4}$)&	0&7569(28)($^{+189}_{-0}$)\\
  1&50&  0&607(3)(2)($^{+0}_{-5}$)& 0&6911(22)($^{+201}_{-0}$)&    0&672(4)(2)($^{+0}_{-4}$)&	0&7519(29)($^{+190}_{-0}$)\\
  1&30&  0&587(3)(2)($^{+0}_{-5}$)& 0&6728(22)($^{+204}_{-0}$)&    0&660(5)(2)($^{+0}_{-4}$)&	0&7409(29)($^{+192}_{-0}$)\\
  1&125& 0&564(3)(2)($^{+0}_{-5}$)& 0&6510(21)($^{+207}_{-0}$)&    0&645(5)(2)($^{+0}_{-4}$)&	0&727(3)($^{+20}_{-0}$)\\  
  1&00&  0&541(3)(2)($^{+0}_{-5}$)& 0&6291(21)($^{+210}_{-0}$)&  0&630(5)(2)($^{+0}_{-4}$)&	0&712(3)($^{+20}_{-0}$)\\  
  0&80&  0&483(3)(3)($^{+0}_{-5}$)& 0&5733(23)($^{+215}_{-0}$)&    0&590(5)(2)($^{+0}_{-3}$)&	0&673(4)($^{+20}_{-0}$)\\  
  0&60&  0&380(3)(3)($^{+0}_{-6}$)& 0&4737(22)($^{+225}_{-0}$)&    0&512(6)(3)($^{+0}_{-4}$)&	0&597(4)($^{+20}_{-0}$)\\  
\end{tabular}
\end{table}

\begin{table}
\caption{\label{chextrtab62}Chiral extrapolation at $\beta = 6.2$. The 
parenthesis gives the statistical uncertainty. The results have been extrapolated to $\kappa_c = 0.135873$ and $\kappa_s = 0.13466$.}
\begin{tabular}{*{5}{r@{.}l}}
\multicolumn{2}{c}{$am_Q$} &
  \multicolumn{4}{c}{$am_{\rm sim,ps}$} &\multicolumn{4}{c}{$am_{\rm sim,v}$}\\
\cline{3-6}\cline{7-10}
\multicolumn{2}{c}{} &
\multicolumn{2}{c}{$\kappa_c$} &\multicolumn{2}{c}{$\kappa_s$} &
\multicolumn{2}{c}{$\kappa_c$} &\multicolumn{2}{c}{$\kappa_s$} 
\\ \hline
6&0 & 0&417(19) & 0&448(10) & 0&383(36) & 0&453(13) \\ 
4&0 & 0&396(16) & 0&439(8)  & 0&392(17) & 0&444(9) \\
2&0 & 0&383(12) & 0&419(6)  & 0&387(17) & 0&429(6) \\
1&3 & 0&341(8)  & 0&381(5)  & 0&352(13) & 0&395(6) \\
1&2 & 0&331(8)  & 0&371(5)  & 0&342(12) & 0&384(6) \\
1&1 & 0&317(8)  & 0&357(5)  & 0&316(15) & 0&370(5) \\
\end{tabular}
\end{table}     

\begin{table}
\caption{\label{shifttab57} Mass shift at $\beta=5.7$.}
\begin{tabular}{*{3}{r@{.}l}}
\multicolumn{2}{c}{$am_Q$} &
\multicolumn{2}{c}{$a\Delta_{\rm rel}$} &
  \multicolumn{2}{c}{$a\Delta_{\rm pert}$} \\ \hline
 20&0&   14&(5)&     18&6(11)\\
 12&5&   9&3(20)&   11&7(7)\\
 10&0&   7&9(13)&    9&4(5)\\
  8&0&   6&7(9)&    7&6(4)\\
  6&0&   5&4(6)&    5&7(3)\\
  5&0&   4&6(4)&     4&83(25)\\
  4&0&   3&96(26)&  3&89(24)\\
  3&5&   3&52(22)&  3&42(22)\\
  3&15&  3&20(19)&  3&10(23)\\
  2&75&  2&85(16)&  2&72(22)\\
  2&45&  2&57(14)&  2&44(21)\\
  2&2&   2&34(12)&  2&21(19)\\
  2&0&   2&16(11)&   2&03(17)\\
  1&7&   1&89(9)&    1&75(10)\\
  1&5&   1&68(6)&    1&56(10)\\
  1&3&   1&51(6)&    1&39(9)\\
  1&125& 1&37(5)&    1&23(7)\\
  1&0&   1&27(5)&    1&14(7)\\
  0&8&   1&16(4)&    1&01(6)\\
  0&6&   1&13(4)&    \void{2}\\
\end{tabular}
\end{table}

\begin{table}
\caption{\label{shifttab62} Mass shift at $\beta=6.2$.}
\begin{tabular}{*{3}{r@{.}l}}
\multicolumn{2}{c}{$am_Q$} &
\multicolumn{2}{c}{$a\Delta_{\rm H}$} &
  \multicolumn{2}{c}{$a\Delta_{\rm pert}$} \\ \hline
6&0&   5&5($^{+5}_{-6}$)   &  5&82(16)\\
4&5&    \void{2}            &  4&40(12) \\
4&0&   4&19($^{+22}_{-31}$) &  3&92(11)\\
2&5&   \void{2}             &  2&49(9)\\
2&0&   2&26($^{+21}_{-26}$) &  2&02(8)\\
1&6&   \void{2}            &  1&64(5) \\
1&44&  \void{2}            &  1&49(4) \\
1&3&   1&28($^{+7}_{-13}$) &  1&36(4)\\
1&2&   1&17($^{+11}_{-12}$) &  1&27(4)\\
1&1&   1&07($^{+14}_{-20}$) &  1&18(3)\\
\end{tabular}
\end{table}

\begin{table}
\caption{\label{deltaperttab} 1-loop coefficient and $q^*$ of the 
perturbative expansion of the mass shift.}
\begin{tabular}{r@{.}lc*{2}{r@{.}l}}
\multicolumn{2}{c}{$am_Q$}& $n$ & \multicolumn{2}{c}{$\Delta^{(1)}$} & 
\multicolumn{2}{c}{$aq_{\Delta}^*$} \\ \hline
  20&00 &  1 &   $-$0&2968(66) &   1&765(71) \\ 
  12&50 &  1 &   $-$0&2605(46) &   1&777(52) \\
  10&00 &  1 &   $-$0&2446(35) &   1&775(45) \\
   7&00 &  1 &   $-$0&1987(37) &   1&807(49) \\
   5&00 &  1 &   $-$0&1522(34) &   1&969(69) \\
   4&00 &  1 &   $-$0&1227(27) &   1&778(56) \\
   4&00 &  2 &   $-$0&1115(23) &   1&686(63) \\
   3&50 &  2 &   $-$0&0889(23) &   1&574(76) \\
   3&00 &  2 &   $-$0&0538(21) &   1&379(90) \\
   2&70 &  2 &   $-$0&0308(20) &   1&27(17)  \\
   2&50 &  2 &   $-$0&0104(23) &   0&43(18)  \\
   2&00 &  2 &    0&0527(23) &   1&25(28)  \\
   1&70 &  2 &    0&1109(25) &   1&743(66) \\
   1&60 &  2 &    0&1346(25) &   1&639(53) \\
   1&50 &  2 &    0&1615(24) &   1&583(42) \\
   1&40 &  3 &    0&2256(25) &   1&858(38) \\
   1&20 &  3 &    0&3276(26) &   1&765(27) \\
   1&00 &  4 &    0&5660(31) &   1&793(20) \\
   0&80 &  5 &    1&0915(42) &   1&805(13)
\end{tabular}
\end{table}

\begin{table}
\caption{\label{strangesplit}Splitting between the strange and
non-strange meson at $\beta=5.7$. The first parenthesis gives the
statistical uncertainty, the second the effect of the different chiral
extrapolations and the third the uncertainty arising from the
$\kappa_s$ determination.}
\begin{tabular}{*{4}{r@{.}l}}
\multicolumn{2}{c}{$am_Q$} &
\multicolumn{2}{c}{$a(m_{{\rm ps},s}-m_{{\rm ps}})$} 
& \multicolumn{2}{c}{$a(m_{{\rm v},s}-m_{{\rm v}})$} & 
\multicolumn{2}{c}{$a(m_{{\rm sav},s}-m_{\rm sav})$} \\ \hline
 20&000&  0&0697(26)($^{+73}_{-0}$)($^{+164}_{-0}$)&  0&0691(27)($^{+63}_{-0}$)($^{+163}_{-0}$)&  0&0693(27)($^{+63}_{-0}$)($^{+163}_{-0}$)\\ 
 12&500&  0&0717(23)($^{+76}_{-0}$)($^{+169}_{-0}$)&  0&0709(23)($^{+61}_{-0}$)($^{+167}_{-0}$)&  0&0711(24)($^{+62}_{-0}$)($^{+167}_{-0}$)\\
 10&000&  0&0727(21)($^{+76}_{-0}$)($^{+171}_{-0}$)&  0&0717(22)($^{+61}_{-0}$)($^{+169}_{-0}$)&  0&0720(22)($^{+62}_{-0}$)($^{+170}_{-0}$)\\
  8&000&  0&0736(20)($^{+76}_{-0}$)($^{+173}_{-0}$)&  0&0725(20)($^{+60}_{-0}$)($^{+171}_{-0}$)&  0&0728(20)($^{+61}_{-0}$)($^{+172}_{-0}$)\\
  6&000&  0&0749(18)($^{+74}_{-0}$)($^{+176}_{-0}$)&  0&0736(19)($^{+57}_{-0}$)($^{+173}_{-0}$)&  0&0739(19)($^{+58}_{-0}$)($^{+174}_{-0}$)\\
  5&000&  0&0758(18)($^{+72}_{-0}$)($^{+179}_{-0}$)&  0&0742(18)($^{+54}_{-0}$)($^{+175}_{-0}$)&  0&0746(18)($^{+56}_{-0}$)($^{+176}_{-0}$)\\
  4&000&  0&0770(16)($^{+70}_{-0}$)($^{+182}_{-0}$)&  0&0751(18)($^{+50}_{-0}$)($^{+177}_{-0}$)&  0&0756(18)($^{+53}_{-0}$)($^{+178}_{-0}$)\\
  3&500&  0&0779(16)($^{+69}_{-0}$)($^{+184}_{-0}$)&  0&0757(18)($^{+48}_{-0}$)($^{+178}_{-0}$)&  0&0762(17)($^{+51}_{-0}$)($^{+180}_{-0}$)\\
  3&150&  0&0786(15)($^{+67}_{-0}$)($^{+185}_{-0}$)&  0&0761(17)($^{+46}_{-0}$)($^{+179}_{-0}$)&  0&0767(17)($^{+49}_{-0}$)($^{+181}_{-0}$)\\
  2&750&  0&0795(15)($^{+65}_{-0}$)($^{+187}_{-0}$)&  0&0767(17)($^{+44}_{-0}$)($^{+181}_{-0}$)&  0&0774(17)($^{+47}_{-0}$)($^{+182}_{-0}$)\\
  2&450&  0&0803(14)($^{+63}_{-0}$)($^{+189}_{-0}$)&  0&0773(18)($^{+42}_{-0}$)($^{+182}_{-0}$)&  0&0780(17)($^{+45}_{-0}$)($^{+184}_{-0}$)\\
  2&200&  0&0811(14)($^{+61}_{-0}$)($^{+191}_{-0}$)&  0&0778(18)($^{+41}_{-0}$)($^{+183}_{-0}$)&  0&0786(17)($^{+44}_{-0}$)($^{+185}_{-0}$)\\
  2&000&  0&0818(13)($^{+59}_{-0}$)($^{+193}_{-0}$)&  0&0783(18)($^{+41}_{-0}$)($^{+185}_{-0}$)&  0&0792(17)($^{+43}_{-0}$)($^{+187}_{-0}$)\\
  1&700&  0&0831(13)($^{+57}_{-0}$)($^{+196}_{-0}$)&  0&0792(18)($^{+40}_{-0}$)($^{+187}_{-0}$)&  0&0802(17)($^{+42}_{-0}$)($^{+189}_{-0}$)\\
  1&500&  0&0842(12)($^{+56}_{-0}$)($^{+198}_{-0}$)&  0&0800(19)($^{+40}_{-0}$)($^{+188}_{-0}$)&  0&0810(17)($^{+42}_{-0}$)($^{+191}_{-0}$)\\
  1&300&  0&0855(12)($^{+56}_{-0}$)($^{+202}_{-0}$)&  0&0808(19)($^{+41}_{-0}$)($^{+190}_{-0}$)&  0&0819(17)($^{+42}_{-0}$)($^{+193}_{-0}$)\\
  1&125&  0&0869(12)($^{+56}_{-0}$)($^{+205}_{-0}$)&  0&0817(19)($^{+42}_{-0}$)($^{+192}_{-0}$)&  0&0830(17)($^{+44}_{-0}$)($^{+196}_{-0}$)\\
  1&000&  0&0882(11)($^{+57}_{-0}$)($^{+208}_{-0}$)&  0&0824(20)($^{+44}_{-0}$)($^{+194}_{-0}$)&  0&0839(17)($^{+45}_{-0}$)($^{+198}_{-0}$)\\
  0&800&  0&0903(12)($^{+55}_{-0}$)($^{+213}_{-0}$)&  0&0827(22)($^{+31}_{-0}$)($^{+195}_{-0}$)&  0&0846(19)($^{+35}_{-0}$)($^{+199}_{-0}$)\\
  0&600&  0&0940(12)($^{+63}_{-0}$)($^{+222}_{-0}$)&  0&0851(24)($^{+38}_{-0}$)($^{+201}_{-0}$)&  0&0874(20)($^{+42}_{-0}$)($^{+206}_{-0}$)\\
\end{tabular}
\end{table}

\begin{table}
\caption{\label{strangesplit6.2}Splitting between the strange and
non-strange meson at $\beta=6.2$. Results are quoted for $\kappa_s$
determined from the $K$ meson. Fixing $\kappa_s$ from the $\phi$ would
lead to an increase by 9\%.  The error bar gives the statistical
uncertainty only. }
\begin{tabular}{*{4}{r@{.}l}}
\multicolumn{2}{c}{$am_Q$} &
\multicolumn{2}{c}{$a(m_{{\rm ps},s}-m_{{\rm ps}})$} 
& \multicolumn{2}{c}{$a(m_{{\rm v},s}-m_{{\rm v}})$} & 
\multicolumn{2}{c}{$a(m_{{\rm sav},s}-m_{\rm sav})$} \\ \hline
6&0 & 0&031(17) & 0&070(29) & 0&049(22) \\
4&0 & 0&043(11) & 0&052(18) & 0&050(15) \\
2&0 & 0&036(8)  & 0&042(12) & 0&041(10) \\
1&3 & 0&039(5)  & 0&042(10) & 0&043(9)  \\
1&2 & 0&040(5)  & 0&042(8)  & 0&043(7)  \\
1&1 & 0&040(5)  & 0&055(12) & 0&051(10) \\
\end{tabular}
\end{table}

\begin{table}
\caption{\label{radial} Radially excited S-wave states at $\beta=5.7$. These 
have been calculated in run S for $\kappa = 0.1400$.}
\begin{tabular}{*{7}{r@{.}l}}
\multicolumn{2}{c}{$am_Q$} & 
   \multicolumn{6}{c}{$am_{\rm sim}(2S)$}& \multicolumn{6}{c}{$am_{\rm sim}(2S)-am_{\rm sim}(1S)$}\\
     \cline{3-8}\cline{9-14}
\multicolumn{2}{c}{}&
  \multicolumn{2}{c}{ps} &   \multicolumn{2}{c}{vector} &   \multicolumn{2}{c}{spin-av} & 
  \multicolumn{2}{c}{ps} &   \multicolumn{2}{c}{vector} &   \multicolumn{2}{c}{spin-av} 
 \\ \hline
  8&0&    1&148(27)& 1&158(26)& 1&155(26)& 0&417(26)&  0&412(25)& 0&413(25)\\
  4&0&    1&21(4)&   1&22(4)&   1&22(4)&   0&48(4)&    0&46(4)&   0&46(3)\\  
  3&15&   1&23(4)&   1&24(4)&   1&24(4)&   0&50(4)&    0&48(4)&   0&48(4)\\  
  2&0&    1&26(5)&   1&27(5)&   1&26(5)&   0&54(5)&    0&51(5)&   0&52(5)\\  
  1&125&  1&23(9)&   1&27(8)&   1&26(8)&   0&58(9)&    0&55(8)&   0&56(8)\\  
  0&8  &  1&17(12)&  1&25(10)&  1&23(10)&  0&60(12)&   0&58(9)&   0&59(9)\\  
\end{tabular}
\end{table}

\begin{table}
\caption{\label{radial6.2} Radially excited S-wave states at
$\beta=6.2$. These have been calculated in run N for $\kappa =
0.1346$. These results are extracted from double exponential
vector-fits to the smeared-local propagators with smearing functions
as listed in the second column.}
\begin{tabular}{r@{.}l c *{6}{r@{.}l}}
\multicolumn{2}{c}{$am_Q$} & smearing &
   \multicolumn{6}{c}{$am_{\rm sim}(2S)$}& \multicolumn{6}{c}{$am_{\rm sim}(2S)-am_{\rm sim}(1S)$}\\
     \cline{4-9}\cline{10-15}
\multicolumn{3}{c}{}&
  \multicolumn{2}{c}{ps} &   \multicolumn{2}{c}{vector} &   \multicolumn{2}{c}{spin-av} & 
  \multicolumn{2}{c}{ps} &   \multicolumn{2}{c}{vector} &   \multicolumn{2}{c}{spin-av} 
 \\ \hline
  2&50  & $\hsmear{g}{1}$, $\hsmear{g}{2}$ &  0&586(32) & 0&586(36) & 0&586(35) & 0&154(34) & 0&145(39) & 0&147(37) \\
  1&44  & $\hsmear{g}{1}$, $\hsmear{e}{1}$ &  0&560(32) & 0&565(34) & 0&564(32) & 0&165(33) & 0&158(35) & 0&160(34)\\
\end{tabular}
\end{table}

\begin{table}
\caption{\label{ptab57} Simulation masses of the $P$-states at
$\beta=5.7$. These have been calculated in run S for $\kappa =
0.1400$. Values given in \textit{italics} are obtained from fits with
low values of $Q$ and disregarded in the further analysis. Therefore
the spin-average has been calculated according to
eq.~(\protect\ref{psaveq57}). The $^3P_1$ and $^1P_1$ operators should both
yield the lightest physical $J^P=1^+$ state.
}
\begin{tabular}{*{8}{r@{.}l}}
\multicolumn{2}{c}{$am_Q$} & 
   \multicolumn{2}{c}{$am(1^3P_0)$}& 
   \multicolumn{2}{c}{$am(1^1P_1)$}& 
   \multicolumn{2}{c}{$am(1^3P_1)$}& 
   \multicolumn{2}{c}{$am(1^3P_2E)$}& 
   \multicolumn{2}{c}{$am(1^3P_2T)$}&
   \multicolumn{2}{c}{$am(1P_{\rm sav})$}&
   \multicolumn{2}{c}{$am(2P_{\rm sav})$}
 \\ \hline
8&0&      1&065(28)&      1&071(26)&  1&072(27)&    1&13(8)&  1&15(7)             &   1&09(7)& 1&57(21)\\
4&0&      1&088(27)&	  1&094(25)&  1&097(26)&    1&09(7)&  1&13(7)             &   1&09(6)& 1&64(20)\\
3&15&     1&094(27)&	  1&103(26)&  1&105(26)&    1&08(7)&  1&12(7)             &   1&09(6)& 1&68(20)\\
2&0&      1&097(27)&      1&115(26)&  1&116(26)&    1&05(7)&  {\it 1}&{\it 11(7)} &   1&09(6)& 1&74(22) \\
1&125&    1&062(28)&	  1&100(26)&  1&094(26)&    1&01(8)&  {\it 1}&{\it 09(7)} &   1&07(6)& 1&73(24) \\
0&8  &    0&995(29)&      1&045(26)&  1&034(25)&    0&97(9)&  {\it 1}&{\it 05(8)} &   1&02(6)& 1&64(24) \\
\end{tabular}
\end{table}

\begin{table}
\caption{\label{ptab62} Simulation masses of the $P$-states at
$\beta=6.2$. These have been calculated in run P for $\kappa =
0.1346$ and $am_Q=1.6$. We report on the $P$-wave ground state and
radially excited state.
The spin-average has been calculated according to
eq.~(\protect\ref{psaveq62}).}
\begin{tabular}{*{7}{r@{.}l}}
   \multicolumn{2}{c}{$am(1^3P_0)$}& 
   \multicolumn{2}{c}{$am(1^1P_1)$}& 
   \multicolumn{2}{c}{$am(1^3P_1)$}& 
   \multicolumn{2}{c}{$am(1^3P_2E)$}& 
   \multicolumn{2}{c}{$am(1^3P_2T)$}&
   \multicolumn{2}{c}{$am(1P_{\rm sav})$}&
   \multicolumn{2}{c}{$am(2P_{\rm sav})$}
 \\ \hline
0&521(17) & 0&560(19) & 0&553(20) & 0&593(22) & 0&588(23) & 0&568(17) & 0&90(8)
\end{tabular}
\end{table}
    	        
\begin{table}
\caption{\label{psplittab57} Splittings of the $P$-states at
$\beta=5.7$. These have been calculated in run S for $\kappa =
0.1400$.}
\begin{tabular}{*{5}{r@{.}l}}
\multicolumn{2}{c}{$am_Q$} & 
   \multicolumn{2}{c}{$am(1P\!-\!1S)_{\rm sav}$} &
   \multicolumn{2}{c}{$am(2P\!-\!1S)_{\rm sav}$} &
   \multicolumn{2}{c}{$am(2P\!-\!1P)_{\rm sav}$} &
     \multicolumn{2}{c}{$am(^3P_{2}E\!-\!{^3P_{0}})$}
 \\ \hline
  8&0&     0&35(7) & 0&83(22) 	& 0&48(19)	  &  0&12(10) \\
  4&0&     0&34(6) & 0&90(20) 	& 0&55(18)	  &  0&03(8)  \\
  3&15&    0&34(6) & 0&93(20) 	& 0&59(18)	  &  0&00(8)  \\
  2&0&     0&35(6) & 0&99(22)	& 0&65(20)	  & $-$0&04(8)  \\
  1&125&   0&36(5) & 1&02(24)	& 0&66(21)	  & $-$0&07(8)  \\
  0&8  &   0&37(6) & 1&00(23)	& 0&63(21)	  & $-$0&05(9)  \\
\end{tabular}		
\end{table}		

\begin{table}
\caption{\label{psplittab62} Splittings of the $P$-states at
$\beta=6.2$. These have been calculated in run P for $\kappa =
0.1346$ and $am_Q = 1.6$.}
\begin{tabular}{*{6}{r@{.}l}}
   \multicolumn{2}{c}{$am(1P\!-\!1S)_{\rm sav}$} &
   \multicolumn{2}{c}{$am(2P\!-\!1S)_{\rm sav}$} &
   \multicolumn{2}{c}{$am(2P\!-\!1P)_{\rm sav}$} &
   \multicolumn{2}{c}{$am(^1P_{1}\!-\! {^3P_{0}})$} &
   \multicolumn{2}{c}{$am(^3P_{1}\!-\! {^3P_{0}})$} &
   \multicolumn{2}{c}{$am(^3P_{2}\!-\! {^3P_{0}})$} \\ \hline
0&152(17) & 0&49(8) & 0&33(8) & 0&039(18) & 0&032(12) & 0&069(25) \\
\end{tabular}		
\end{table}

\begin{table}		
\caption{\label{hpftab57} The hyperfine splitting $m_{\rm hpf}$ at
$\beta = 5.7$.  The directly measured results are obtained from the
difference of the results in table~\ref{swavetab57}. The results for
$\kappa_c$ and $\kappa_s$ are extracted from linear fits to all
three simulation results.  The first parenthesis gives the statistical
error. At $\kappa_c$ the second parenthesis gives the uncertainty of
the chiral extrapolation.  We quote the strange results for $\kappa_s$
from the $K$ and the second parenthesis gives the deviation of the
result for $\kappa_s$ from the $\phi$.}
\begin{tabular}{*{6}{r@{.}l}}
\multicolumn{2}{c}{$am_Q$} &
  \multicolumn{6}{c}{$am_{\rm hpf}$ directly measured} &
  \multicolumn{4}{c}{$am_{\rm hpf}$ extrapolated/interpolated} \\
\cline{3-8} \cline{9-12}
\multicolumn{2}{c}{} &
\multicolumn{2}{c}{$\kappa=0.1380$} &\multicolumn{2}{c}{$\kappa=0.1390$} &\multicolumn{2}{c}{$\kappa=0.1400$} &\multicolumn{2}{c}{$\kappa_c$} &\multicolumn{2}{c}{$\kappa_s$} \\ \hline
 20&0&   0&0054(3)&  0&0056(4)&  0&0058(4)&   0&0060(6)($^{+14}_{-0}$)&  0&0057(4)($^{+0}_{-1}$)\\     
 12&5&   0&0087(4)&  0&0089(5)&  0&0092(5)&   0&0093(8)($^{+21}_{-0}$)&  0&0090(5)($^{+0}_{-1}$)\\ 
 10&0&   0&0108(5)&  0&0110(5)&  0&0114(6)&   0&0115(9)($^{+23}_{-0}$)&  0&0111(6)(0)\\ 
  8&0&   0&0135(5)&  0&0137(6)&  0&0141(7)&   0&0142(10)($^{+24}_{-0}$)& 0&0138(7)($^{+0}_{-1}$)\\ 
  6&0&   0&0177(6)&  0&0180(6)&  0&0185(8)&   0&0187(11)($^{+26}_{-0}$)& 0&0181(7)($^{+0}_{-1}$)\\ 
  5&0&   0&0210(6)&  0&0213(7)&  0&0219(8)&   0&0224(12)($^{+27}_{-0}$)& 0&0216(8)($^{+0}_{-2}$)\\ 
  4&0&   0&0256(7)&  0&0261(8)&  0&0268(9)&   0&0276(13)($^{+28}_{-0}$)& 0&0265(9)($^{+0}_{-3}$)\\ 
  3&5&   0&0288(8)&  0&0294(8)&  0&0301(10)&  0&0312(14)($^{+30}_{-0}$)& 0&0298(10)($^{+0}_{-3}$)\\
  3&15&  0&0316(8)&  0&0322(9)&  0&0330(10)&  0&0343(15)($^{+30}_{-0}$)& 0&0327(10)($^{+0}_{-4}$)\\
  2&75&  0&0355(8)&  0&0363(10)& 0&0372(11)&  0&0387(16)($^{+31}_{-0}$)& 0&0368(11)($^{+0}_{-4}$)\\
  2&45&  0&0392(9)&  0&0400(10)& 0&0410(12)&  0&0427(17)($^{+31}_{-0}$)& 0&0406(11)($^{+0}_{-5}$)\\
  2&2&   0&0429(9)&  0&0437(11)& 0&0448(13)&  0&0468(18)($^{+30}_{-0}$)& 0&0444(12)($^{+0}_{-6}$)\\
  2&0&   0&0464(10)& 0&0473(11)& 0&0484(13)&  0&0506(19)($^{+30}_{-0}$)& 0&0480(12)($^{+0}_{-6}$)\\
  1&7&   0&0529(11)& 0&0539(12)& 0&0552(15)&  0&0578(20)($^{+28}_{-0}$)& 0&0547(13)($^{+0}_{-7}$)\\
  1&5&   0&0584(11)& 0&0595(13)& 0&0609(16)&  0&0639(22)($^{+26}_{-0}$)& 0&0604(14)($^{+0}_{-8}$)\\
  1&3&   0&0655(12)& 0&0668(14)& 0&0683(17)&  0&0717(23)($^{+26}_{-0}$)& 0&0678(16)($^{+0}_{-10}$)\\
  1&125& 0&0731(14)& 0&0746(16)& 0&0763(19)&  0&0803(25)($^{+24}_{-0}$)& 0&0758(17)($^{+0}_{-11}$)\\
  1&0&   0&0799(15)& 0&0816(17)& 0&0833(20)&  0&0879(27)($^{+22}_{-0}$)& 0&0828(19)($^{+0}_{-12}$)\\
  0&8&   0&0945(17)& 0&0965(20)& 0&0998(25)&  0&106(3)($^{+4}_{-0}$)&  	 0&0990(23)($^{+0}_{-15}$)\\
  0&6&   0&1185(24)& 0&1210(26)& 0&124(3)&    0&131(4)($^{+4}_{-0}$)&  	 0&1230(29)($^{+0}_{-18}$)\\
\end{tabular}
\end{table}

\begin{table}
\caption{\label{hpftab62} The hyperfine splitting $m_{\rm hpf}$ at
$\beta = 6.2$.  The results are obtained from ratio-fits to the 
propagators with source and sink smearing.
The parenthesis gives the statistical
uncertainty.}
\begin{tabular}{*{4}{r@{.}l}}
\multicolumn{2}{c}{$am_Q$} &
\multicolumn{2}{c}{$\kappa=0.1346$} &\multicolumn{2}{c}{$\kappa=0.1351$}
 &\multicolumn{2}{c}{$\kappa=0.1353$} \\ \hline
6&0 & $-$0&0016(22) & $-$0&0026(30) & $-$0&004(4) \\
4&0 &  0&0026(23) &  0&0024(30) &  0&0024(33) \\
2&0 &  0&0077(25) &  0&0072(32) &  0&007(4) \\
1&3 &  0&0130(24) &  0&0121(36) &  0&012(4) \\
1&2 &  0&0141(24) &  0&0132(36) &  0&013(4) \\
1&1 &  0&0154(24) &  0&0146(36) &  0&014(4) \\ \hline
4&5 &  0&0040(5)  &  \void{2} & \void{2} \\
2&5 &  0&0070(7)  &  \void{2} & \void{2} \\
1&44&  0&0120(8)  &  \void{2} & \void{2} \\ \hline
1&6 &  0&0123(18) &  \void{2} & \void{2} \\
\end{tabular}\end{table}

\begin{table}
\caption{\label{hqettab} Dependence of the splittings $\Delta m$ on
the spin-averaged meson mass for $\beta = 5.7$. We report the constant
and linear coefficient of the dependence on $1/m_{\rm sav}$. 
For the strange splittings, the number in parenthesis give
the statistical error, the uncertainty from the chiral extrapolation,
the value of $\kappa_s$ and the systematic uncertainty of the
$a$-value. The last column reports the \emph{experimental} slope from
the difference of the splitting in the $B$ and $D$ system. Here we used
the spin-average of the $j_l = \frac{3}{2}$ states for the $P$-state.
Uncertainties which do not apply or have not been evaluated for 
reasons detailed in the text, are marked with (--).
}
\begin{tabular}{c *3{r@{.}l} }
Splitting & \multicolumn{4}{c}{lattice} & \multicolumn{2}{c}{experiment} \\
\cline{2-5} \cline{6-7}
&\multicolumn{2}{c}{$\Delta m(m_{\rm sav}^{-1}= 0)$ in GeV}
& \multicolumn{2}{c}{$\frac{\partial (\Delta m)}{\partial m^{-1}_{\rm sav}}|_{m^{-1}_{\rm sav}= 0}$ in GeV$^2$}
& \multicolumn{2}{c}{$\frac{\partial (\Delta m)}{\partial m^{-1}_{\rm sav}}$ in GeV$^2$}
\\ \hline
$m_{{\rm ps},s}-m_{{\rm ps}}$ &  0&079(3)($^{+7}_{-0}$)($^{+19}_{-0}$)($^{+4}_{-0}$) & 0&037(7)(--)($^{+9}_{-0}$)($^{+4}_{-0}$) & 0&028(7)\\
$m_{{\rm v},s}-m_{{\rm v}}$ & 0&078(3)($^{+6}_{-0}$)($^{+19}_{-0}$)($^{+4}_{-0}$) & 0&027(8)(--)($^{+6}_{-0}$)($^{+3}_{-0}$) & 0&035(12)\\
$m_{{\rm sav},s}-m_{{\rm sav}}$ & 0&079(3)($^{+6}_{-0}$)($^{+19}_{-0}$)($^{+4}_{-0}$) & 0&029(8)(--)($^{+7}_{-0}$)($^{+3}_{-0}$) & 0&033(9)\\ \hline
$m_{{\rm ps},s}(2S)-m_{{\rm ps},s}(1S)$   & 0&40(3)(--)(--)($^{+2}_{-0}$)  & 0&57(16)(--)(--)($^{+6}_{-0}$) & \void{2} \\
$m_{{\rm v},s}(2S)-m_{{\rm v},s}(1S)$     & 0&40(3)(--)(--)($^{+2}_{-0}$)  & 0&42(16)(--)(--)($^{+4}_{-0}$) & \void{2} \\
$m_{{\rm sav},s}(2S)-m_{{\rm sav},s}(1S)$ & 0&40(3)(--)(--)($^{+2}_{-0}$)  & 0&45(15)(--)(--)($^{+5}_{-0}$) & \void{2} \\ \hline
$m_{{\rm sav},s}(1P)-m_{{\rm sav},s}(1S)$ & 0&36(8)(--)(--)($^{+2}_{-0}$)  & 0&07(12)(--)(--)($^{+1}_{-0}$) & 0&11(5) \\ \hline
$m_{\rm v}-m_{\rm ps}$        &  $-$0&0001(7)($^{+5}_{-0}$)(--)(0) & 0&151(10)($^{+16}_{-0}$)(--)($^{+16}_{-0}$) & 0&297(1) \\
$m_{\rm v,s}-m_{\rm ps,s}$    &  $-$0&0002(4)(--)(0)(0) & 0&144(6)(--)($^{+0}_{-2}$)($^{+15}_{-0}$) & 0&304(17)\\
\end{tabular}
\end{table}

\begin{table}
\caption{\label{Bsplittab} Meson masses and splittings in the $B$ system for $\beta=5.7$. Overlines denote spin-averaged states.}
\begin{tabular}{l r@{.}l cccccc r@{.}l}
Splitting & \multicolumn{2}{c}{Value} & \multicolumn{6}{c}{Uncertainties}                                   &\multicolumn{2}{c}{Experiment}\\ 
\cline{4-9}
&\multicolumn{2}{c}{}  & stat & chiral          &  strange & shift & $a$-stat & $a$-chiral & \multicolumn{2}{c}{}\\ \hline \hline
$B_s-B$  &    85&6~MeV   & (20) & ($^{+78}_{-0}$) & ($^{+202}_{-0}$) &(3)&(9) & ($^{+46}_{-0}$)& 90&2(22)~MeV \\
$B^*_s -B^*$& 83&6~MeV   & (20) & ($^{+56}_{-0}$) & ($^{+198}_{-0}$) &(2)&(9) & ($^{+44}_{-0}$)& 91&4(38)~MeV \\
$\bar B_{s} - \bar B$& 84&1~MeV & (20) & ($^{+59}_{-0}$) & ($^{+200}_{-0}$) &(2)&(9)& ($^{+45}_{-0}$)& 91&(3)~MeV \\ \hline
$B^{(*)}(2S)-B^{(*)}(1S)$ & \void{2}     & --- & ---          &  ---     & ---   & --- & ---             & 580&(10)~MeV \protect\cite{bradref1,bradref2}\\
$B_s(2S)-B_s(1S)$ & 526&~MeV & (38)& ---          &  ---     & (7)   & (7) & ($^{+34}_{-0}$) &  \void{2} \\
$B_s^*(2S)-B_s^*(1S)$ & 509&~MeV & (38)& ---          &  ---     & (6)   & (7) & ($^{+32}_{-0}$) &  \void{2} \\
$\bar B_s(2S)-\bar B_s(1S)$ & 513&~MeV & (37)& ---          &  ---     & (6)   & (7) & ($^{+33}_{-0}$) &  \void{2} \\ \hline
$B^*_{sJ}(5850) - \bar B_s(1S)$ & \void{2}     & --- & ---          &  ---     & ---   & --- & ---             & 448&(15)~MeV \\
$\bar B_s(1P) - \bar B_s(1S)$ & 385&~MeV & (70) &---&---& (0) & (4)& ($^{+19}_{-0}$) & \void{2} \\ \hline
$\bar B_s(2P) - \bar B_s(1P)$ & 610&~MeV & (200) &---&---& (9) & (6) & ($^{+40}_{-0}$) & \void{2} \\ \hline
$B^*-B$  &    29&5~MeV   & (15) & ($^{+31}_{-0}$) &   ---    & (16)  & (6) & ($^{+31}_{-0}$) & 45&78(35)~MeV\\     
$B^*_s - B_s$ & 28&3~MeV & (10) & ---      &  ($^{+0}_{-3}$) & (15)  & (6) & ($^{+30}_{-0}$) & 47&0(26)~MeV\\ \hline
$B^*_{s2} - B^*_{s0}$ & 41&~MeV & (94) & --- & --- & (11) & (3) & ($^{+14}_{-0}$) & \void{2} \\
\end{tabular}
\end{table}

\begin{table}
\caption{\label{Bsplittab6.2} Meson masses and splittings in the $B$
system for $\beta=6.2$.  
The radially excited $S$-wave states are
extracted from fits with low $Q$-values compared to the other results. 
We quote them in italics.}
\begin{tabular}{l r@{.}l ccccc r@{.}l}
Splitting & \multicolumn{2}{c}{Value} & \multicolumn{5}{c}{Uncertainties}                                   &\multicolumn{2}{c}{Experiment}\\ 
\cline{4-8}
\multicolumn{3}{c}{ }           & stat &  strange & shift & $a$-stat & $a$-chiral & \multicolumn{2}{c}{ }\\ \hline \hline
$B_s-B$  &     96&~MeV          & (17) & ($^{+9}_{-0}$)  & --- &($^{+3}_{-4}$) & ($^{+4}_{-0}$)& 90&2(22)~MeV \\
$B^*_s -B^*$& 109&~MeV          & (26) & ($^{+10}_{-0}$) & --- &($^{+3}_{-4}$) & ($^{+4}_{-0}$)& 91&4(38)~MeV \\
$\bar B_{s} - \bar B$& 109&~MeV & (23) & ($^{+10}_{-0}$) & --- &($^{+3}_{-5}$) & ($^{+4}_{-0}$)& 91&(3)~MeV \\ \hline
$B^{(*)}(2S)-B^{(*)}(1S)$ & \void{2} & --- &  ---  & ---  & --- & ---  & 580&(10)~MeV \protect\cite{bradref1,bradref2}\\
$B_s(2S)-B_s(1S)$           & {\it 420}&~MeV & (85)& ---  & (3) & ($^{+12}_{-20}$) & ($^{+17}_{-0}$) &  \void{2} \\
$B_s^*(2S)-B_s^*(1S)$       & {\it 400}&~MeV & (90)& ---  & (3) & ($^{+12}_{-20}$) & ($^{+17}_{-0}$) &  \void{2} \\
$\bar B_s(2S)-\bar B_s(1S)$ & {\it 405}&~MeV & (90)& ---  & (3) & ($^{+12}_{-20}$) & ($^{+17}_{-0}$) &  \void{2} \\ \hline
$B^*_{sJ}(5850) - \bar B_s(1S)$ & \void{2}     & --- & ---              & ---   & --- & ---             & 448&(15)~MeV \\
$\bar B_s(1P) - \bar B_s(1S)$ & 395&~MeV & (45) &---&---& ($^{+9}_{-15}$) & ($^{+14}_{-0}$) & \void{2} \\ \hline
$\bar B_s(2P) - \bar B_s(1P)$ & 855&~MeV & (210) &---&---& ($^{+20}_{-33}$) & ($^{+30}_{-0}$) & \void{2} \\ \hline
$B^*_s - B_s$ & 27&3~MeV         & (20)  & ---             & (8) & ($^{+15}_{-22}$)  & ($^{+22}_{-0}$)& 47&0(26)~MeV \\ \hline 
$B^*_{s2} - B^*_{s0}$ & 179&~MeV & (65) & --- & --- & ($^{+4}_{-7}$) & ($^{+6}_{-0}$) & \void{2} \\
\end{tabular}
\end{table}

\begin{table}
\caption{\label{Bsplittabsum} Summary of the results on the $B$-meson
spectrum.  The table gives the average of our results and the result
of \protect\cite{arifalat98}. Errors exclude quenching effects but
include residual lattice spacing artifacts of ${\cal
O}(\alpha_sa,a^2)$. Again overlines denote spin-averaged states.  }
\begin{tabular}{l r@{.}l r@{.}l}
Splitting                        & \multicolumn{2}{c}{Value} &\multicolumn{2}{c}{Experiment}\\ \hline \hline
$B_s-B$                          & 90&(10)~MeV               & 90&2(22)~MeV \\   
$B^*_s -B^*$                     & 90&(10)~MeV  	     & 91&4(38)~MeV \\
$\bar B_{s} - \bar B$            & 90&(10)~MeV               & 91&(3)~MeV \\ \hline
$B^{(*)}(2S)-B^{(*)}(1S)$        & \void{2}                  & 580&(10)~MeV \protect\cite{bradref1,bradref2}\\
$B(2S)-B(1S)$                    & 600&(90)~MeV              & \void{2} \\ \hline
$B_s(2S)-B_s(1S)$                & 540&(60)~MeV              & \void{2} \\
$B_s^*(2S)-B_s^*(1S)$            & 525&(80)~MeV              & \void{2} \\
$\bar B_s(2S)-\bar B_s(1S)$      & 530&(80)~MeV              & \void{2} \\ \hline
$B^*_{J}(5732) - \bar B(1S)$     & \void{2}                  & 385&(12)~MeV \\
$\bar B(1P) - \bar B(1S)$        & 455&(50)~MeV              & \void{2} \\ \hline
$B^*_{sJ}(5850) - \bar B_s(1S)$  & \void{2}                  & 448&(15)~MeV \\
$\bar B_s(1P) - \bar B_s(1S)$    & 411&(45)~MeV              & \void{2} \\ \hline
$\bar B_s(2P) - \bar B_s(1P)$    & 730&(200)~MeV             & \void{2} \\ \hline
$B^*-B$                          & 29&(5)~MeV                & 45&78(35)~MeV\\      
$B^*_s - B_s$                    & 28&5(31)~MeV		     & 47&0(26)~MeV\\ \hline
$B^*_{2} - B^*_{0}$              & \multicolumn{2}{c}{0 -- 250~MeV} & \void{2} \\ 
$B^*_{s2} - B^*_{s0}$            & \multicolumn{2}{c}{0 -- 250~MeV} & \void{2} \\ 
\end{tabular}
\end{table}

\begin{table}
\caption{\label{Dsplittab} Meson masses and splittings in the $D$
system for $\beta=5.7$. Overlines denote spin-averaged states.}
\begin{tabular}{l r@{.}l cccccc r@{.}l}
Splitting & \multicolumn{2}{c}{Value} & \multicolumn{6}{c}{Uncertainties}                                   &\multicolumn{2}{c}{Experiment}\\ 
\cline{4-9}
\multicolumn{3}{c}{ }            & stat & chiral          &  strange & shift & $a$-stat & $a$-chiral & \multicolumn{2}{c}{ }\\ \hline \hline
$D_s-D$  &    99&9~MeV   & (13) & ($^{+62}_{-0}$) & ($^{+234}_{-0}$) &(8)&(14) & ($^{+69}_{-0}$)& 99&2(5)~MeV \\ 
$D^*_s-D^*$ & 92&2~MeV   & (23) & ($^{+40}_{-0}$) & ($^{+218}_{-0}$) &(1)&(10) & ($^{+50}_{-0}$)& 102&4(9)~MeV \\
$\bar D_s-\bar D$ & 94&1~MeV & (20) & ($^{+44}_{-0}$) & ($^{+222}_{-0}$) &(2)&(11) & ($^{+53}_{-0}$)& 101&6(8)~MeV \\ \hline
$D^*(2S)-D^*(1S)$ & \void{2}     & --- & ---          &  ---     & ---   & --- & ---             & 629&(2)(6)~MeV \cite{Dradial}\\
$D_s(2S)-D_s(1S)$ & 665&~MeV & (130)& ---          &  ---     & (5)   & (9) & ($^{+45}_{-0}$) &  \void{2} \\
$D_s^*(2S)-D_s^*(1S)$ & 640&~MeV & (80)& ---          &  ---     & (9)   & (11) & ($^{+54}_{-0}$) &  \void{2} \\
$\bar D_s(2S)-\bar D_s(1S)$ & 645&~MeV & (100)& ---          &  ---     & (8)   & (11) & ($^{+51}_{-0}$) &  \void{2} \\ \hline
$\bar D^{**}_{s}(j_l=\frac{3}{2}) - \bar D_s $ & \void{2}     & --- & ---          &  ---     & ---   & --- & ---             & 483&(1)~MeV \\
$\bar D_s(1P) - \bar D_s(1S)$ & 411&~MeV & (61) &---&---& (3) & (4) & ($^{+27}_{-0}$) & \void{2} \\ \hline
$\bar D_s(2P) - \bar D_s(1P)$ & 710&~MeV & (230) &---&---& (7) & (7) & ($^{+36}_{-0}$) & \void{2} \\ \hline
$D^*-D$     & 110&~MeV           & (3) & ($^{+3}_{-0}$) &   ---    & (6)  & (5) & ($^{+22}_{-0}$) & 140&64(10)~MeV \\
$D^*_s-D_s$ & 103&~MeV           & (2) & ---      &  ($^{+0}_{-2}$)& (6)  & (4) & ($^{+20}_{-0}$) & 143&8(4)~MeV \\     
\end{tabular}
\end{table}

\newpage

\newpage

\newpage

\newpage

\newpage

 \newpage

\newpage

\newpage

\newpage

\newpage

\newpage

\newpage

\newpage

\newpage

\begin{figure}
\centerline{\epsfig{file=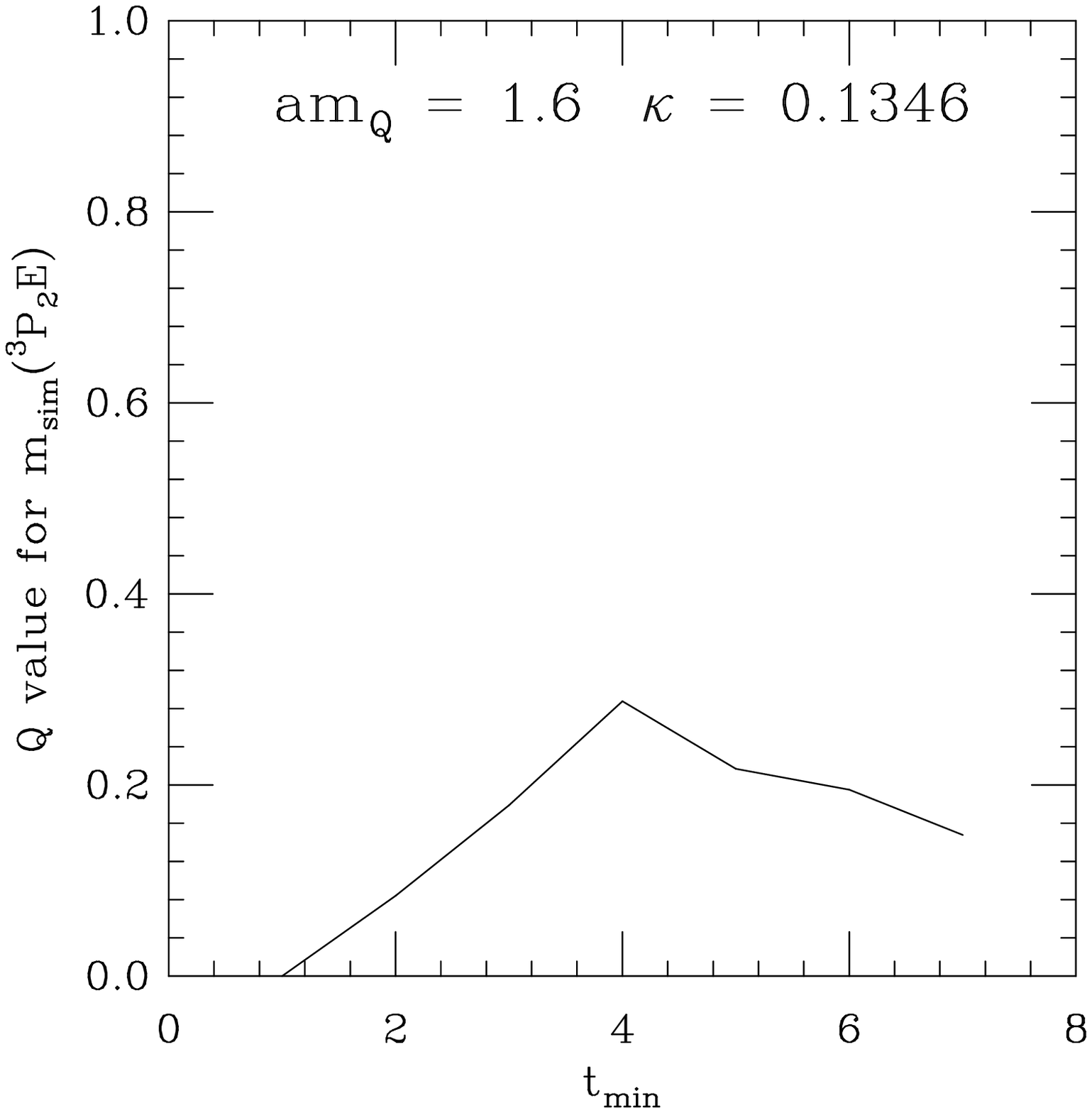,width=8cm}
\epsfig{file=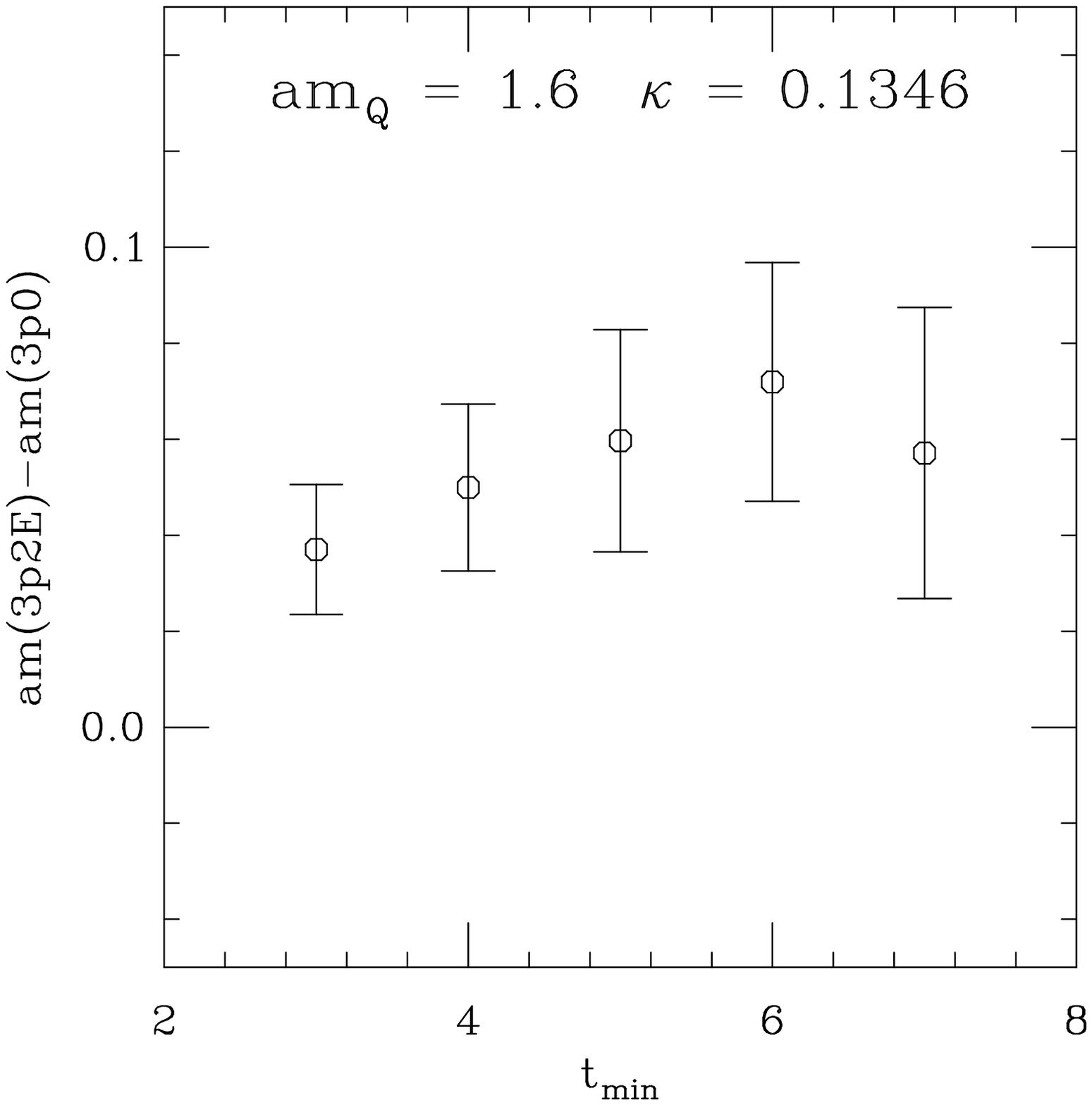,width=8cm}}
\caption{\label{pfinetminplot62} Fit range dependence of the
$m(^3P_2E)-m(^3P_0)$ $P$-state fine structure from the double exponential
vector fit at $\beta=6.2$. The left hand side displays the $Q$-value of
the fit to the $^3P_2E$-state. The fit to the $^3P_0$-state gives
even higher $Q$-values. The right hand side gives the splitting
as determined from the difference of the individual fit results.}
\end{figure}\newpage

\begin{figure}
\centerline{\epsfig{file=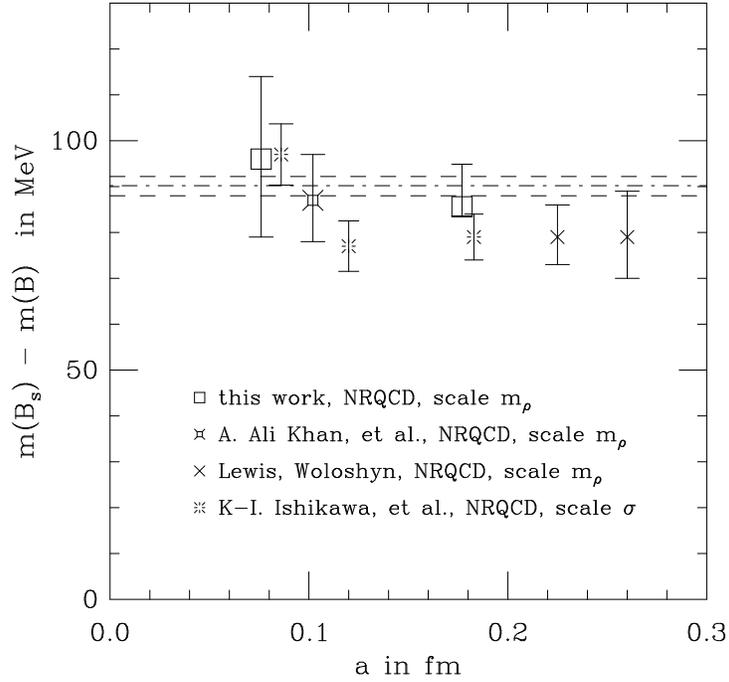,width=10cm}}
\caption{\label{bscalefig} Compilation of results for the
$B_s-B$ splitting.
Additional results are from
\protect\cite{arifalat98,lewiswolo,ishikawa}.  Results are for 
$\kappa_s$ fixed from $K/\rho$ ratio, apart from the crosses. The latter
use the $K^*/K$ ratio to fix $\kappa_s$, which tends to
shift them downwards. Please
note the bursts and crosses do not contain all sources of uncertainty
included in the squares and fancy squares.  The horizontal lines give
the experimental result from \protect\cite{pdg}.}
\end{figure}\newpage

\begin{figure}
\centerline{\epsfig{file=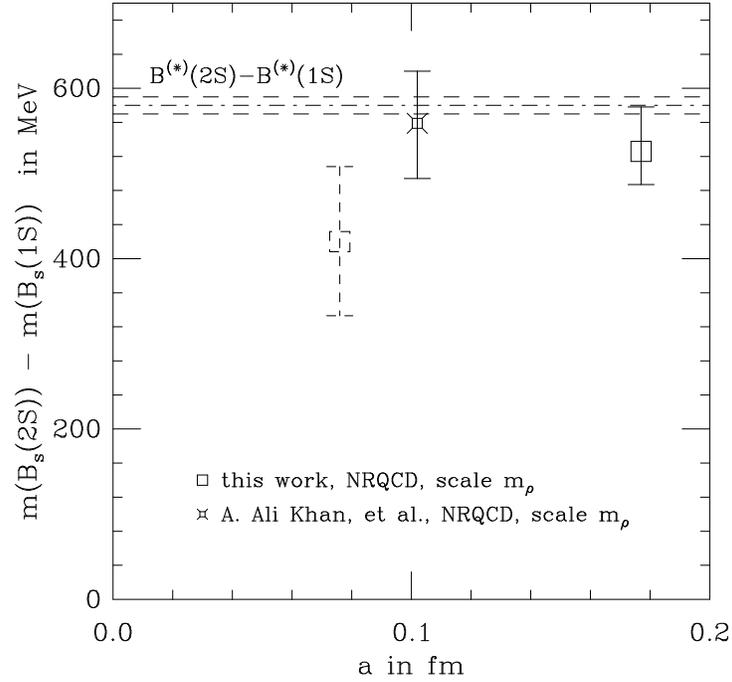,width=10cm}}
\caption{\label{bradscalefig}Scaling of the $B'_s - B_s$
splitting. The squares give our results and the fancy square the one
from \protect\cite{arifalat98}. On our finest lattice the extraction of
the result turned out to be substantially more difficult than elsewhere,
so we give this result with dashed lines, see text.
The horizontal lines give a preliminary experimental result for an 
admixture of the non-strange $B'-B$ and $B^*{}'-B^*$ splitting from 
the DELPHI-collaboration \protect\cite{bradref1,bradref2} for comparison.  }
\end{figure}\newpage

\begin{figure}
\centerline{\epsfig{file=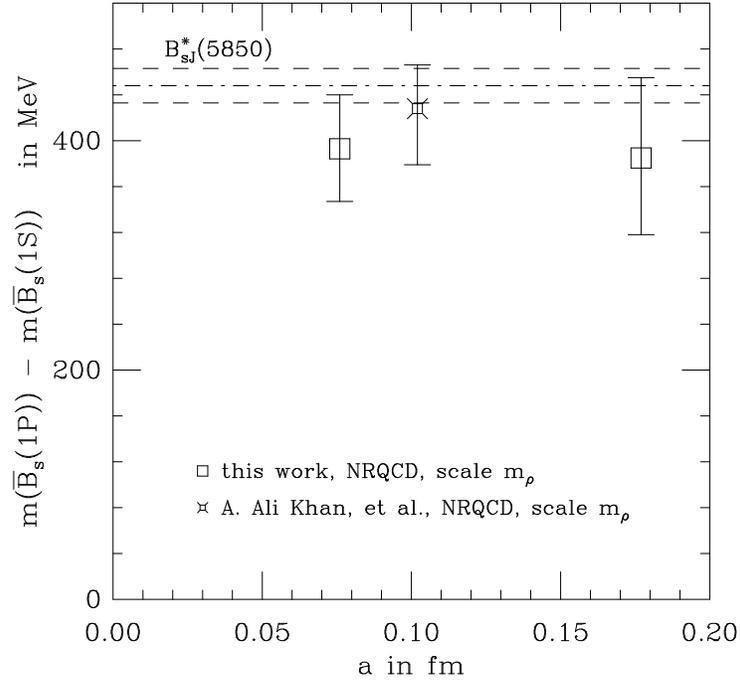,width=10cm}}
\caption{\label{borbscalefig} Scaling of the spin-averaged $\bar
B_s(1P)-\bar B_s(1S)$ splitting. We include the result of
\protect\cite{arifalat98}.  The horizontal lines give the
$B^*_{sJ}(5850)$ resonance, which is expected to be a superposition of
the two $j_l = \frac{3}{2}$ states.}
\end{figure}\newpage

\begin{figure}
\centerline{\epsfig{file=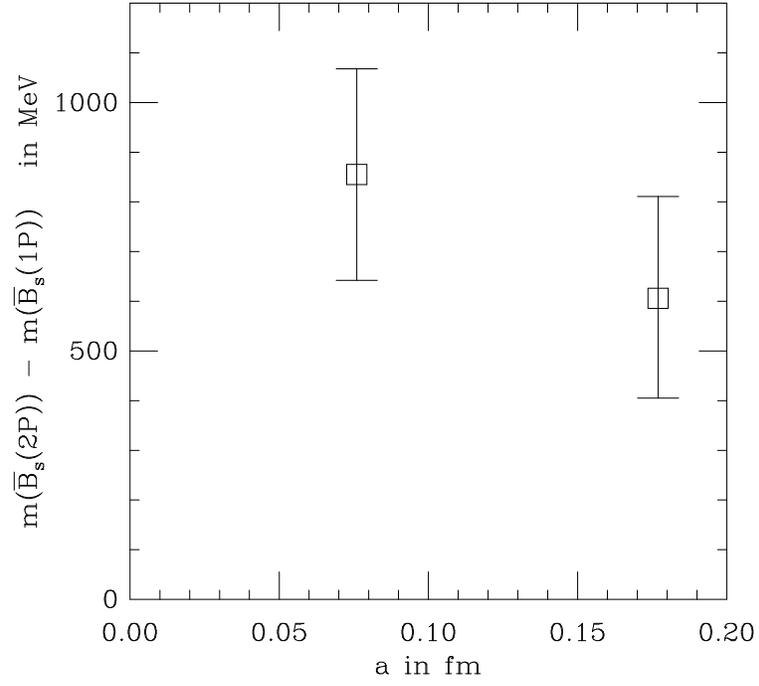,width=10cm}}
\caption{\label{borbradscalefig} The spin-averaged $\bar B_s(2P)-\bar
B_s(1P)$ splitting for two different values of the lattice spacing
$a$.}
\end{figure}\newpage

\begin{figure}
\centerline{\epsfig{file=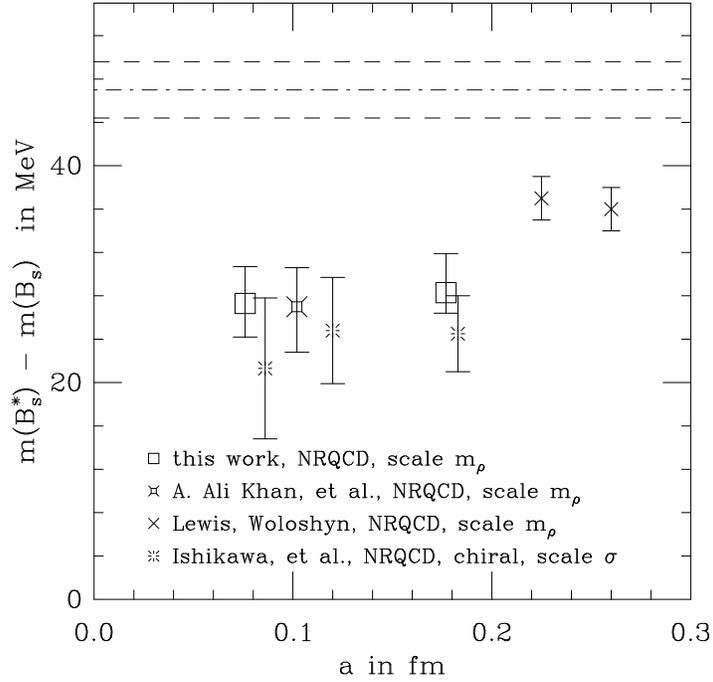,width=10cm}}
\caption{\label{bhypscalefig}Results for the hyperfine splitting
$B^*_s-B_s$ for different $a$-values. Results from
\protect\cite{arifalat98,lewiswolo,ishikawa} are included into the
plot. The bursts give results for the chirally extrapolated $B^*-B$
splitting.  Their error bar gives only statistical errors. Crosses
omit the uncertainties from the determination of the bare $b$-mass.
The horizontal lines give the experimental result from
\protect\cite{pdg}.  }
\end{figure}\newpage

\begin{figure}
\centerline{\epsfig{file=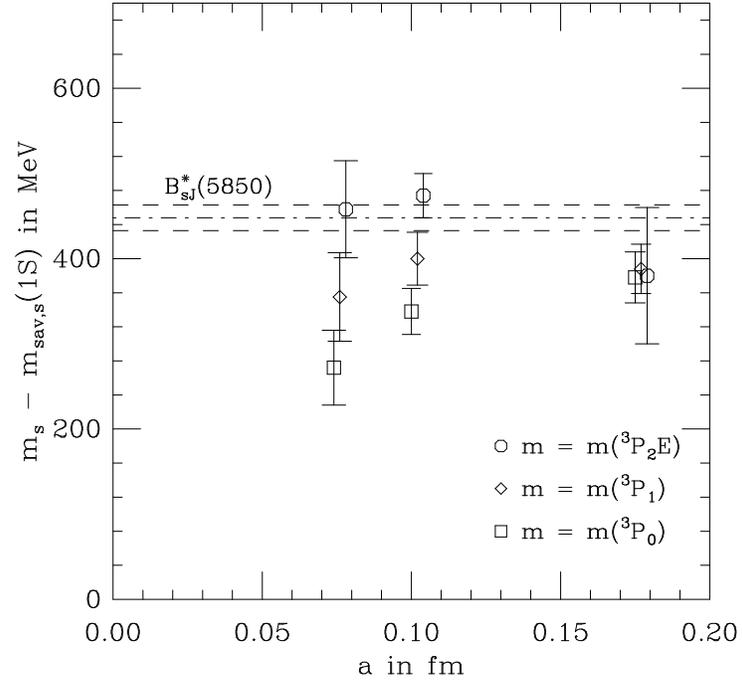,width=10cm}}
\caption{\label{pfinescale} Scaling of the $P$-state fine structure of
the $B_s$. Results at $a=0.102$~fm are from
\protect\cite{arifalat98}. The squares and octagons are displaced for
clarity. }
\end{figure}\newpage

\begin{figure}
\centerline{\epsfig{file=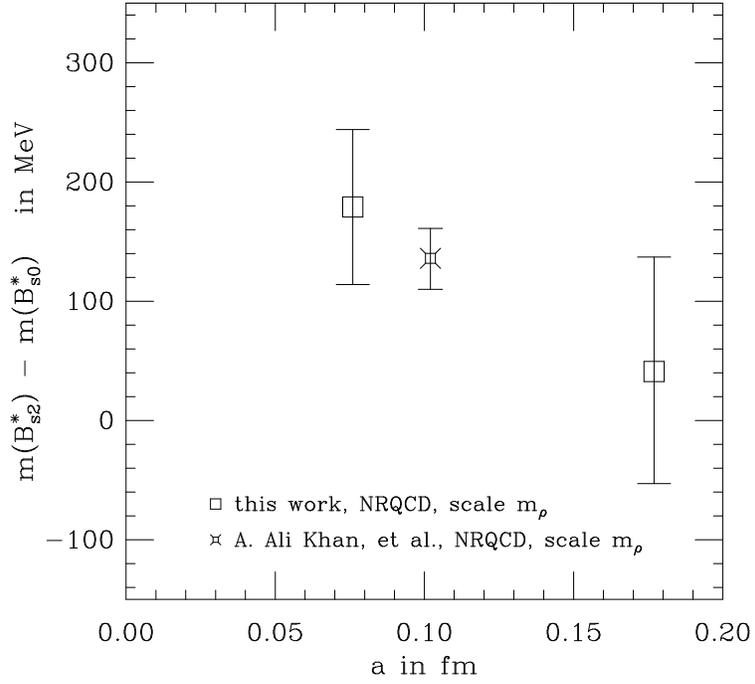,width=10cm}}
\caption{\label{bpfinescalefig}Results for the 
$P$-state fine structure $B^*_{s2}-B^*_{s0}$ for three different
$a$-values. The middle point has been taken from
\protect\cite{arifalat98}. Experimentally
this splitting is so far unobserved.}
\end{figure}\newpage

\begin{figure}
\centerline{\epsfig{file=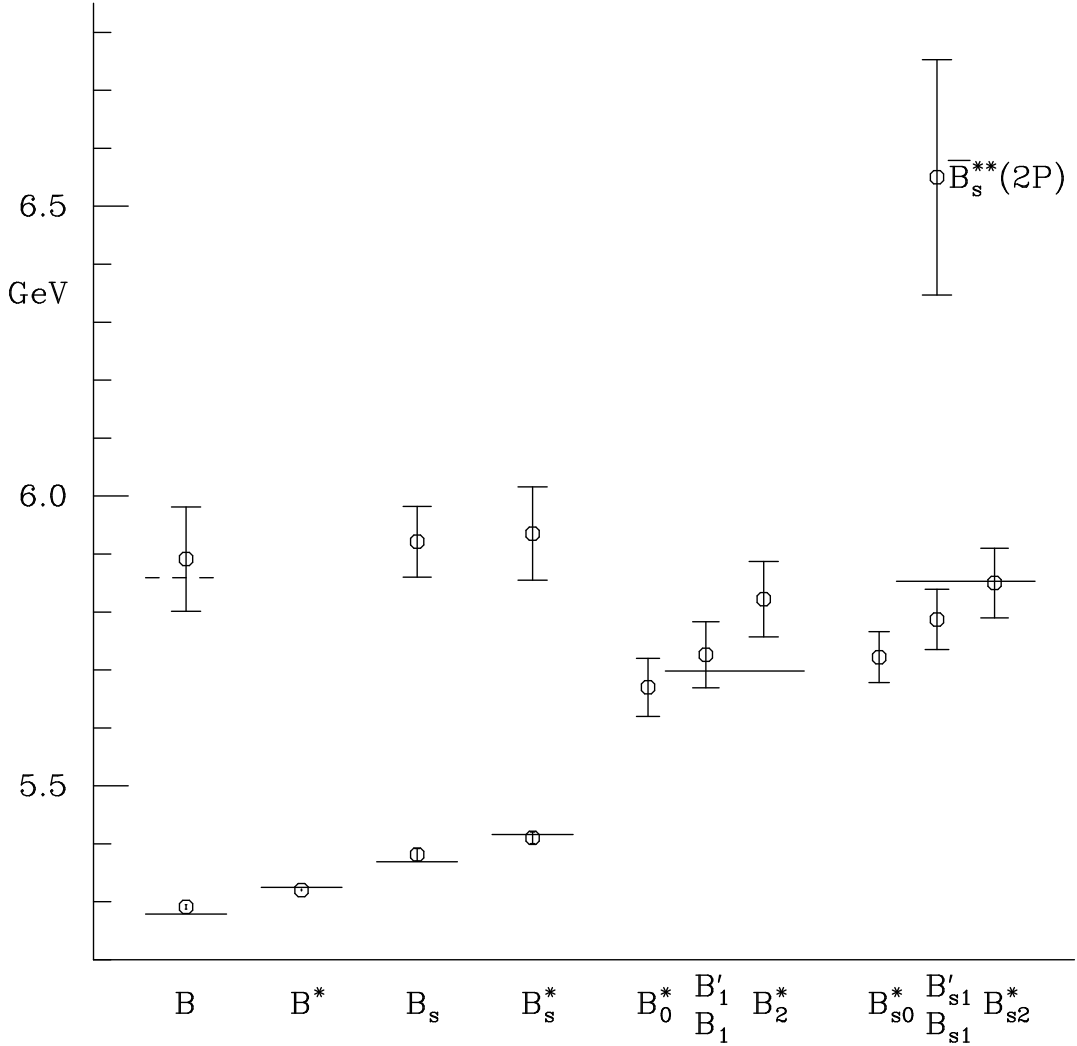,width=15cm,
}}
\caption{\label{bspectrumfig}Spectrum of $B$-mesons, summarising our
results and those of \protect\cite{arifalat98}. As before, the lattice
results, given by the octagons, give the splitting with respect to the
spin-averaged $1S$-state $\bar B$.  The experimental results from
\protect\cite{pdg} are included by horizontal lines.  The dashed line
displays a result from the DELPHI collaboration
\protect\cite{bradref1,bradref2}, interpreted to be the $B^{(*)}{}'$.
The $P$ states are compared to the experimental result for the
$B_J^*(5732)$ and $B_{sJ}^*(5850)$.}
\end{figure}\newpage

\begin{figure}
\centerline{\epsfig{file=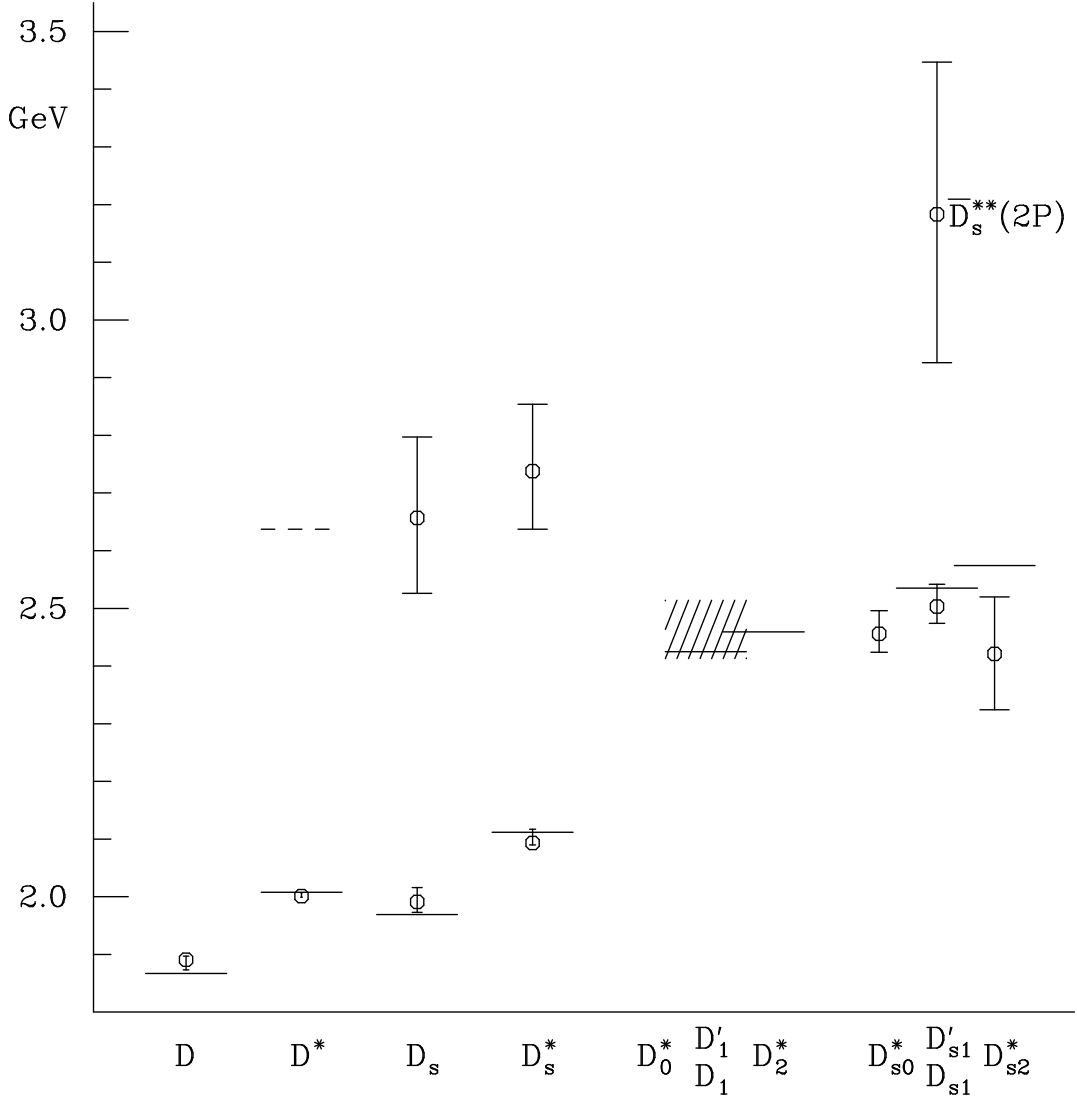,width=15cm,
}}
\caption{\label{dspectrumfig}Spectrum of $D$-mesons from our results
at $\beta=5.7$. Lattice results are given by octagons, experimental
results from \protect\cite{pdg} by horizontal lines. The lattice results
give the splitting with respect to the spin-average $\bar D$ of the $1S$-wave.
The dashed line
displays a result from the DELPHI collaboration
\protect\cite{Dradial}, interpreted to be the ${D^*}'$.  There are two
non-degenerate $P$-states with $J^P$ quantum numbers $1^+$
corresponding to the $j_l=\frac{1}{2}$ and $\frac{3}{2}$ state.  We
denote these by $D_1$ and $D'_1$ respectively and similarly for the $D_s$.  The
CLEO collaboration reported preliminary results for the $D_1$
corresponding to $j_l=\frac{1}{2}$ \protect\cite{D1wide,D1wide2}.  The
shaded area gives this result.}

\end{figure}\newpage

\begin{figure}
\centerline{\epsfig{file=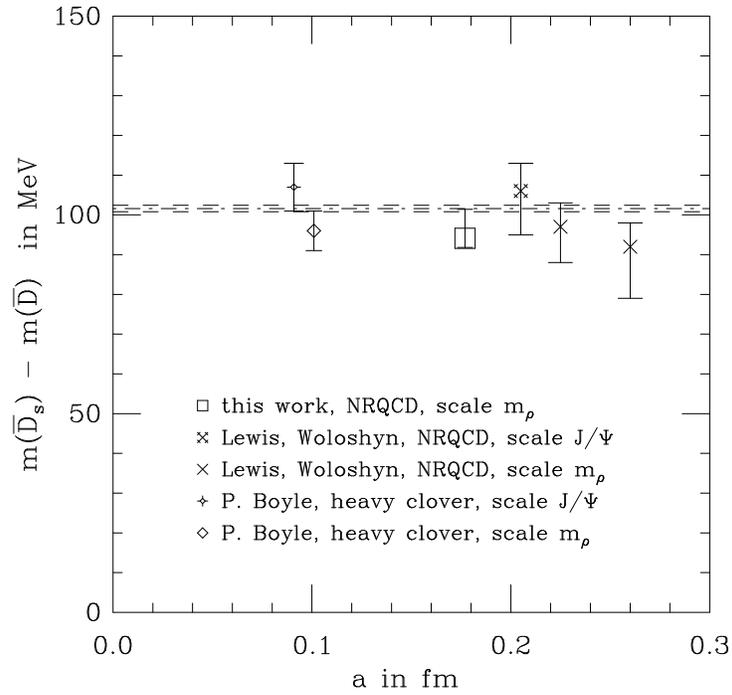,width=10cm}}
\caption{\label{dscalefig} Comparison of recent lattice results for
the strange to non-strange splitting for the spin-average of the $D$
and $D^*$-mesons. The square and both of the diamonds
\protect\cite{peterhl} use the $K/\rho$ mass-ratio to fix $\kappa_s$,
whereas the crosses \protect\cite{lewiswolo} use the $K^*/K$ ratio.
The horizontal line gives the experimental result from
\protect\cite{pdg}.}
\end{figure}\newpage

\begin{figure}
\centerline{\epsfig{file=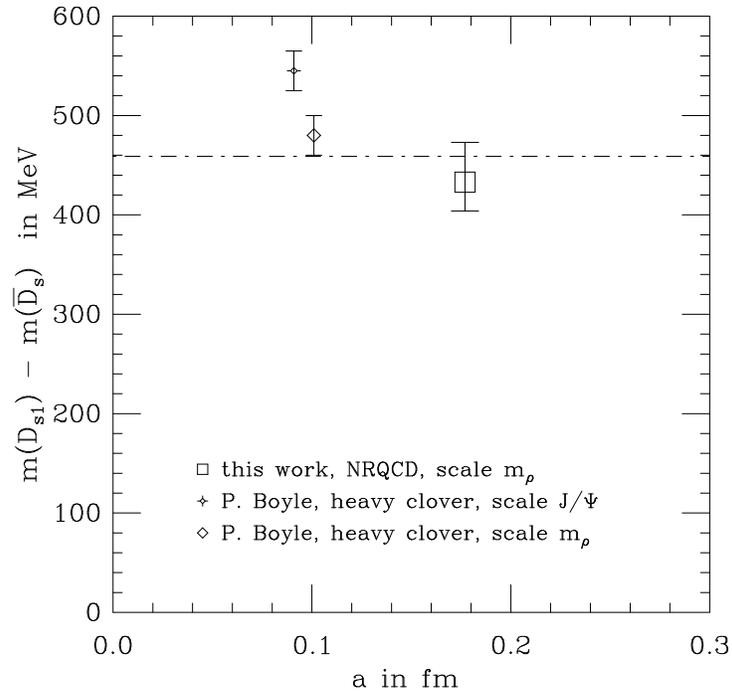,width=10cm}}
\caption{\label{dpscalefig} Comparison of the $D_{s1} - \bar D_s$ to
the result of \protect\cite{peterhl}. The experimental result for the
$D_{s1}$ is again from \protect\cite{pdg}.}
\end{figure}\newpage

\begin{figure}
\centerline{\epsfig{file=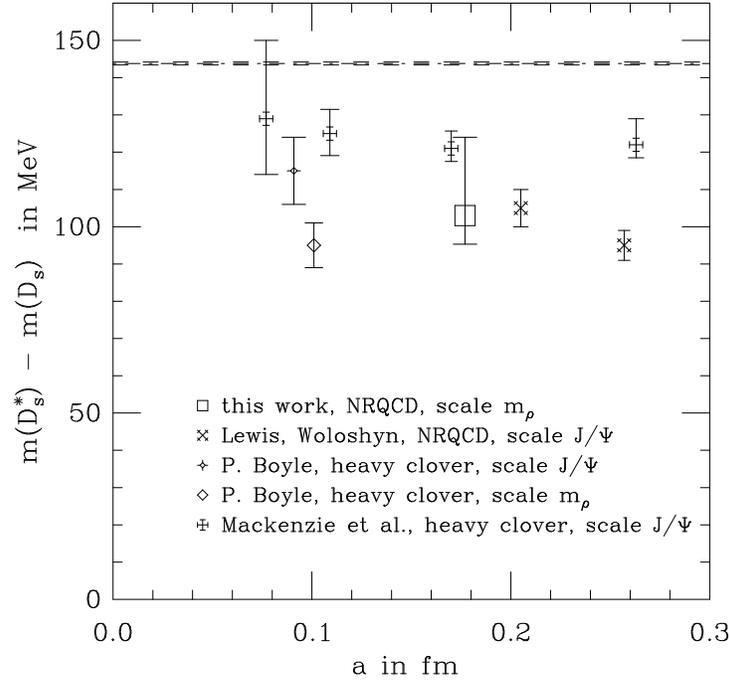,width=10cm}}
\caption{\label{dhypscalefig} Comparison of our $D_s$-meson hyperfine
splitting to the findings of
\protect\cite{lewiswolo,peterhl,fermispec}. Please note that these
other results do not necessarily include all the sources of
uncertainty, that we have included, see
table~\protect\ref{Dsplittab}. They therefore may have underestimated
error bars. Again, we quote the experimental result from
\protect\cite{pdg}. See text for a discussion.}
\end{figure}\newpage

\end{document}